Strong Fiber from Uniaxial Fullerene Supramolecules Aligned with Carbon Nanotubes

John Bulmer[1,2], Michelle Durán-Chaves[3], Daniel M. Long[4,5], Jeremiah Lipp[4,5], Steven Williams[3], Mitchell Trafford[3], Anthony Pelton[4,5], Jared Shank[4,5], Benji Maruyama[4], Larry Drummy[4], Matteo Pasquali[3], Hilmar Koerner[4], Timothy Haugan[1]

1 Aerospace Systems Directorate, Air Force Research Laboratory, Wright-Patterson Air Force Base, Ohio, USA, 45433

2 National Research Council, Washington, D.C. 20001, USA

3 Department of Chemical & Biomolecular Engineering, Department of Chemistry, Department of Materials Science & NanoEngineering, The Smalley-Curl Institute, Rice University, Houston, TX, 77005, USA

4 Air Force Research Laboratory, Materials and Manufacturing Directorate, AFRL/RX, Wright-Patterson Air Force Base, OH, 45433, USA

5 UES, Inc. 4401 Dayton Xenia Rd, Dayton, OH, 45432, USA

Carbon nanotube (CNT) wires approach copper's specific conductivity[1] and surpass carbon fiber's strength [2,3,4,5], with further improvement anticipated with greater aspect ratios [6,7] and incorporation of dopants with long-range structural order[8]. Fullerenes assemble into multitudes of process-dependent supramolecular crystals[9,10,11,12,13] and, while initially insulating, they become marginally conductive (up to 0.05 MSm$^{-1}$) and superconductive ($T_c$=18 K with K and 28 K with Rb)[14,15,16,17] after doping. These were small (100's μm long), soft (hardness comparable to indium[18]), and typically unaligned (non-free-standing exceptions, see [19,20]), which hindered development of fullerene based wires. Individual fullerenes were previously incorporated into CNT fibers, although randomly without self-assembly into supramolecules[21,22]. Here, a simple variation in established CNT acid extrusion creates a fiber composed of uniaxial chains of aligned fullerene supramolecules, self-assembled between aligned few-walled CNT bundles. This will provide a testbed for novel fullerene wire transport and prospects in CNT wire advancement.

A leading CNT fiber manufacturing method[6] involves dissolving high quality CNTs into a superacid (typically chlorosulfonic acid (CSA) though less hazardous acids were recently shown effective [23]) forming a liquid crystalline CNT solution where CNTs protonate, electrostatically repel each other, and align. After extrusion into a coagulant, a dense CNT fiber is made with highly aligned microstructure. In this letter, we explored adding CNT and C60 powder together within the CSA with the following mass loadings: Low load (0.2% C60, 2% CNT, 97.8% CSA); High Load (2% C60, 2% CNT, 96% CSA); and neat (2% CNT, 98% CSA). Under extended reaction times in CSA (several hours at 50 °C or longer at room temperature), C60 is known to undergo significant chlorination[24,25]. On shorter time scales, C60 in CSA may be described by a complex multi-step equilibrium between C60 and its chloronium ion, C60Cl$^+$. We found C60 alone in CSA to quickly form a purple solution (see supplemental figure 1), which continued to react over the course of two hours (see supplemental figure 2). The C60 powder mixed with the CNTs in CSA without complications and extrusion was carried out as normal. Brown precipitants emerging from the fiber approximately 1 cm from the extrusion point indicate the C60 mass loadings are not fully preserved in the fiber. CNT fiber made with CSA extrusion is well known to contain derivatives of CSA[6,7], which could later react with intentionally added dopants. Further, it has been shown that C60 crystal size and structure are dictated by solvent conditions[9,10,11,12]. For these reasons, we report on the C60 CNT fiber as is, as well as after an annealing at 300 °C in Ar. We found these conditions removed the residual acid and increased C60 crystallization, although was gentle enough to not hamper C60 Raman signatures.

Transmission wide angle X Ray diffraction (WAXS) of the fiber (Figure 1) shows two mutually aligned crystalline phases, with specific characteristics dependent on processing. For as-is high loading, there are two features from the aligned CNT fiber diffraction, lobes for aligned van der Waals spacing at 4 Å and carbon-carbon ring at 2.2 Å, superimposed with another complementary 'fiber-diffraction like' pattern with hkl diffraction peaks from a uniaxial aligned unit cell (see supplemental section 4). Annealing increases order and crystallinity, while the d-spacing is not disturbed (supplemental figure 5). This complementary diffraction pattern coincides with indexed hkl peaks of face centered cubic (FCC) C60 crystals in literature, as well as our WAXS measurements on neat C60 powder (supplemental figure 10). The separation of hkl peaks into arcs and layer lines indicate preferred orientation with respect to the CNT fiber and drawing direction. The as-is low loading shows a weakly ordered complementary phase (supplemental figure 6). After annealing, this crystallizes, improves in order (narrower radial peaks) and increases in orientation (narrower azimuthal arcs). Additional lobes appearing off the meridional and equatorial are indicate of 2D or 3D structure (supplemental figure 7). The CNT carbon-carbon spacing, van der Waals spacing, and alignment (Herman orientation of 0.8) did not change appreciably between loadings, indicative of separated domains of $C_{60}$ and CNT in the fiber architecture. The $C_{60}$ alignment could not be calculated unambiguously, although was subjectively lower than the CNTs (estimated Herman orientation parameter of < 0.5).

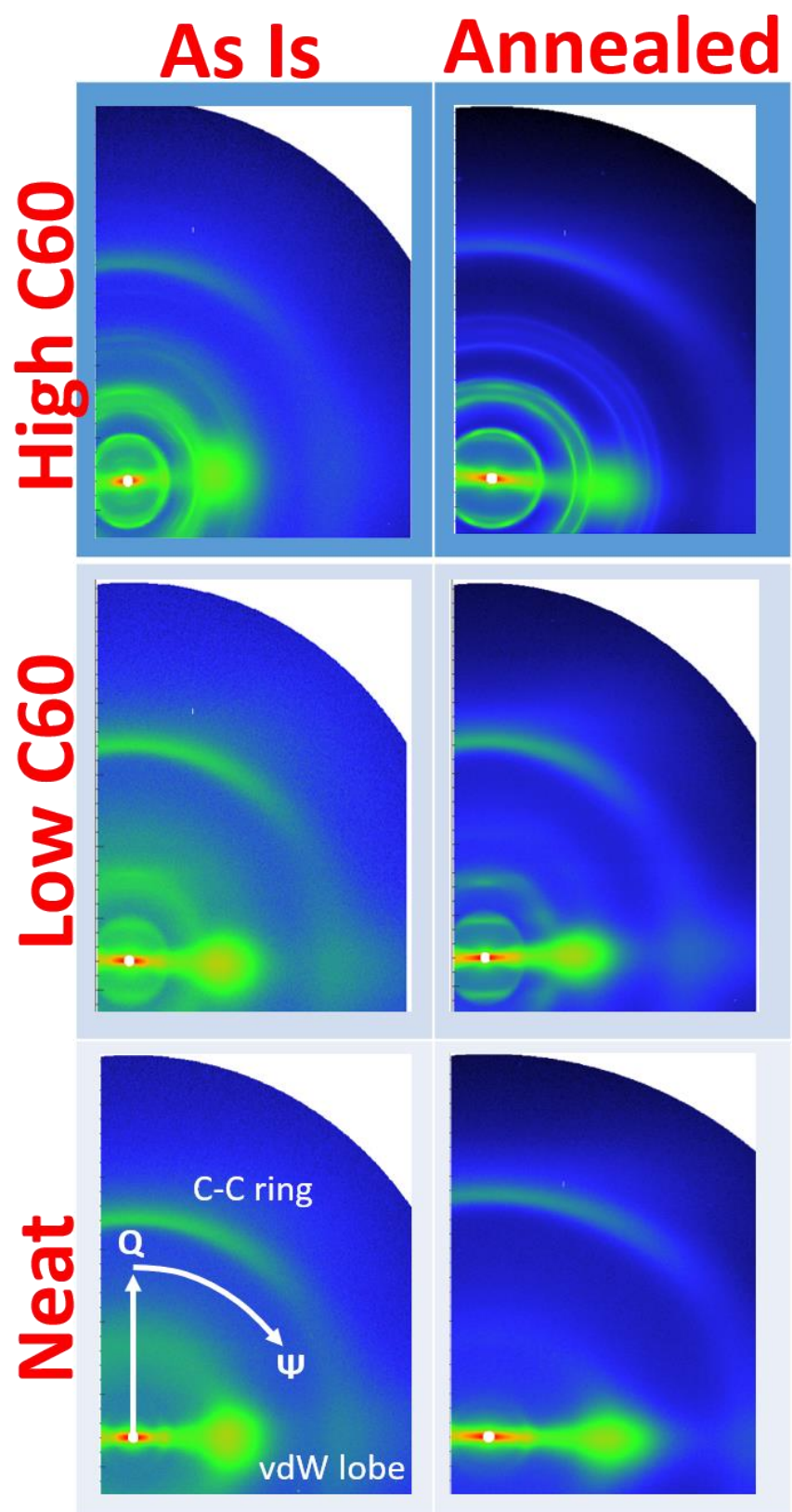

Figure 1| Transmission WAXS of the high and low loading C60 CNT fiber, as well as a neat CNT fiber, for as is and annealed states. Color represents intensity and is a function of scattering vector Q and the azimuthal angle Ψ. For the as is neat CNT fiber, the typical van der Waals spacing lobe (4 Å) for aligned CNTs and carbon carbon ring (2.2 Å) are annotated. Additional arcs appearing in the other plots are C60 agglomerations with different process-dependent degrees of order and orientation.

In terms of as-is physical properties, the electrical conductivity (in absolute and weight specific terms) was similar between the neat and low load fiber (both ranging 6- 8 MSm$^{-1}$, with specific conductivity ranging 55%-70% of copper). The low load fiber was notably the mechanically strongest (1.6-2 GPa, with specific strength 1-1.1 N Tex$^{-1}$), relative to the neat fiber (1.1-1.3 GPa, with specific strength 0.6-0.69 N Tex$^{-1}$). CNT fibers fail by the smooth surfaces of CNT structures sheering past each other and it is conceivable that the corrugated surface of the C60 molecules increases friction between CNT structures[7]. Absolute conductivity and tensile strength of the high loading fiber were approximately half that of the neat fibers. After factoring in weight, the specific strength did not change and specific conductivity was 60-70% of the neat fiber. The high loading fiber increased in diameter (the neat and low load fiber diameters were 70% of the high load) and lowered in density (the high load fiber was 67-80% the density of the neat). Annealing overall decreased density 9%- 14% and increased specific strength 15%- 30%. See supplemental section 3 for detailed physical properties.

Scanning electron microscopy (SEM) shows spheroidal particulates on the surface of the low load fiber (Figure 2b). The high load fiber has a coating with similar spheroidal particulates appearing on top the coating (Figure 2c). Energy Dispersive X-ray Spectroscopy (EDS) shows that the spheroidal particulates and coating are carbon, similar to the signal of the neat CNT fiber (supplemental figure 15-25). There are trace sulfur and chlorine signals uniformly throughout all as-is samples from the acid processing, although these trace signals go away after annealing as expected. Raman spectroscopy (figure 2d) with laser polarization perpendicular to the fiber alignment returns a spectra with a superposition of CNT and C60 features and, when parallel, returns a more typical CNT spectra from the established antenna amplification effect when polarization aligns with CNTs[26]. The G peak of the neat CNT fiber (1590 cm$^{-1}$) and the Raman $A_g$ mode of the C60 powder (1467 cm$^{-1}$) did not shift appreciably in the annealed high load fiber.

Nano resolution computed tomography (NanoCT) operating in absorption mode maps the 3D internal density structure (large field of view with nominal spatial resolution of 150 nm, with a voxel size of ~65 nm), whereas phase mode dramatically displays the interface contrast and is particularly suited for low Z materials like carbon. Comparing the results from both modes reveals that the neat CNT fiber and low loading fiber have a similar and relatively homogeneous density distributions with some striations running down the fiber and few voids (figure 2 e,f and supplemental figure 32 and 34). In contrast, the high loading fiber has a complex structure with a multitude of interfaces oriented in the direction of the fiber (figure 2 g, h, I and supplemental figure 36). Comparing with absorption mode data (supplemental figure 32 - 38), the green and blue oriented interfacial colors correspond to voids. Segmentation of the absorption contrast data using ORS Dragonfly determined these voids comprise around 10% of the fiber by volume and are internally connected. The annealed high loading fiber is qualitatively similar (supplemental figure 38) with a void volume also around 10% by volume.

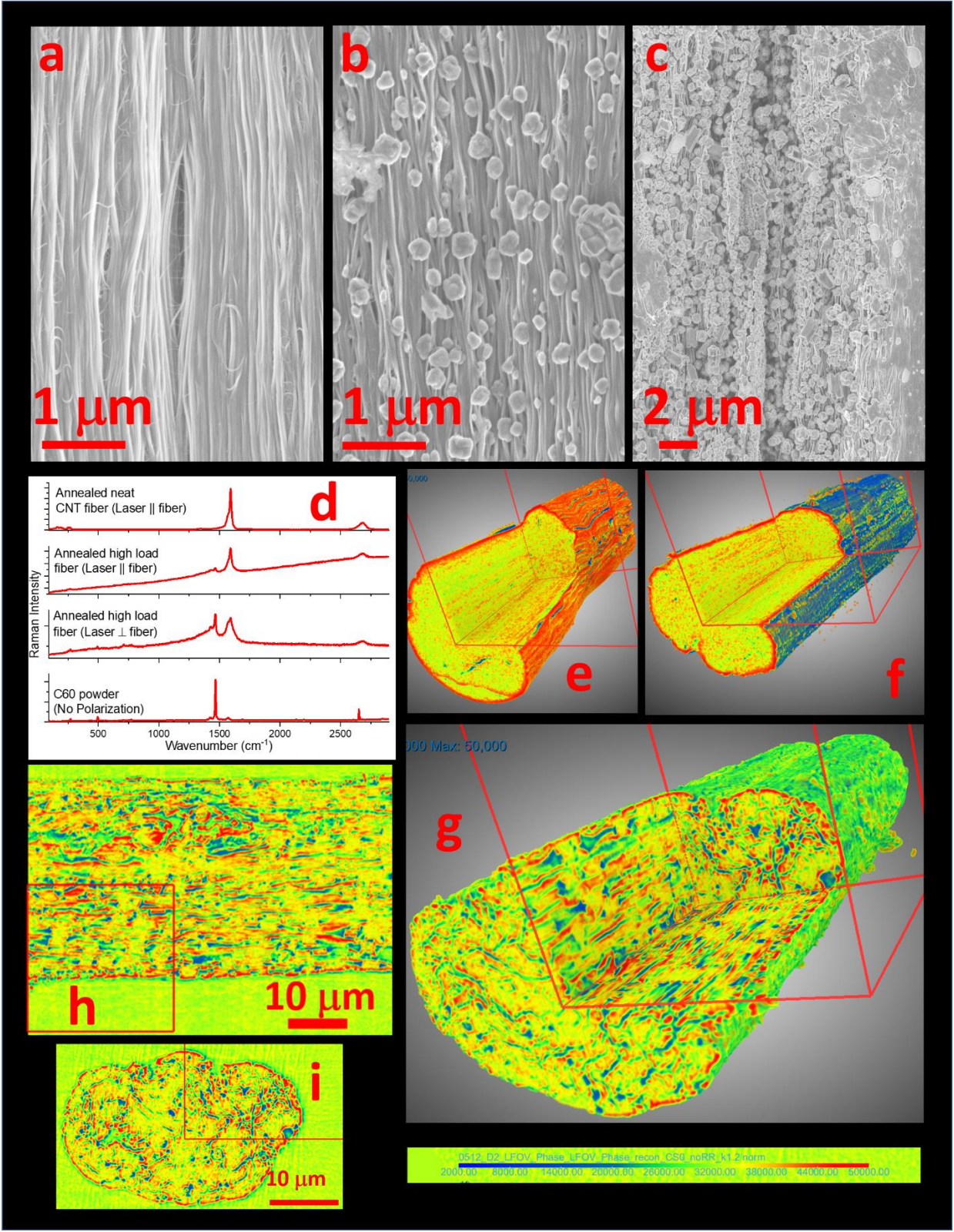

Figure 2| Surface and internal structure of the C60 CNT fiber. SEM images are shown for a, the neat CNT fiber; b, low loading fiber; and c, high loading fiber. d, typical Raman spectra of annealed high loading

fiber, showing the effect between laser polarization and fiber alignment. Neat CNT fiber and C60 powder are provided for comparison. Tick marks between Raman plots represent the same arbitrary value. NanoCT scan in phase mode indicates interfaces of density variation for e, the as is neat CNT fiber; f, as is low loading fiber; and g, as is high loading fiber. A section is removed for internal viewing as depicted by the red box (25 μm x 25 μm x 25 μm). h, a two-dimensional panel of this as is high loading fiber and i, its cross section. These 2D sections comprise the faces of the red box intersecting the sample in the 3D image. At the bottom is a color scale, where the color scale is normalized between samples.

Transmission electron microscopy (TEM) images and diffraction patterns of focused ion beam-prepared cross-sections further elucidate the internal structure of the fibers. The as-is low load fiber (figure 3 a,b) shows small oval inclusions (selected inclusions ranged between 140 to 170 nm in length) oriented in the direction of CNT alignment and lined up in rows. These oval inclusions are highlighted by the white angular voids where the CNTs flanking the inclusion were milled away during FIB preparation. Most high resolution TEM images indicates that these inclusions are largely amorphous (supplemental figure 39-45) with limited signatures of short-range order (supplemental figure 52). After annealing, TEM shows clearer signatures of short-range order (supplemental figure 46- 52). HAADF STEM showed that the density and thickness between CNTs and C60 inclusions are similar (supplemental figures 49 and 52). The high load fiber (figure 3c,d) still has these small oval inclusions, although now have large crystalline granular inclusions roughly aligned in the direction of the CNTs. Length of some selected granular inclusions ranged along the fiber axis 3 to 9 μm; selected widths span 0.75 to 2 μm. Figure 3 e shows high resolution TEM of a single crystal region of a larger granular inclusion, showing hexagonal packing and alignment of the C60 into linear rows. The electron diffraction pattern from this grain (figure 3f ) has an outer hexagonal ring of diffraction points (spacing 4.8-5.0 Å), which belong the {220} family of planes of the face centered cubic (FCC) structure, the typical configuration for C60 nano-whiskers[9][10][11][12]. Diffraction from the {110} is forbidden in FCC, so the inner two diffraction points (spacing of ~8.4-9.3 Å) are possibility from an oriented super cell, although the inner points are not aligned with the outer hexagonal ring. These TEM results are consistent with the WAXS and NanoCT. TEM never revealed the voids found in NanoCT, but It is possible the FIB redeposition debris filled them while machining the lamella. More detailed analysis and additional figures are provided in the supplemental.

This analysis shows the C60 forms a coating over the CNT fiber, although more importantly, the C60 self assembles in an internal network of crystalline C60 supramolecules where its crystal structure and outer surface are oriented in the direction of CNTs. The concentration of added C60 effects the size, quality, and distribution of the C60 agglomerations. Annealing improves the crystallinity and orientation. A rich experimental space is ahead to determine how different processes effect the internal structure. It is interesting that the CNT bundling, alignment, and physical properties were minimally impacted, if not enhanced in the low loading fiber. Optimizing a low fullerene concentration may be an expedient route for further improvement of CNT wires. The high loading fiber with the high percentage of voids and expanded diameter had generally worse physical properties, though no worse than a factor of two. The CNT network itself maintained a high degree of alignment despite the obstructions. Note that the neat CNT, low loading, and high loading fibers all had similar resistance responses with temperature (300 K to 1.9 K, supplemental figure 65), indicating they have similar intrinsic and extrinsic transport characteristics. At this point the C60 is expected to be electrically insulating, so it appears the conductive CNT network is not impeded by the C60 agglomerations. The crystallinity of the high loading granular inclusions was much better than the much smaller low loading granular inclusions. Raman and WAXS peaks associated with C60 did not shift from the presence CNTs, indicating the CNTs were not applying pressure on the C60 supramolecular crystals, which is an established means at changing critical temperatures in alkali-metal doped fullerene superconductors[27]. Tradeoffs between the percolative path, fullerene aspect ratio, crystallinity, and internal molecular pressure will be relevant for novel transport phenomena.

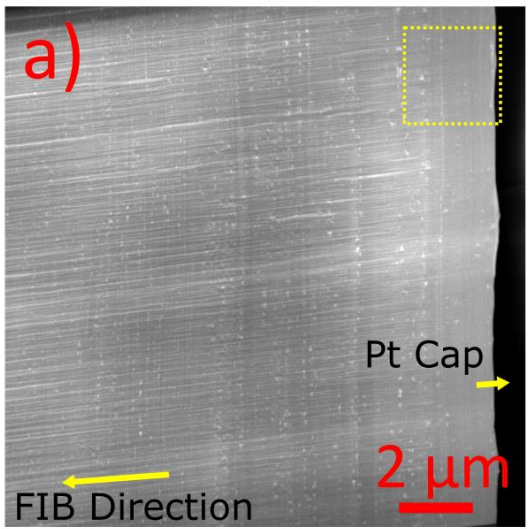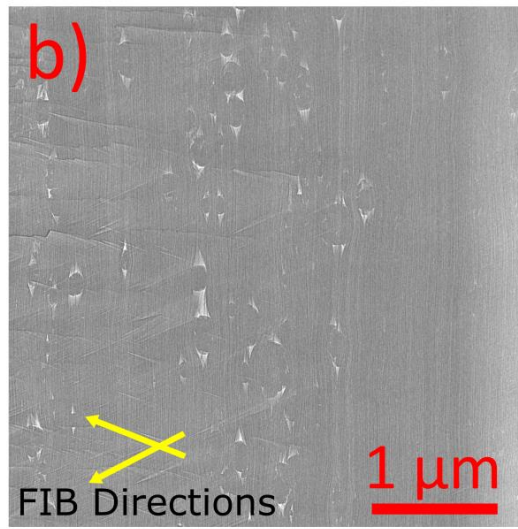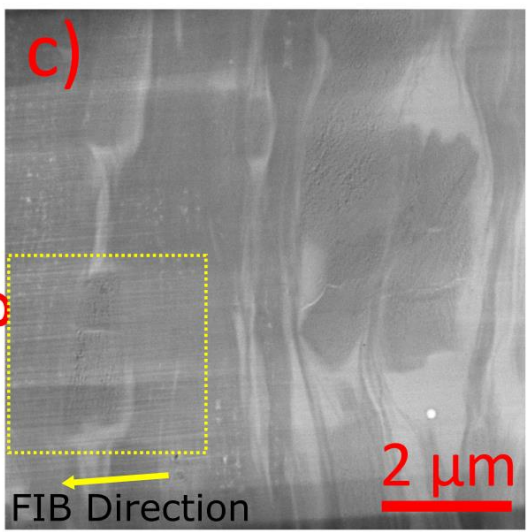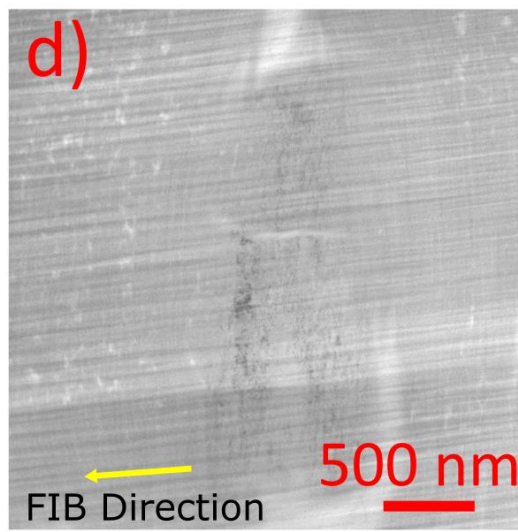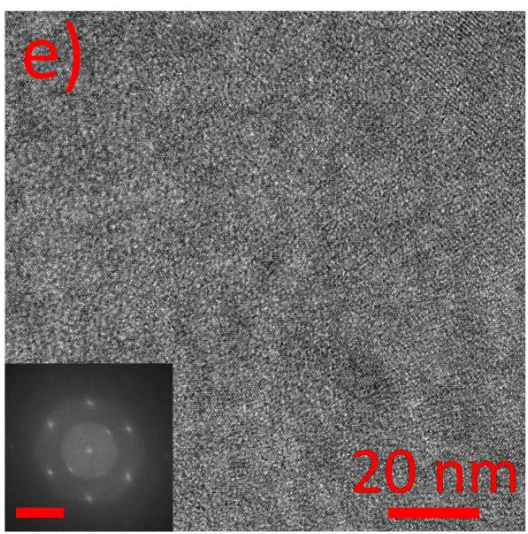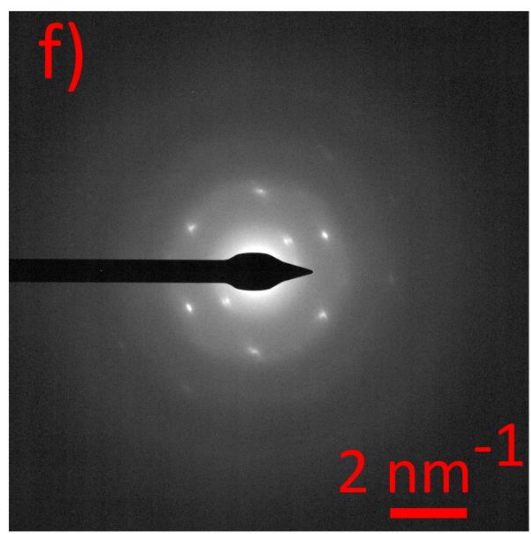

Figure 3| Transmission electron microscopy of the as is C60 CNT fibers. All images were aligned to make the fiber direction vertical and the FIB ion impingement directions are indicated to identify contrast from strong oriented curtaining perpendicular to the fiber direction. The dashed yellow box indicates the location of the higher magnification images in each fiber. (a) Low and (b) high magnification images of the low C60 loading fiber show the prevalence and shape of the oval inclusions throughout the fiber. (c) Low magnification image of the high loading fiber shows a microstructure disrupted by large granular inclusions while (d) the high magnification image shows the less disrupted regions still contain oval inclusions and high contrast in the granular inclusions from their crystalline structure. (e) High resolution TEM imaging shows hexagonal ordering, consistent with the FCC structure of C60 nanowhiskers. The FFT of the HRTEM image (inset, scale bar 2 nm$^{-1}$) confirms FCC packing and also shows a set of larger lattice spacings that are not consistent with FCC C60. (f) Selected area diffraction of this granule confirms the crystalline nature of the inclusion and agrees well with the structure determined from HRTEM. These images confirm that crystalline C60 supramolecules exist within the fiber and show that their outer dimensions (in addition to the alignment of their internal structure) are oriented in the direction of CNT alignment.

Methods

CNT/C60 fiber spinning. The CNTs use in our study were produced by Meijo Nano Carbon Co DX302, no purification was required. Aspect ratio, as determined by extensional rheometery, was 5600. CNTs and C60 (Sigma-Aldrich, 98 %) were dissolved in chlorosulfonic acid (CSA, Sigma-Aldrich, 99%) with a FlackTek speed mixer until a homogeneous solution was obtained . This solution was filtered and then extruded into an acetone bath, acting as the coagulant. A rotating drum collected the fiber, as previously described[6]. The spinning parameters were optimized for each solution to ensure fiber drawing while maintaining a constant draw ratio (1.35) across all samples. For the neat fiber: extrusion rate: 1 m min$^{-1}$ collection rate: 1.35 m min$^{-1}$; for the low load fiber: extrusion rate: 2 m min$^{-1}$ collection rate: 2.7 m min$^{-1}$; for the high load fiber: extrusion rate: 2 m min$^{-1}$ collection rate: 2.7 m min$^{-1}$.

WAXS. Transmission WAXS measurements were performed on a Xenocs Xeuss 3.0 using the Cu Kα wavelength of 1.5406 Å and a Pilatus 300 K detector. All samples were run in the line eraser mode for 10800 s per exposure. Line eraser mode is used to remove horizontal bars that occur in the data due to the design of the detector. A total of two exposures per sample were collected at a sample to detector distance of 55 mm using the standard configuration. The detector was offset from center (35 mm in x and 45 mm in z) in order to collect one full quadrant of the pattern.

NanoCT. X-ray 3D tomographic imaging was accomplished with a Zeiss Xradia Ultra 810 NanoCT microscope, operating in the large field of view mode (65 μm) with a nominal spatial resolution of 150 nm (65 nm wide voxels). A chromium anode generated the quasi-monochromated X-ray beam (5.4 keV). The typical tomography consisted of 361 projections; some samples with larger cross-sections utilized 701 projections. The typical collection times for the phase and absorption contrast modes were 75 s and 70 s, respectively; the one exception was the as is high loading fiber with collection times of 90 s and 120 s. Zeiss XMReconstructor was used to build the tomography data into a 3D volume. Segmentation was performed using ORS Dragonfly. For the phase contrast data, the background residual signal around the fiber was segmented and removed to more clearly visualize features within the fiber.

Focused ion beam and TEM. The sample for TEM was prepared by first manually dicing 1 mm fiber segments with a razor blade which were then carefully placed onto a molybdenum TEM mesh grid that was coated in EpoTek 353ND epoxy. The grid and sample were allowed to cure overnight at room temperature after which a few nanometers of Ir were deposited onto the grid for conductivity. The diced fiber segment was then thinned to electron transparency in a Tescan Lyra2 Ga+ FIB at 30 kV and finished with a 5 kV final polish. A Thermo Fisher Scientific Talos operating at 200 kV was used to image the lamella and take selected area diffraction patterns.

Acknowledgements

We gratefully acknowledge useful and important conversations with Juan Vilatela of Imdea Institute Madrid, Spain and Denis Arcon of University of Ljubljana, Slovenia.

Supplemental Material for

Strong Fiber from Uniaxial Fullerene Supramolecules Aligned with Carbon Nanotubes

John Bulmer, Michelle Durán-Chaves, Daniel M. Long, Jeremiah Lipp, Steven Williams, Mitchell Trafford, Anthony Pelton, Jared Shank, Benji Maruyama, Larry Drummy, Matteo Pasquali, Hilmar Koerner, Timothy Haugan

Supplemental Section 1: Mixing C60, CNTs, an CSA together

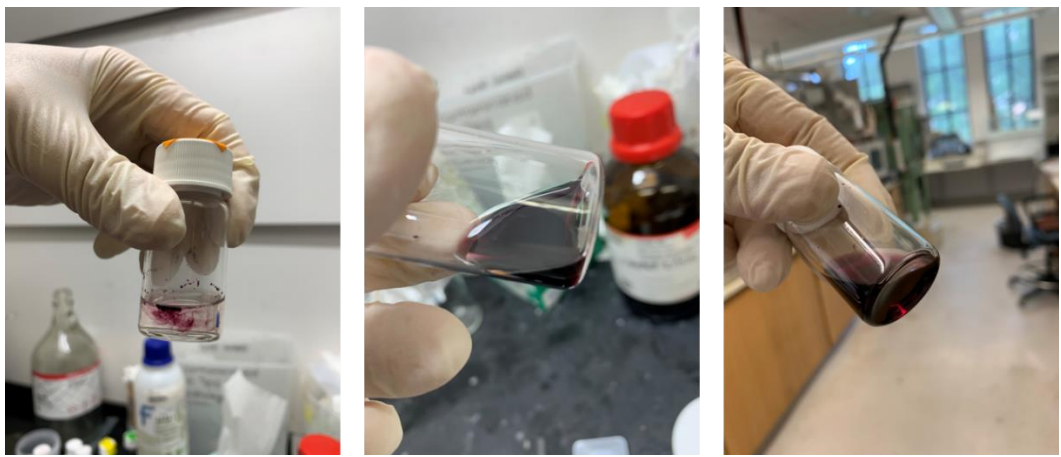

Supplemental Figure 1. Images showing C60 mixing with CSA and turning into a purple solution.

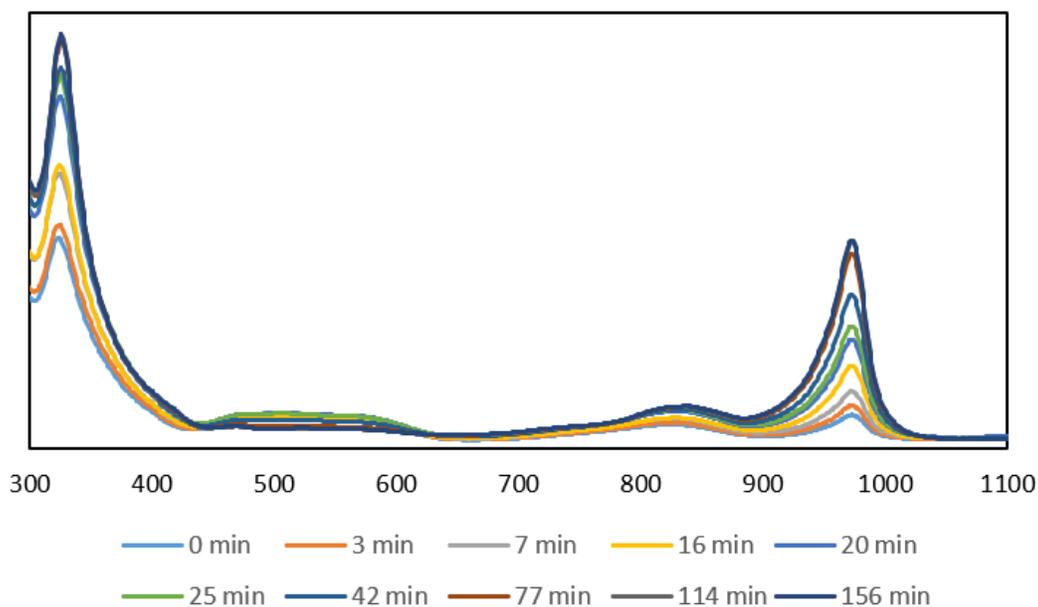

Supplemental Figure 2. UV/vis results showing a gradual change of the C60 in the CSA over a span of approximately two- and one-half hours, consistent with chlorination. Once mixed, the C60 CSA solution was placed into the UV/Vis spectrometer at time zero.

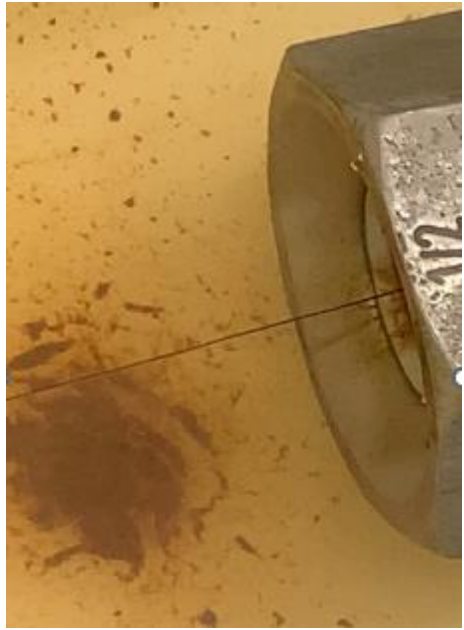

Supplemental Figure 3. Photograph showing brown precipitants emerging from the fiber approximately 1 cm from the extrusion point.

Supplemental Section 2: WAXS

Transmission wide angle XRD (WAXS) data for the CNT C60 fibers. The 2D pattern on the left represents the count intensity (as displayed in the color bar) versus the azimuthal-angle-dependent scattering vector. The plot on the right is the integrated intensity over the available azimuthal angle and Q range.

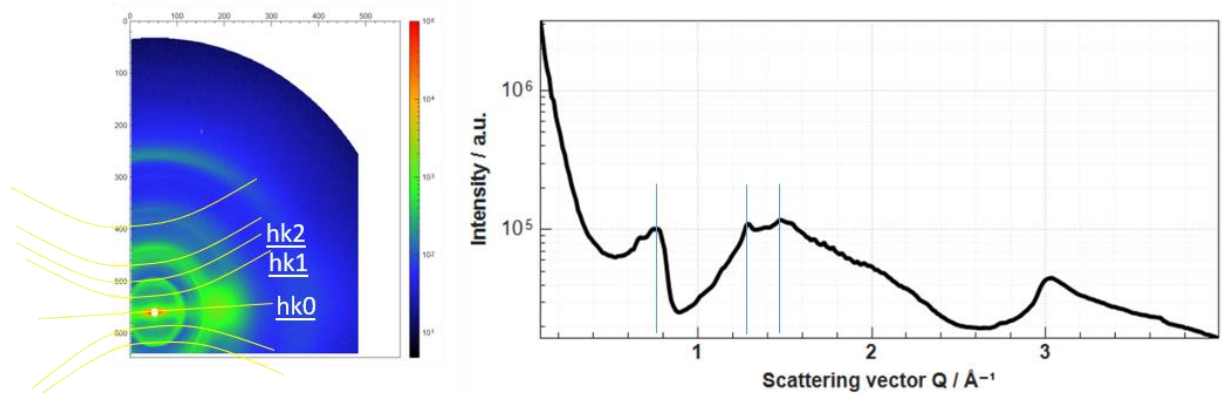

Supplemental Figure 4. As is High C60 CNT fiber. In addition to the CNT fiber peaks, sharp peaks occur that indicate a fiber-diffraction like orientation with hkl diffraction peaks from a uniaxial aligned unit

cell. Layer lines are visible confirming the uniaxial alignment. The 1D pattern shows additional peaks at Q=0.77 Å$^{-1}$ (8.14 Å), Q=1.29 Å$^{-1}$ (4.87 Å) and Q=1.47 Å$^{-1}$ (4.27 Å), representing hkl indices 111, 220 and 311 respectively. These three peaks match diffraction peaks observed for C60 powder.

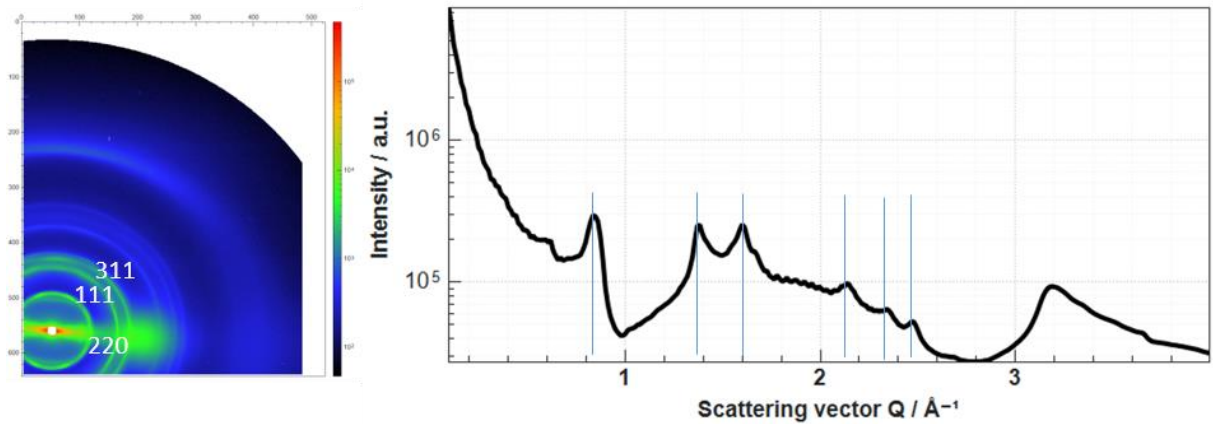

Supplemental Figure 5. Annealed High C60 CNT fiber. The crystalline peaks in the 2D pattern sharpen (increased order and crystallinity) and retain their uniaxial orientation. D-spacings are the same as in the as-is sample in the figure above.

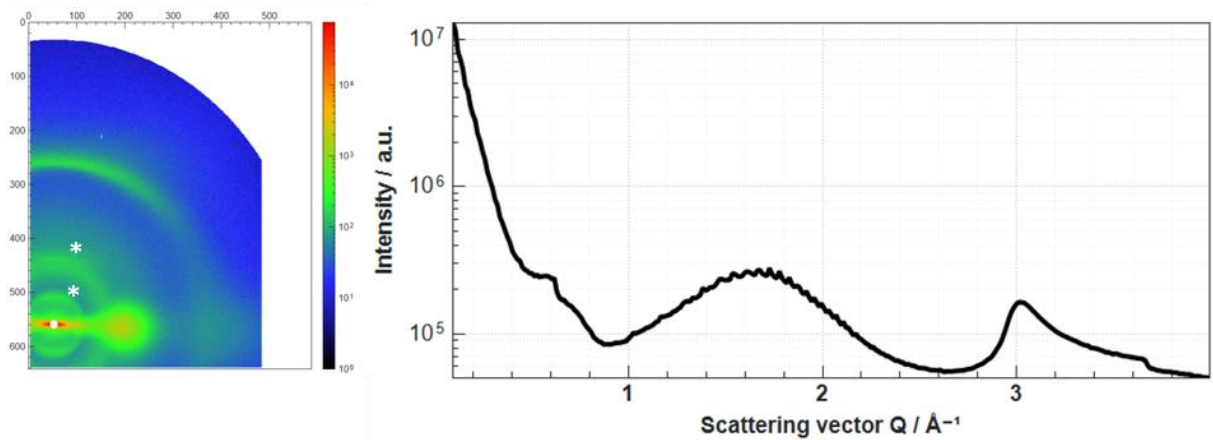

Supplemental Figure 6. As is low C60 CNT fiber. Only a weak meridional peak (*) is observed in the 2D pattern that is indicative of some weakly ordered additional phase. Oscillations in the 1D pattern between Q=1.2 Å$^{-1}$ are artifacts.

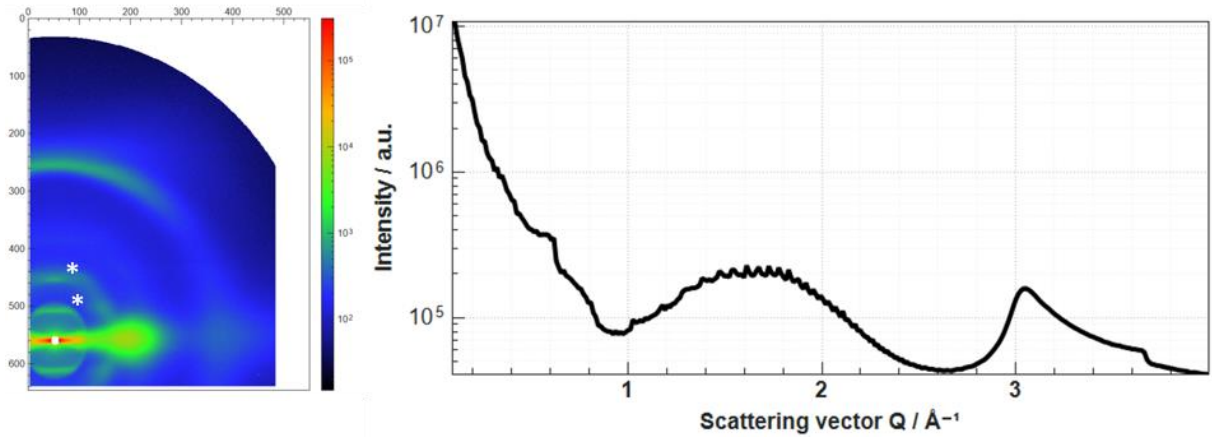

Supplemental Figure 7. Annealed low C60 fiber. Meridional peaks (*) sharpen (improved order) and are narrower (better orientation) and additional lobes occur off meridional and equatorial indicative of 2D or 3D structure. These peaks are weak and do not occur on the 1D pattern.

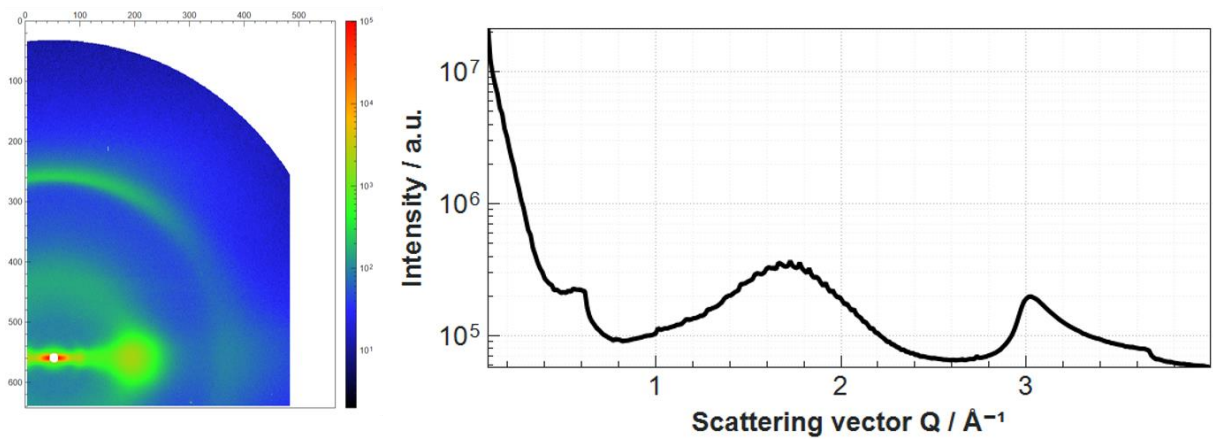

Supplemental figure 8. As-is neat CNT fiber. Expected highly oriented fiber pattern from aligned CNTs.

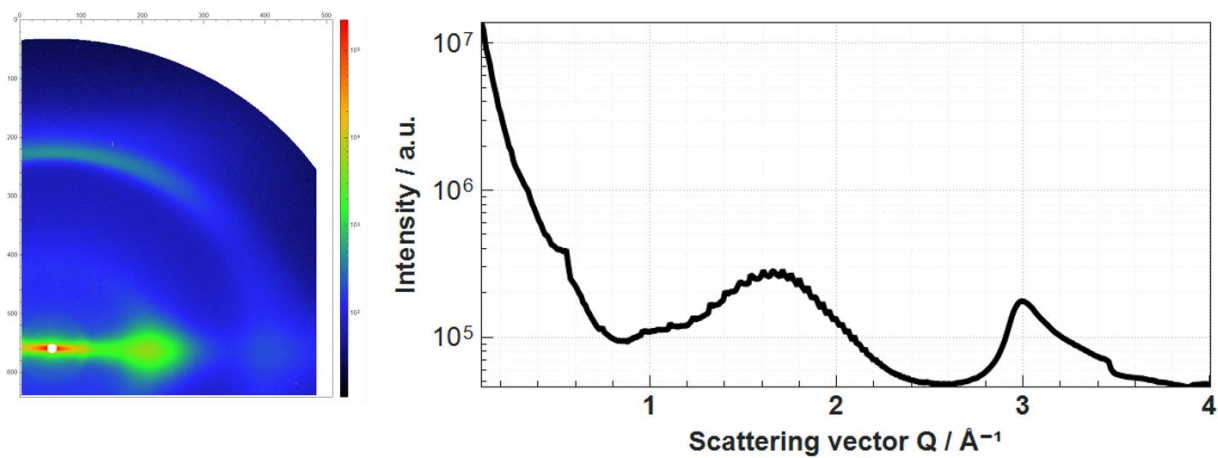

Supplemental figure 9. Annealed neat CNT fiber. Improved alignment after heat treatment seen in the narrowing of the peak at Q = 1.75 Å$^{-1}$.

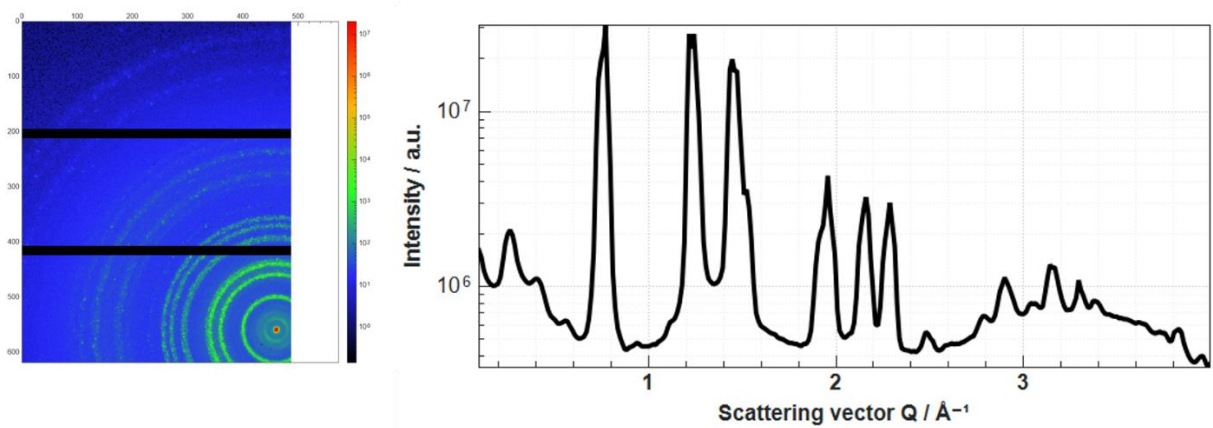

Supplemental figure 10. As-is C60 powder. Highly crystalline material with powder pattern (rings). Several of the peaks match reported values for C60 in whiskers (see below).

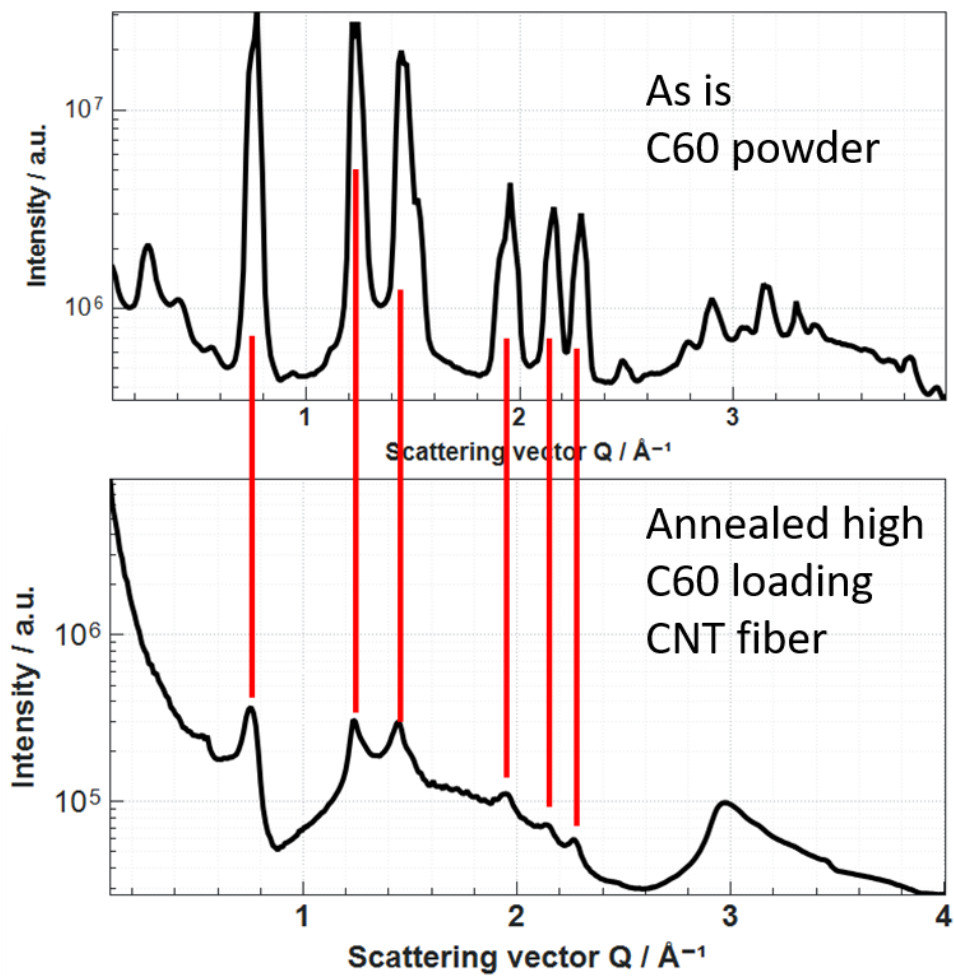

Supplemental figure 11. Comparing diffraction peaks between the as-is C60 powder and annealed high load fiber. Here we see the peaks present in the measured as-is C60 power correspond with the complimentary diffraction peaks in the annealed high yield fiber. Corresponding peaks with d-spacings are listed in table 1 below, in addition to C60 nano-whisker values from literature.

Table 1. Comparison of our WAXS peaks from the high load fiber, with our measured pure C60 powder, with that from C60 literature (J. Minato, K. Miyazawa, "Solvated structure of C60 nanowhiskers" Carbon 43, 2837-2841 (2005).

| peak | as-is high load ($A^{-1}$) | | d (A) | C60 ($A^{-1}$) | Literature* ($A^{-1}$) |
|---|---|---|---|---|---|
| A | 0.77 | 111 | 8.14 | 0.77 | 0.76 |
| B | 1.29 | 220 | 4.87 | 1.22 | 1.25 |
| C | 1.47 | 311 | 4.27 | 1.45 | 1.47 |

Supplemental Section 3: Physical Properties

Physical properties were measured redundantly at two different facilities (Houston and Dayton). Resistance was measured by standard four probe techniques along different sections of the fiber (three or more different pieces, each typically approximately one hundred millimeters long). Tensile strength was measured using tensile testers with gauge lengths 20 mm long (2 mm/min strain rate). Fiber diameter was estimated in a scanning electron microscope in multiple fiber locations. Fiber weight was measured with a microbalance. Below we show the results from both locations that qualitatively agree with each other. Small differences are expected from the degree of inhomogeneity in the fiber, differences with estimating cross section diameter with SEM, differences in instrumentation, and the fact the physical properties can settle over time. In the paper, we reported various physical properties as a range between testing locations. Tables 2, 3, and 4 provide detailed information. Tables 5 and 6 show physical differences before and after annealing.

Table 2. Processing conditions and density

| Name | CNT (wt %) | C60 (wt %) | extrusion rate (m/min) | collection rate (m/min) | draw ratio | Houston diameter (um) | Dayton diameter (um) | Dayton diameter error (um) | Houston linear density (dtex) | Dayton linear density (dtex) | Houston density (g/cm$^3$) | Houston density error | Dayton density (g/cm$^3$) | Dayton density error (g/cm$^3$) |
|---|---|---|---|---|---|---|---|---|---|---|---|---|---|---|
| Neat | 2 | 0 | 1 | 1.35 | 1.35 | 22.49 | 19.98 | 1.08 | 7.14 | 5.73 | 1.80 | 0.40 | 1.83 | 0.22 |
| Low Load | 2 | 0.4 | 2 | 2.7 | 1.35 | 23.3 | 19.84 | 1.19 | 5.64 | 5.70 | 1.32 | 0.14 | 1.84 | 0.25 |
| High Load | 2 | 2 | 2 | 2.7 | 1.35 | 30.78 | 28.03 | 3.53 | 8.93 | 9.14 | 1.20 | 0.33 | 1.48 | 0.38 |

Table 3. Electrical Conductivity

| name | Houston Conductivity (MS/m) | Houston Conductivity error | Dayton Conductivity (MS/m) | Dayton Conductivity Error | Houston Specific Conductivity (Sm²/kg) | Houston Specific conductivity error | Dayton Specific Conductivity (Sm²/kg) | Dayton Specific Conductivity Error |
|---|---|---|---|---|---|---|---|---|
| Neat | 6.5 | 0.9 | 8.06 | 0.88 | 3631 | 78 | 4413 | 254 |
| Low Load | 6.1 | 0.7 | 7.60 | 0.95 | 4644 | 120 | 4126 | 273 |
| High Load | 3 | 0.8 | 4.23 | 1.07 | 2514 | 52 | 2855 | 84 |

Table 4. Mechanical Strength

| name | Houston tensile strength (GPa) | Houston tensile strength error | Dayton tensile strength (GPa) | Dayton tensile strength error | Dayton strain at break | Dayton strain at break error | Houston specific strength (N/tex) | Houston specific strength error | Dayton specific strength (N/tex) | Dayton specific strength error |
|---|---|---|---|---|---|---|---|---|---|---|
| Neat | 1.1 | 0.3 | 1.26 | 0.10 | 0.02 | 0.005 | 0.62 | 0.08 | 0.69 | 0.07 |
| Low Load | 1.6 | 0.2 | 2.01 | 0.22 | 0.04 | 0.01 | 1.2 | 0.1 | 1.09 | 0.13 |
| High Load | 0.7 | 0.2 | 0.90 | 0.09 | 0.03 | 0.006 | 0.6 | 0.06 | 0.61 | 0.06 |

Table 5. Comparison of diameter and density before and after annealing

| Name | Before Annealing Diameter (um) | Before Annealing Diameter error (um) | After Annealing Diameter (um) | After Annealing Diameter error (um) | Before Annealing Linear Density (dtex) | After Annealing Linear Density (dtex) | Before Annealing Density (g/cm3) | Before Annealing Density Error (g/cm3) | After Annealing Density (g/cm3) | After Annealing Density Error (g/cm3) |
|---|---|---|---|---|---|---|---|---|---|---|
| Neat | 19.98 | 1.08 | 19.64 | 2.62 | 5.72 | 5.06 | 1.83 | 0.22 | 1.67 | 0.45 |
| Low Load | 19.84 | 1.19 | 19.31 | 0.5 | 5.7 | 4.65 | 1.84 | 0.25 | 1.59 | 0.09 |
| High Load | 28.03 | 3.53 | 27.06 | 3.26 | 9.14 | 7.79 | 1.48 | 0.38 | 1.35 | 0.33 |

Table 6. Comparison of mechanical properties before and after annealing

| Name | Before Annealing Tensile Strength (GPa) | Before Annealing Tensile strength error | After Annealing Tensile Strength (GPa) | After Annealing Tensile strength error | Before Annealing Strain at Break | Before Annealing Strain at Break Error | After Annealing Strain at Break | After Annealing Strain at Break Error | Before Annealing Specific Strength (N/tex) | Before Annealing Specific Strength Error | After Annealing Specific Strength (N/tex) | After Annealing Specific Strength Error |
|---|---|---|---|---|---|---|---|---|---|---|---|---|
| Neat | 1.26 | 0.1 | 1.49 | 0.41 | 0.024 | 0.005 | 0.026 | 0.005 | 0.69 | 0.07 | 0.89 | 0.07 |
| Low Load | 2.01 | 0.22 | 2 | 0.16 | 0.035 | 0.01 | 0.033 | 0.006 | 1.09 | 0.13 | 1.26 | 0.08 |
| High Load | 0.9 | 0.09 | 0.95 | 0.24 | 0.035 | 0.006 | 0.03263 | 0.007 | 0.61 | 0.06 | 0.7 | 0.06 |

Supplemental Section 4: SEM

The following are selected representative SEM images of the as-is fibers. The SEM was a Zeiss Gemini with relevant SEM parameters provided in the photographs.

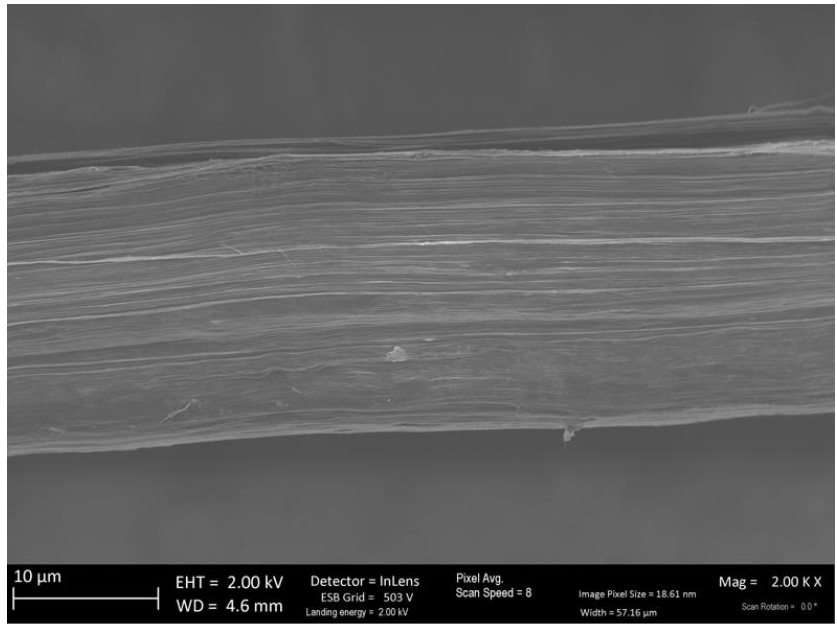
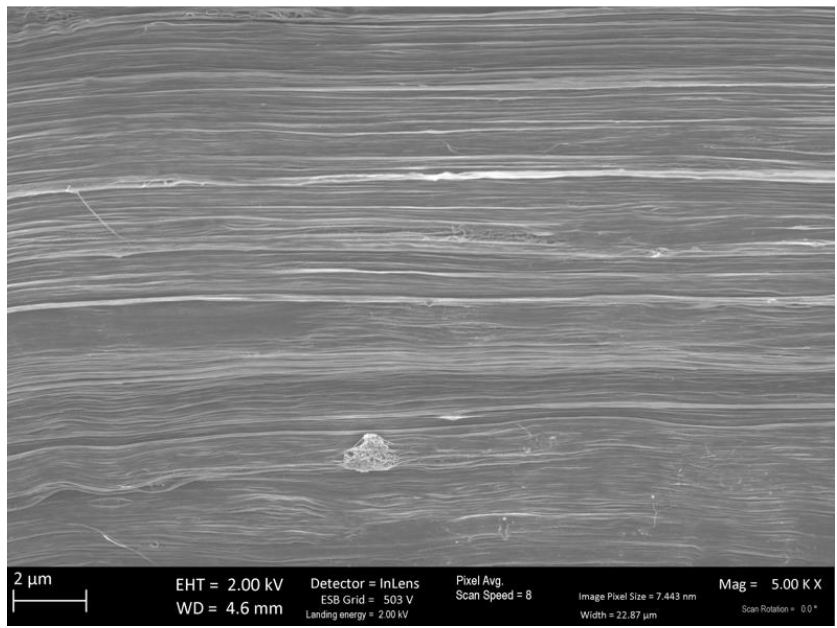

Supplemental figure 12. SEM image of the as-is neat CNT fiber.

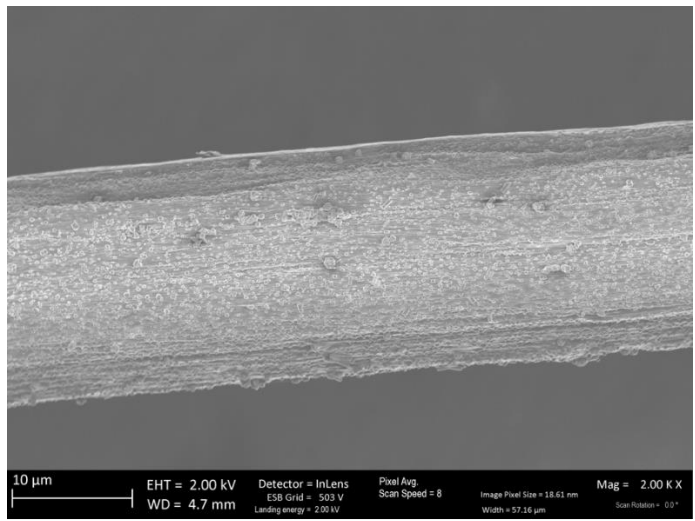
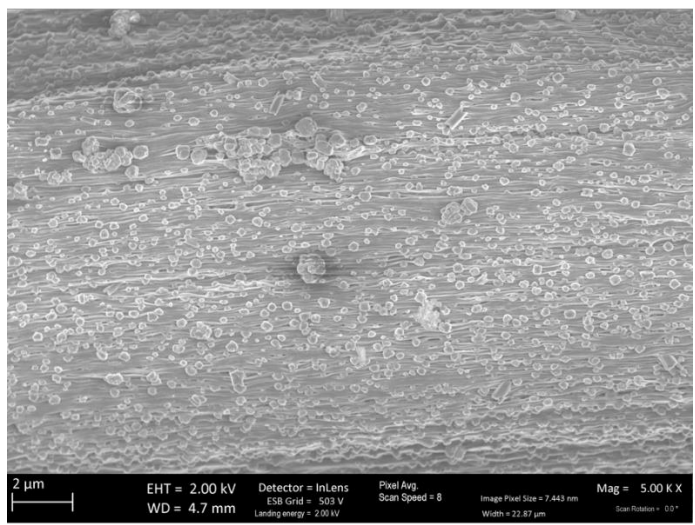
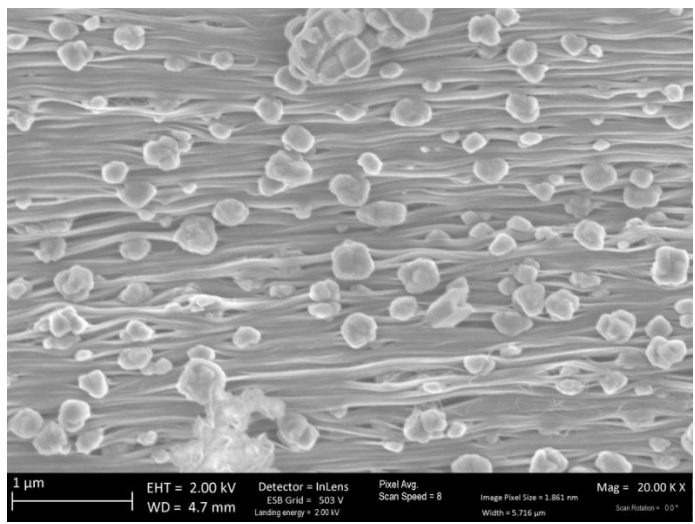

Supplemental figure 13. SEM image of the as-is low load fiber.

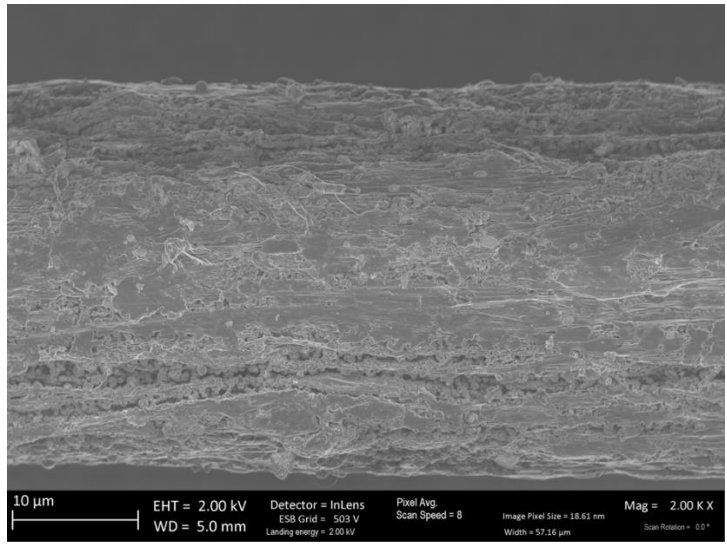
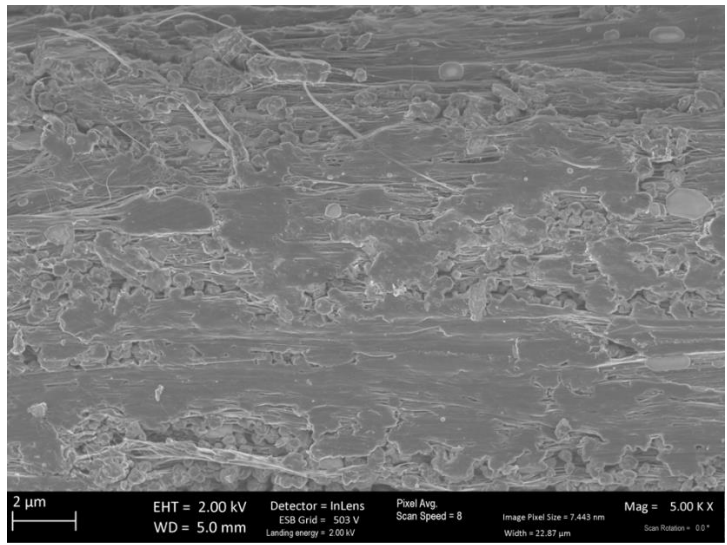
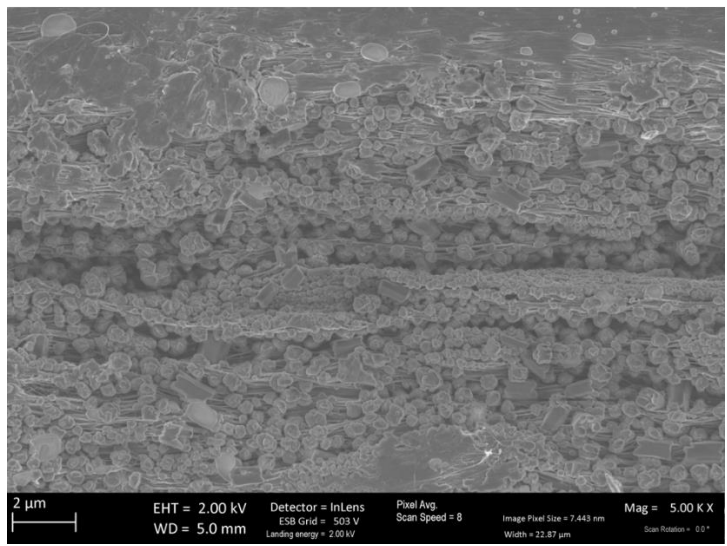

Supplemental figure 14. SEM image of the as-is high load fiber.

The following is Energy-dispersive X-ray spectroscopy (EDS) for the as-is low loading fiber. This was conducted with an Oxford Instrument X-Max Extreme EDS at 5 kV and a 1-minute collection time. As depicted below, scans were taken on the fibrous part as well as the spherical particulates.

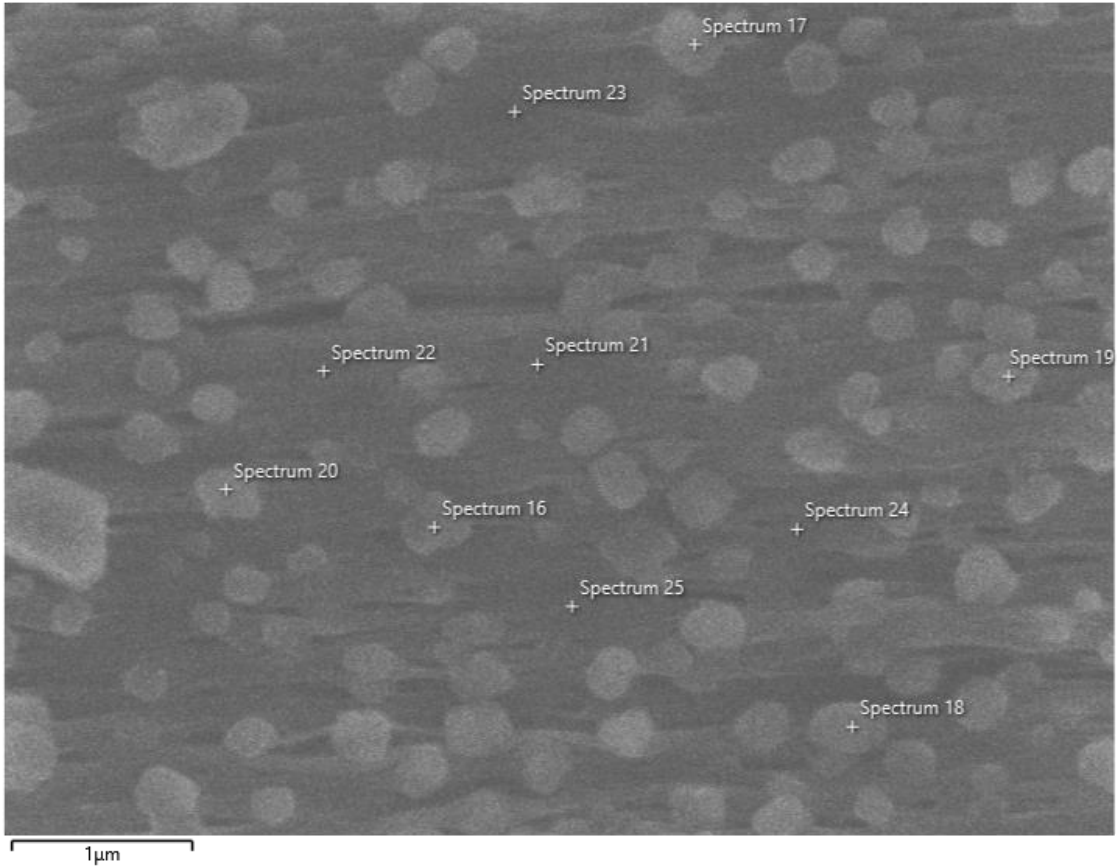

Supplemental figure 15. SEM image of the as-is low loading fiber, showing where the EDS spectrums were collected. These spectrums are below.

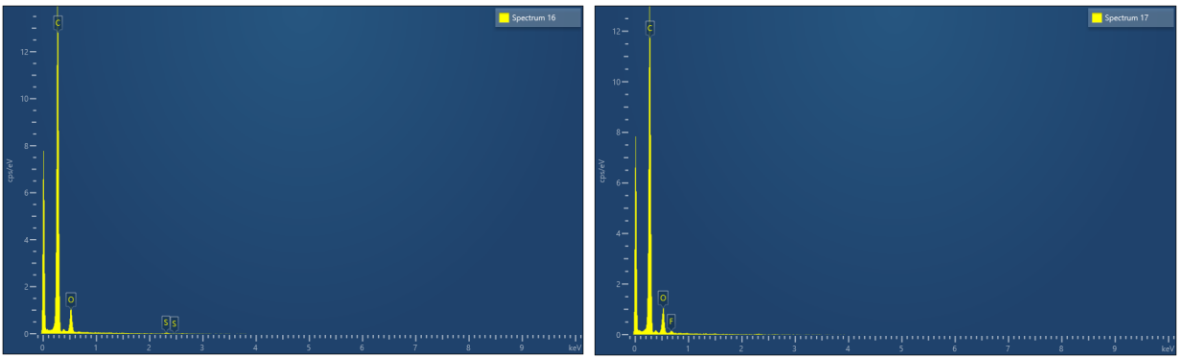

Supplemental figure 16. EDS spectra for as-is low loading fiber.

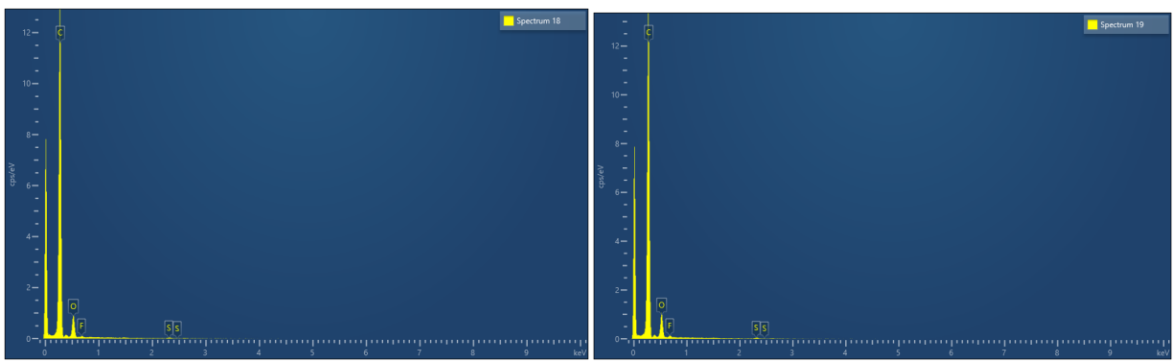

Supplemental figure 17. EDS spectra for as-is low loading fiber.

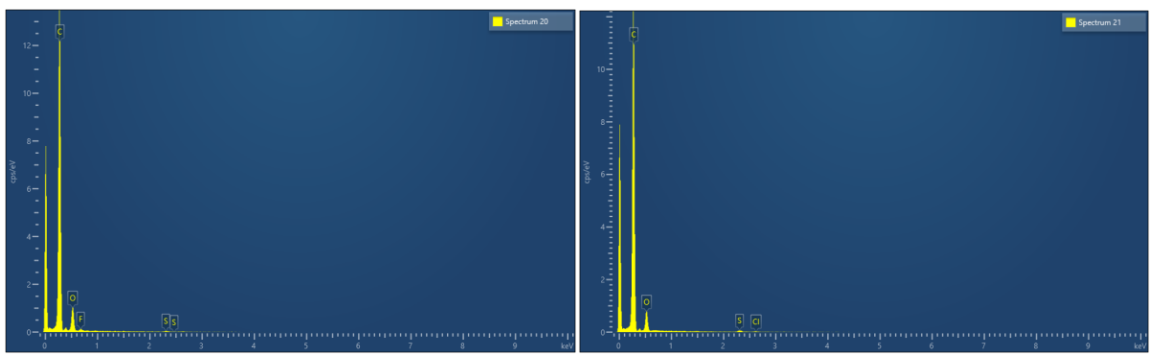

Supplemental figure 18. EDS spectra for as-is low loading fiber.

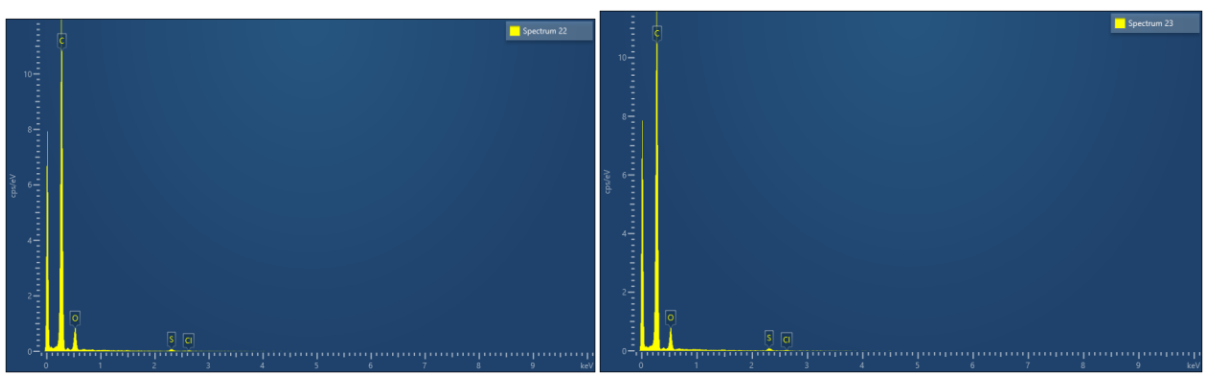

Supplemental figure 19. EDS spectra for as-is low loading fiber.

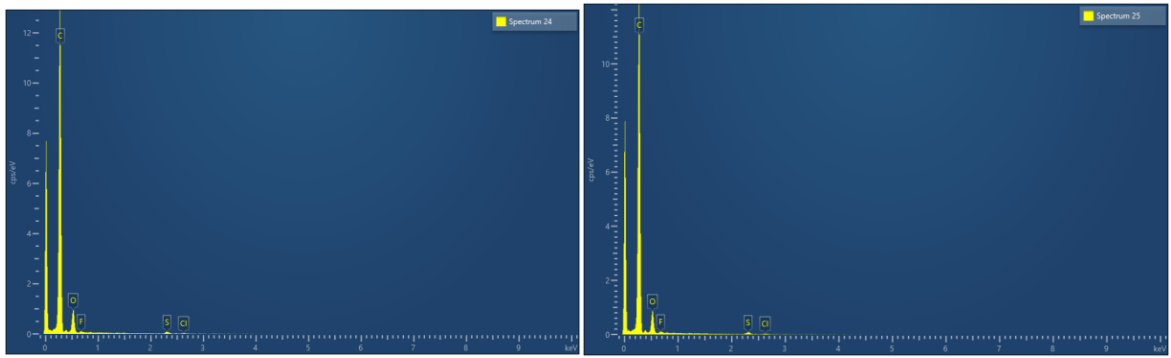

Supplemental figure 20. EDS spectra for as-is low loading fiber.

Below are the EDS results of annealed low loading fiber (5 kV, 1-minute collection). This shows that the signatures of chlorine and sulfur have nearly vanished from the annealing process.

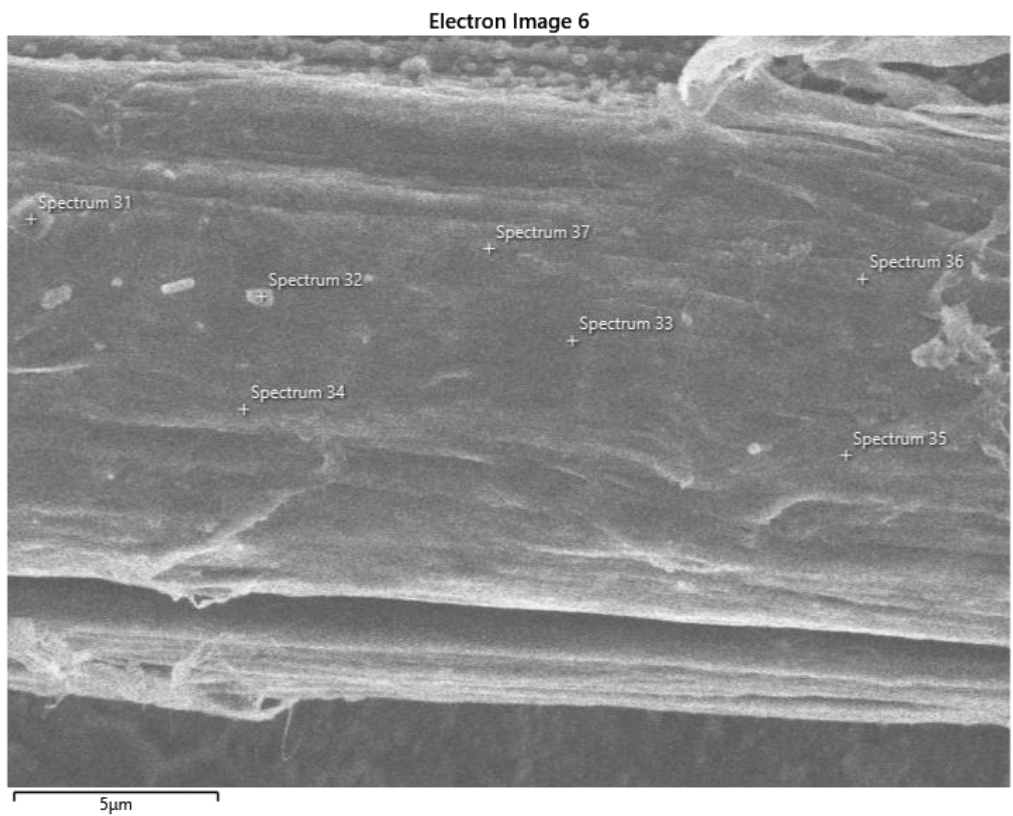

Supplemental figure 21. SEM for annealed low loading fiber, showing where the EDS scans were conducted.

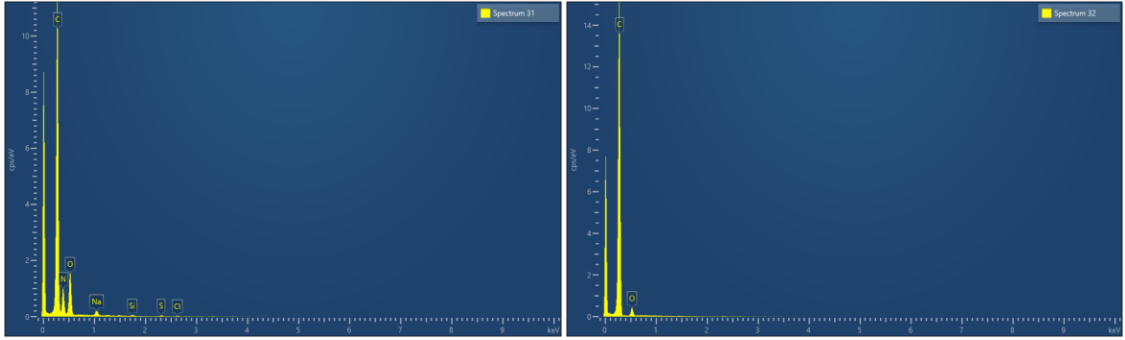

Supplemental figure 22. EDS scan for the annealed low loading fiber.

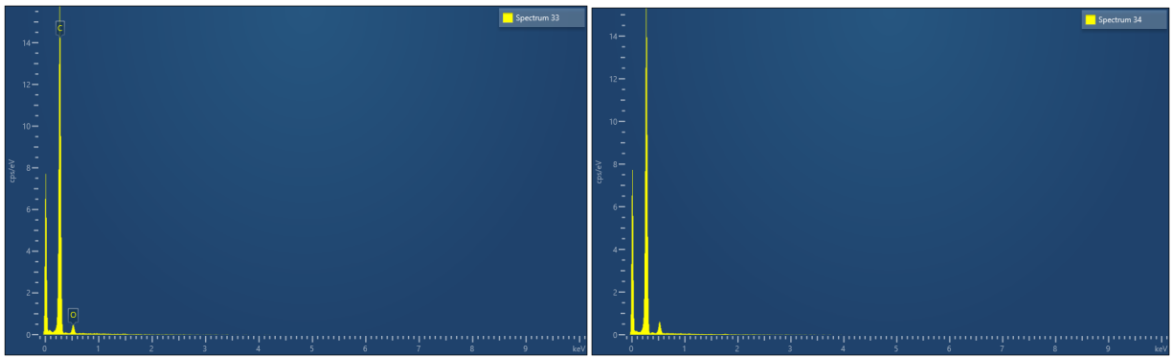

Supplemental figure 23. EDS scan for the annealed low loading fiber.

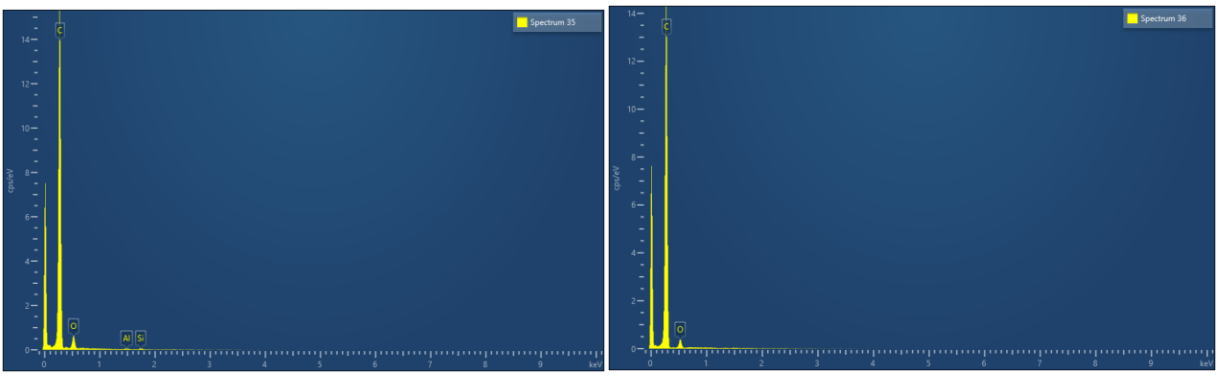

Supplemental figure 24. EDS scan for the annealed low loading fiber.

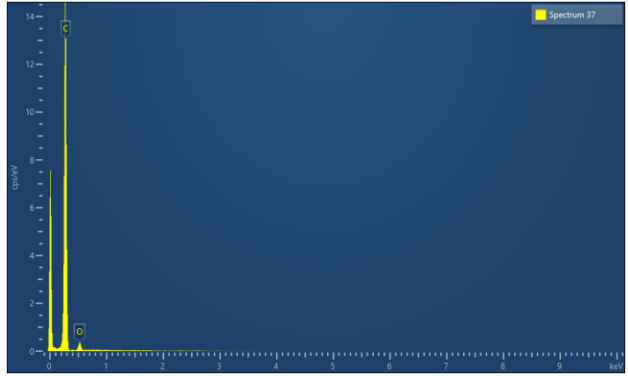

Supplemental figure 25. EDS scan for the annealed low loading fiber.

Supplemental Section 5: Raman

Polarized Raman spectra was obtained by a Renshaw Raman spectrometer with 785 nm, 633 nm, and 514 nm wavelengths under a 50x objective. The Raman was calibrated with a silicon standard beforehand and the laser was linearly polarized, which is either parallel to the fiber microstructure alignment or perpendicular to it. For 514 nm and 633 nm, the laser polarization could be flipped 90 ° without moving the sample; for 785 nm, the actual sample itself had to flipped 90°. Typically, one 10 s accumulation was obtained and laser power was varied to acquire a sufficient signal to noise ratio. Raman laser power settings were not changed when comparing the effects of laser polarization across any particular sample. Below are typical results for 514 nm for neat CNT fiber, C60, and C60 CNT fiber (in the as-is and annealed state). This 514 nm laser line had the most prominent spectral features of both C60 and CNTs, without broad background features. The complete Raman results are in the database.

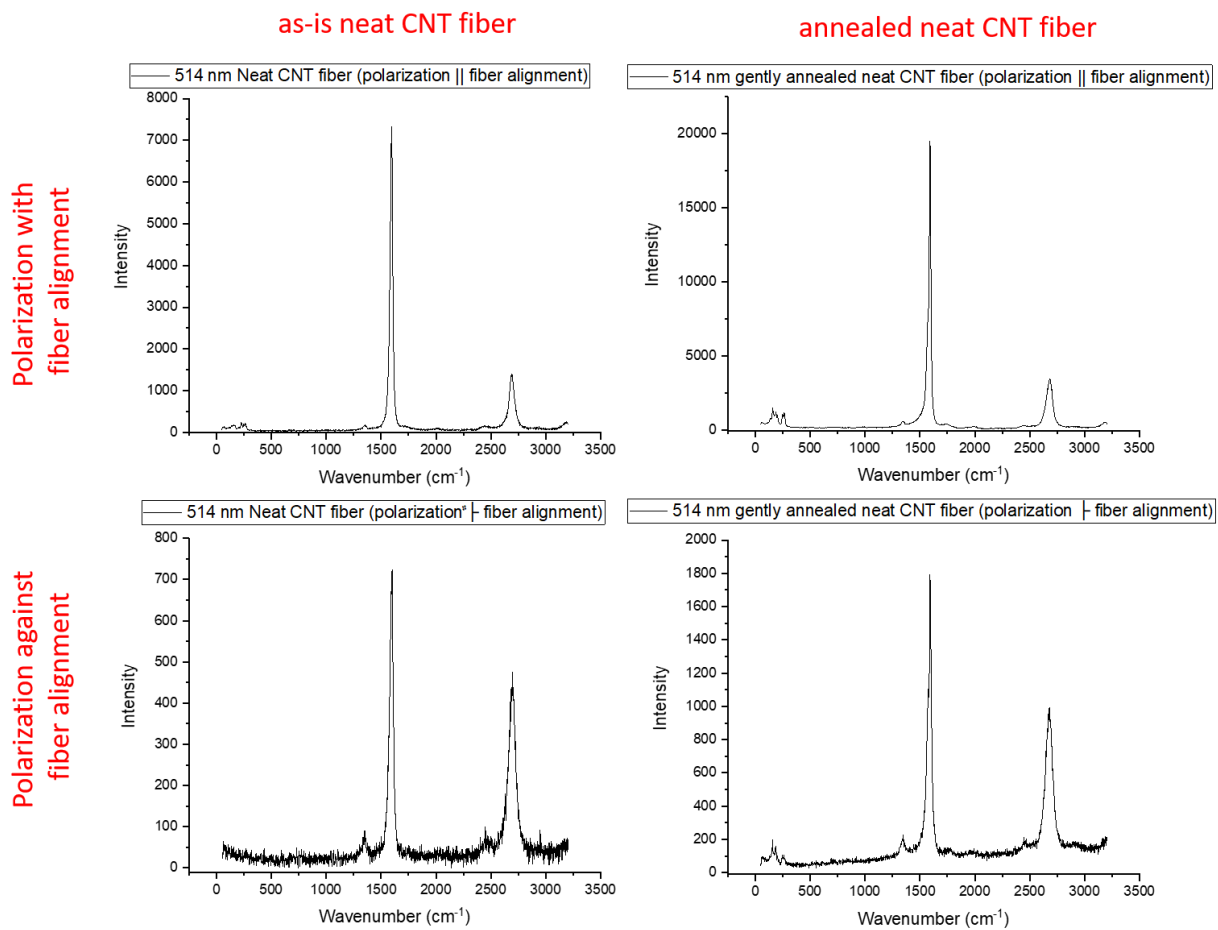

Supplemental figure 26. 514 nm Raman of neat CNT fiber: as-is and annealed, aligned with and against the laser polarization.

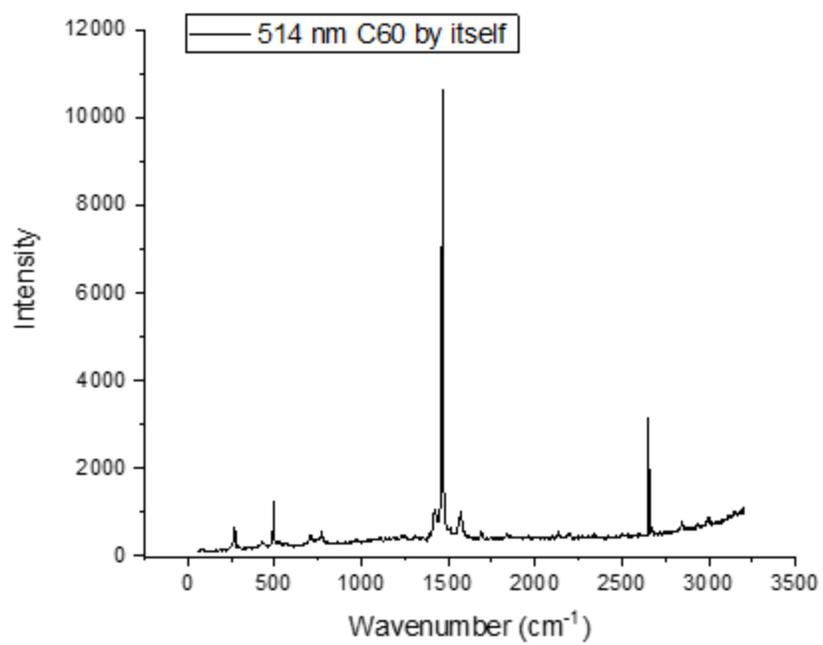

Supplemental figure 27. 514 nm Raman spectra of C60 flakes, as received.

| as-is low load C60 CNT fiber | as-is high load C60 CNT fiber |
|---|---|
| 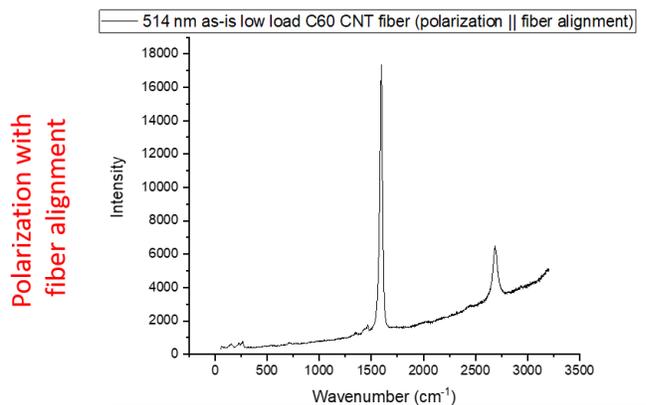 | 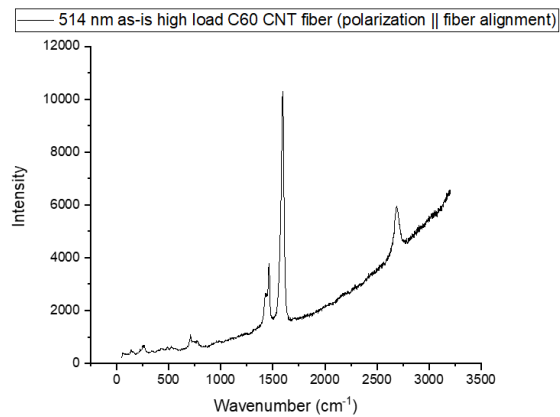 |
| 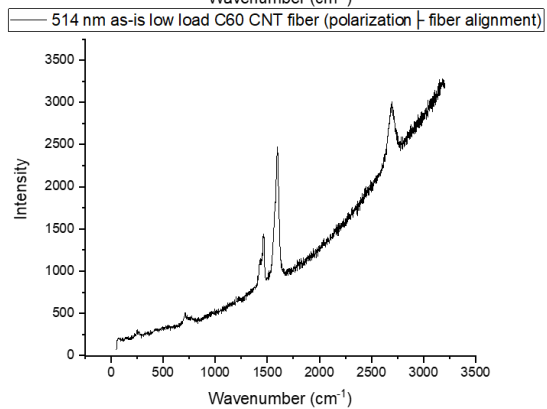 | 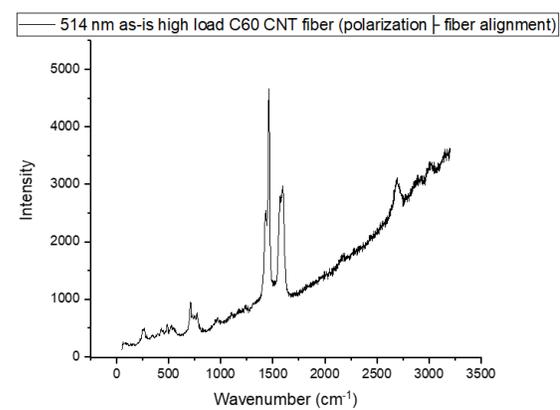 |

Polarization with fiber alignment (top row); Polarization against fiber alignment (bottom row).

Supplemental figure 28. 514 nm Raman spectra of as-is light and heavy load C60 CNT fiber, with and against the Raman laser polarization.

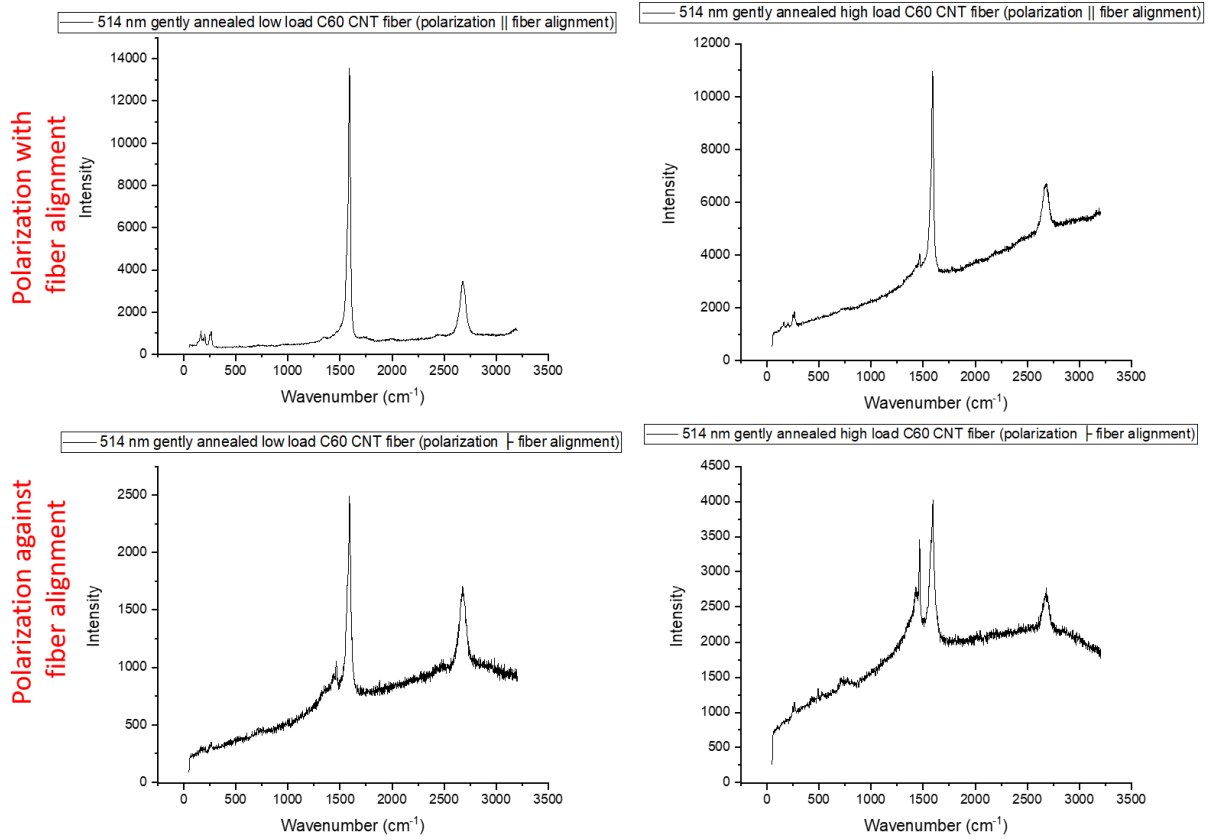

Supplemental figure 29. 514 nm Raman spectra of annealed light and heavy load C60 CNT fiber, with and against the Raman laser polarization.

Supplemental Section 6: NanoCT Scan

The Zeiss Xradia Ultra 810 NanoCT is an x-ray 3D tomographic microscope that provides 3D images of the surface and internal structure of the fiber. In this work we utilized its large field of view (65 µm), though it is capable of even higher resolution in the high resolution modes (50 nm spatial resolution, with 16nm voxel size). Because the diffraction hardware used to focus the beam makes the beam quasi-monochromatic, phase contrast imaging that dramatically improves the contrast of interfacial features in low-absorbing materials (e.g. carbon) is possible. In the absorption contrast mode, the grayscale intensity and corresponding assigned color of a voxel is proportional to the amount of X-ray absorption, which is a function of the atomic number and density. The user selected transfer function sets the exact color scheme and thresholds between the displayed color and X-ray absorption value. In the phase mode, however, the interpretation between intensity/color and material properties is not as straightforward. Below are the results of as-is high load C60 CNT fiber, in both absorption and phase mode.

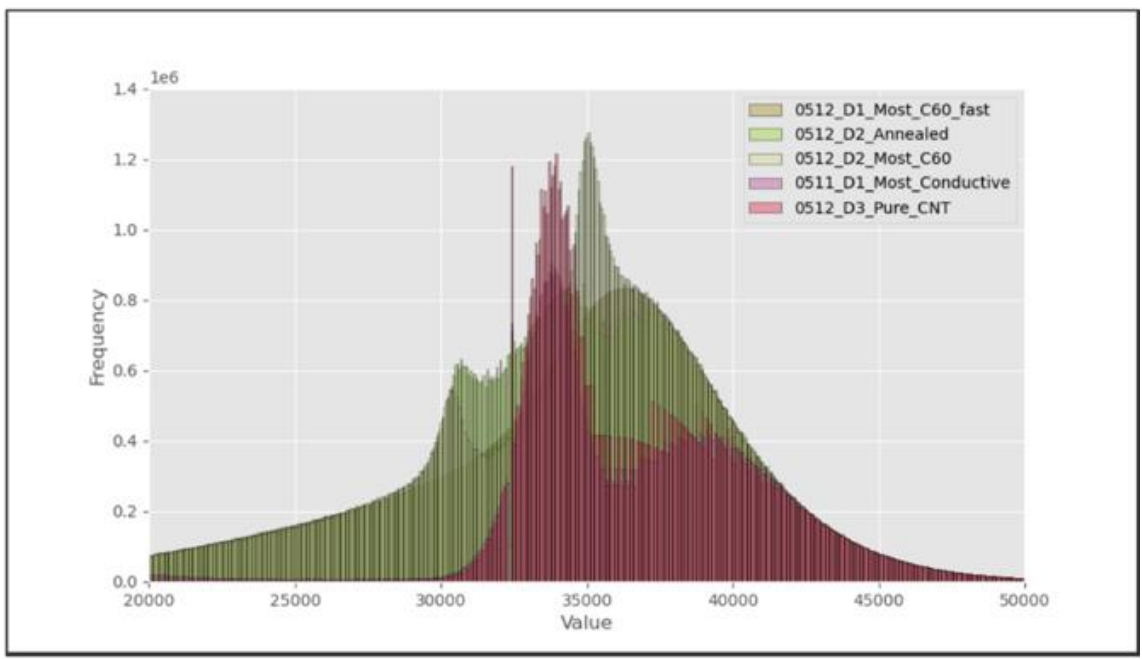

Supplemental figure 30. For the phase mode, these are the histograms of the normalized samples after removing the background.

Below is the absorption and phase mode data for the as is neat CNT fiber.

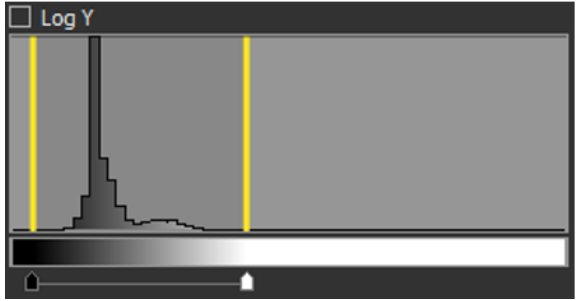 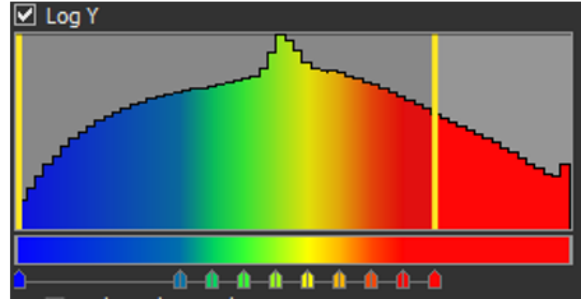

Supplemental figure 31. Histograms for the 2D absorption (grayscale) and 2D phase contrast data, with yellow bars indicating the display range (3,000 to 28,000 for absorption, 2,000 to 50,000 for phase). For each imaging mode, voxels with intensities less than the stated range are not displayed while voxels with intensities greater than the stated range are displayed with the same grayscale / color value as the highest value in the range.

Absorption | Phase

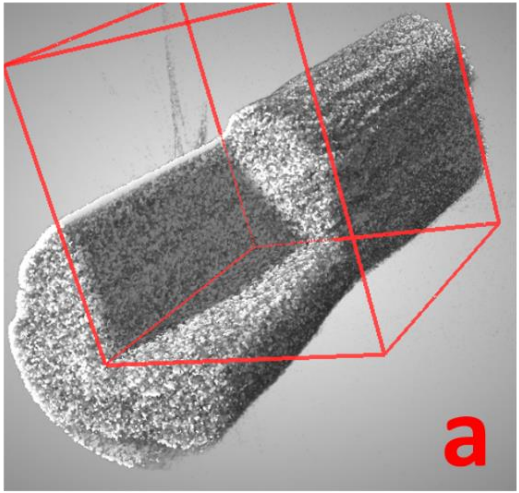 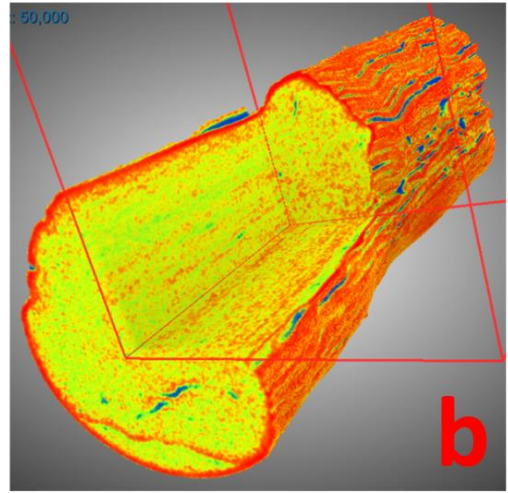

a | b

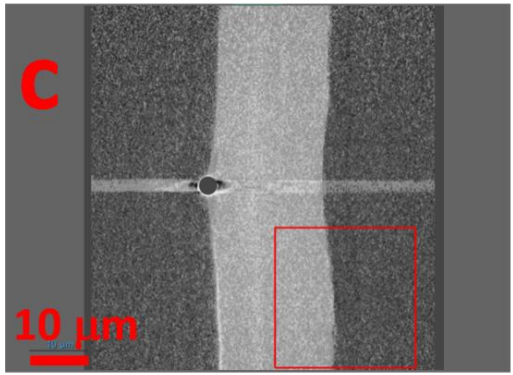 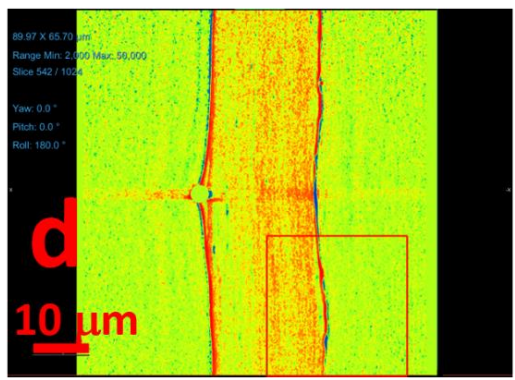

10 μm | 10 μm

c | d

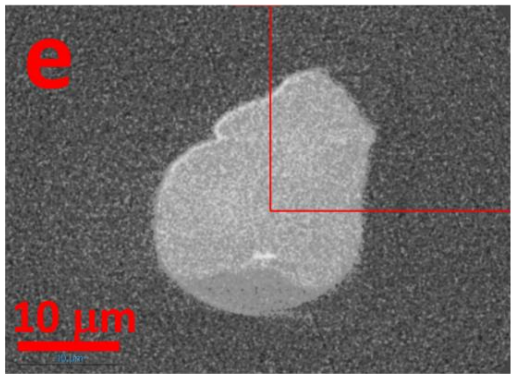 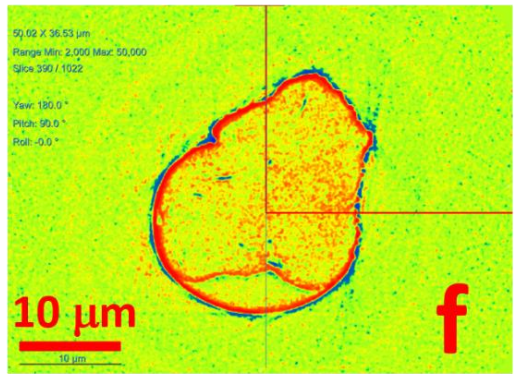

10 μm | 10 μm

e | f

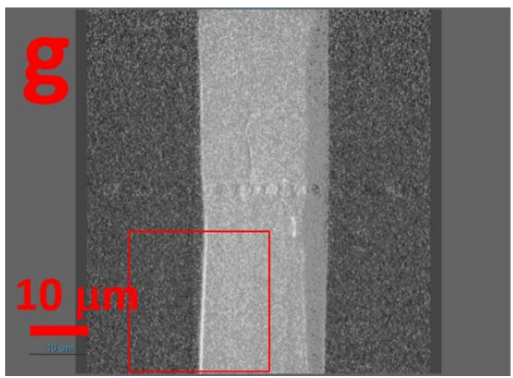 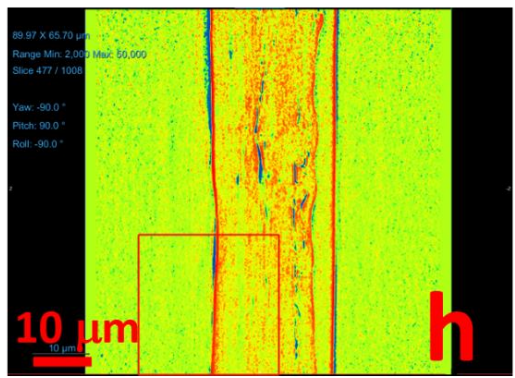

10 μm | 10 μm

g | h

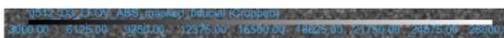 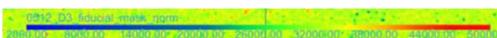

Supplemental figure 32. For the as is neat CNT fiber, absorption (a, c, e, g ) and phase (b,d,f,h) NanoCT scan images. For clarity, the phase contrasting 3D image (b) has the background subtracted and in both 3D images the red box (25 μmx 25 μm x25 μm) cuts away data to better view the internal structure. Faces of the red box correspond to the 2D cross sections (c,d,e,f,g,h) where the background has not been subtracted out. At the bottom of the picture are the color scales, the colors normalized to represent the same values across all samples. A gold fiduciary particle was placed on the side of the fiber to ensure image alignment during reconstruction; the brightest voxels in the fiducial were reassigned values of the average background intensity prior to normalization.

Below is the absorption and phase data for the as is low loading fiber.

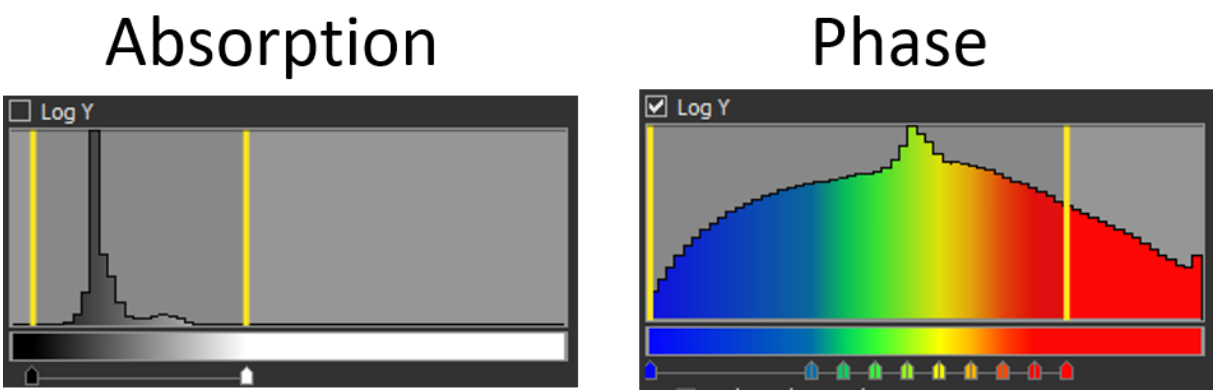

Supplemental figure 33. For the as is low loading fiber, absorption and phase mode histogram showing that the color scheme for the 2D images (yellow bars) is set to the standard 3000 to 28000 (grayscale, for absorption) and 2000 to 50,000 (rainbow, for phase).

| Absorption | Phase |
|---|---|
| 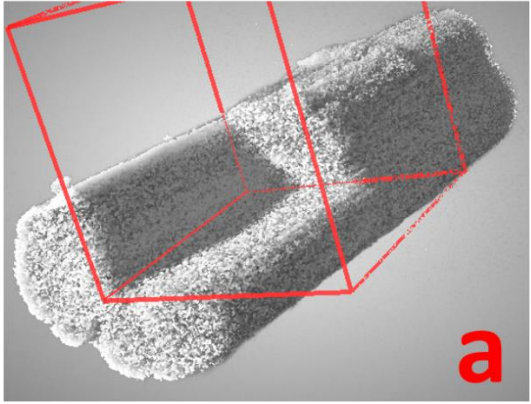 a | 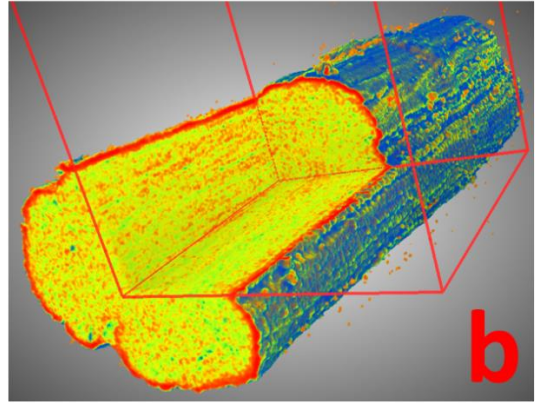 b |
| 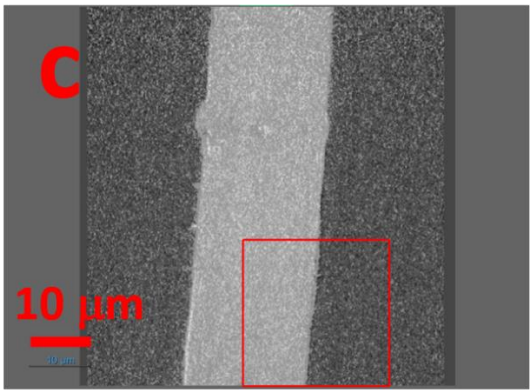 c | 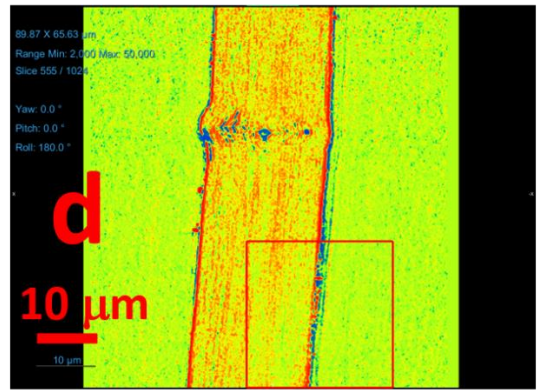 d |
| 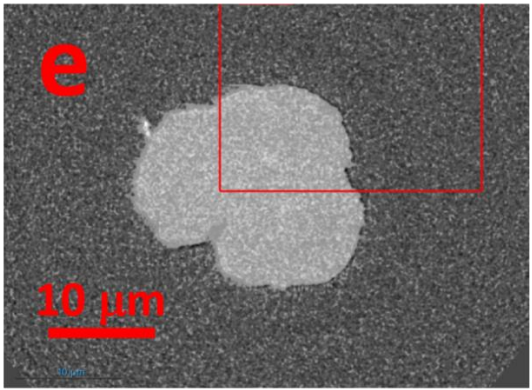 e | 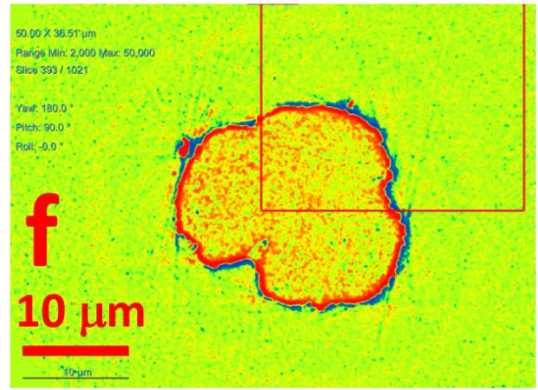 f |
| 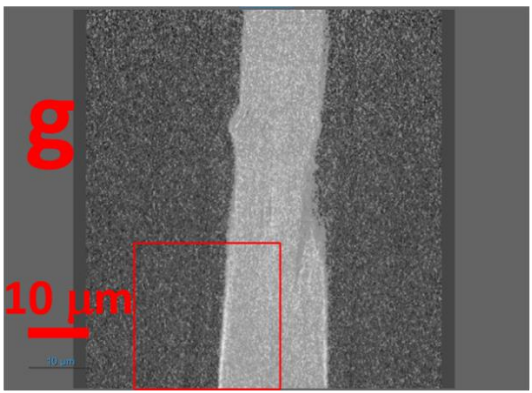 g | 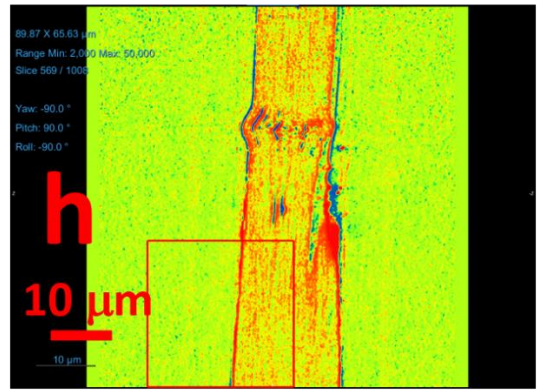 h |

Supplemental figure 34. For the as is low loading fiber, absorption (a, c, e, g ) and phase (b,d,f,h) 2D cross sections and 3D reconstruction. For clarity, the phase contrasting 3D image (b) has the background subtracted and in both 3D images the red box (25 μmx 25 μm  x25 μm) cuts away data to better view the internal structure. Faces of the red box correspond to the 2D cross sections (c,d,e,f,g,h) where the background has not been subtracted out. At the bottom of the picture are the color scales, the colors normalized to represent the same values across all samples.

Below is the phase mode data for the as is high loading fiber.

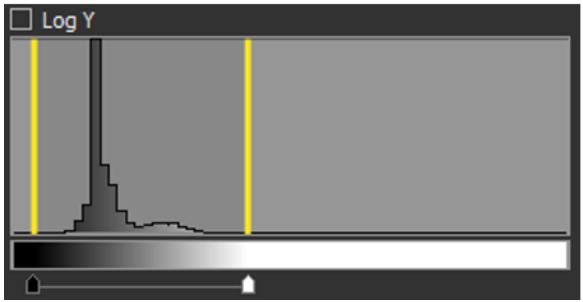
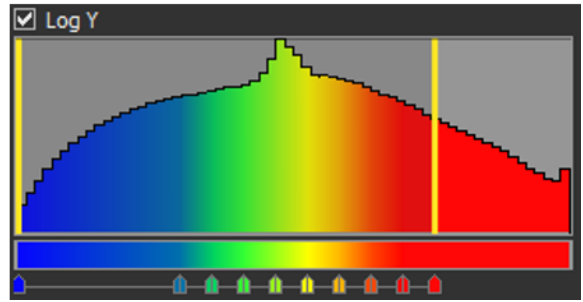

Supplemental figure 35. For the as is high loading fiber, absorption and phase mode histogram showing that the color scheme for the 2D images (yellow bars) is set to the standard 3000 to 28000 (grayscale, for absorption) and 2000 to 50,000 (rainbow, for phase).

| Absorption | Phase |
|---|---|
| 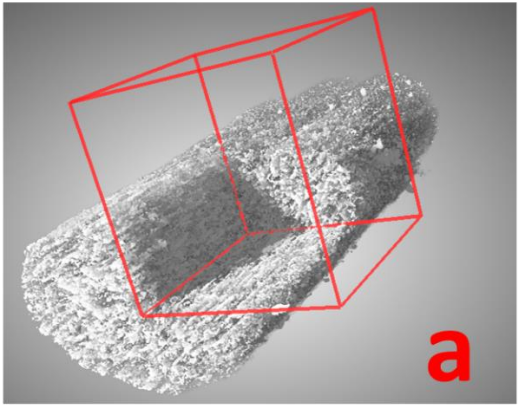 a | 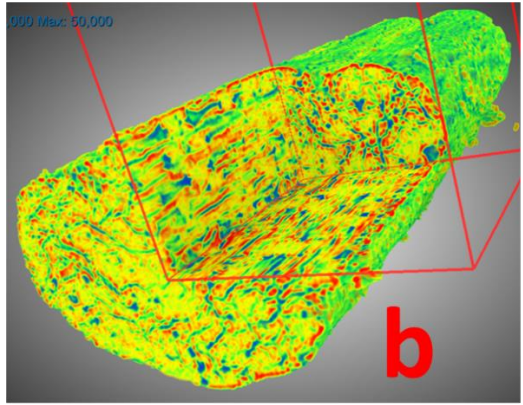 b |
| 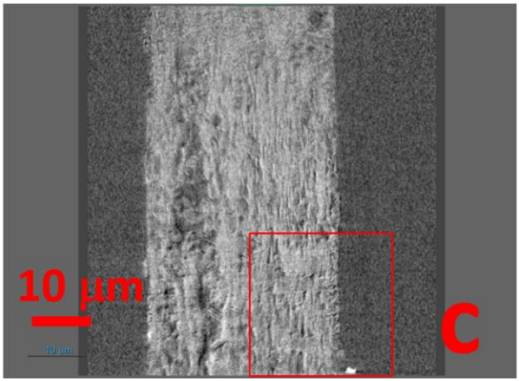 c | 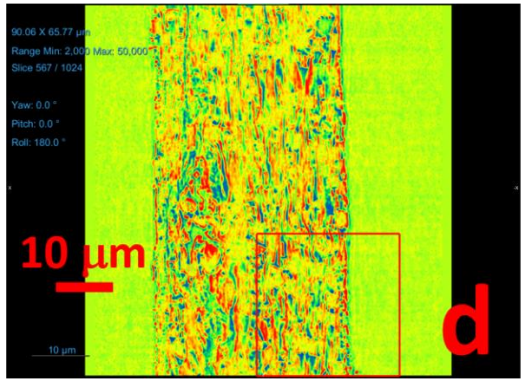 d |
| 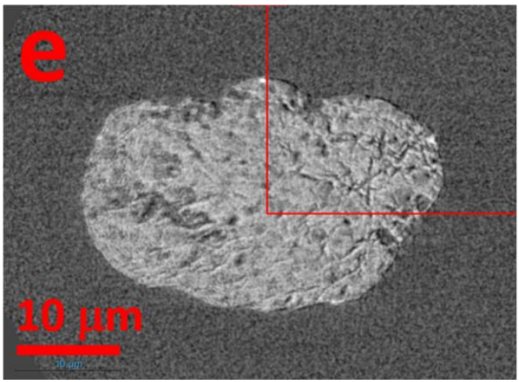 e | 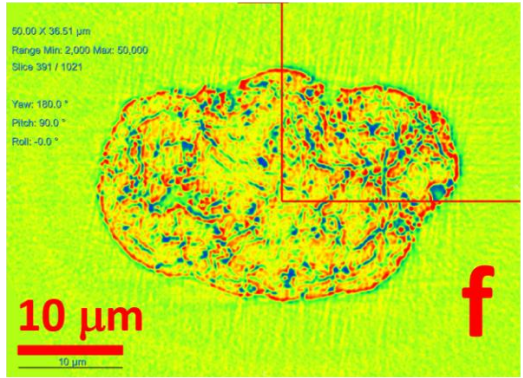 f |
| 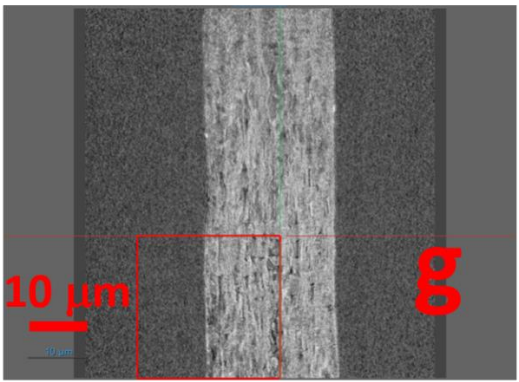 g | 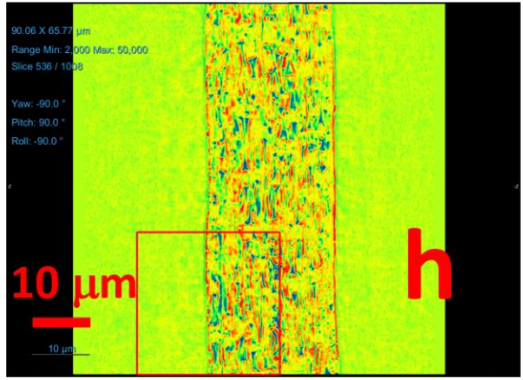 h |

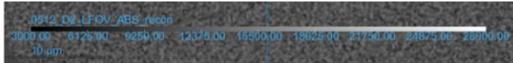
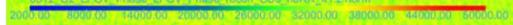

Supplemental figure 36. For the as is high loading fiber, absorption (a, c, e, g ) and phase (b,d,f,h) 2D cross sections and 3D reconstruction. The phase contrasting 3D image (b) has the background subtracted and in both 3D images the red box (25 μmx 25 μm  x25 μm) cuts away data to better view the internal structure. Faces of the red box correspond to the 2D cross sections (c,d,e,f,g,h) where the background has not been subtracted out. At the bottom of the picture are the color scales, the colors normalized to represent the same values across all samples.

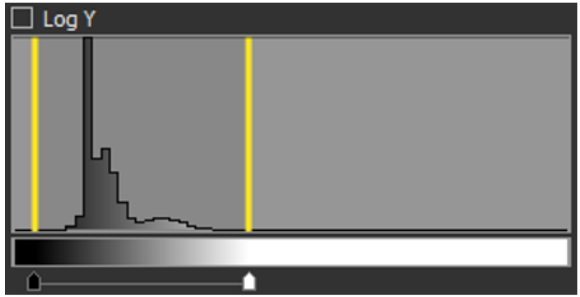
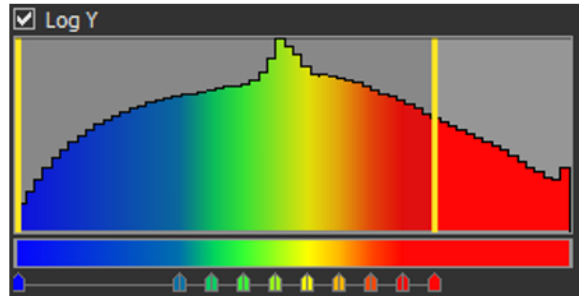

Supplemental figure 37. For the annealed high loading fiber, absorption and phase mode histograms showing that the 2D color scheme (yellow bars) is set to the standard 3000 to 28000 (grayscale, for absorption) and 2000 to 50,000 (rainbow, for phase).

Absorption | Phase

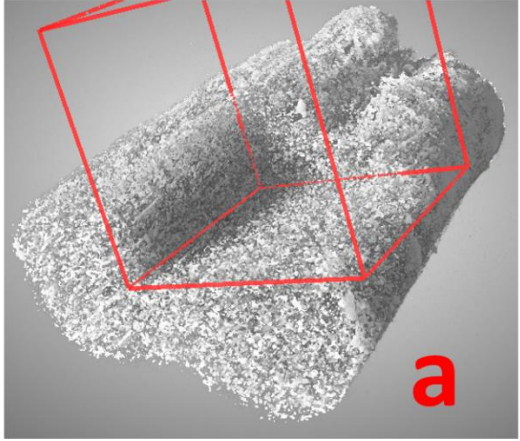
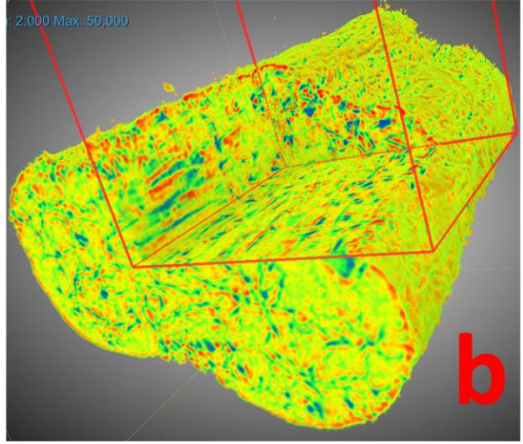
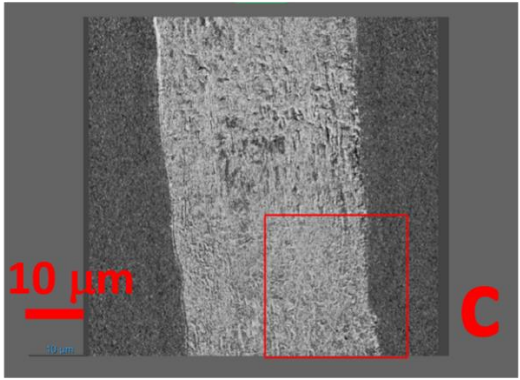
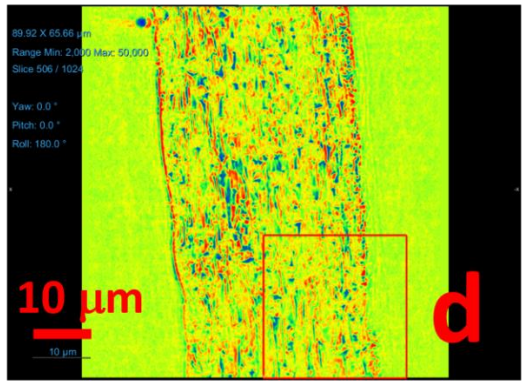
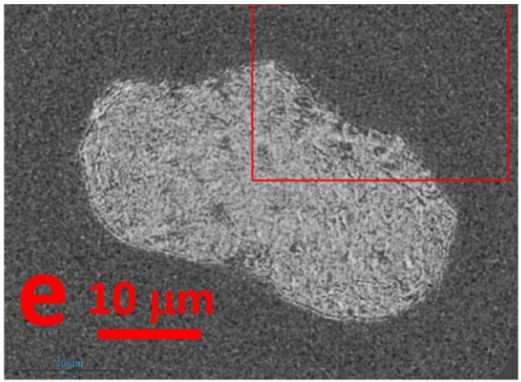
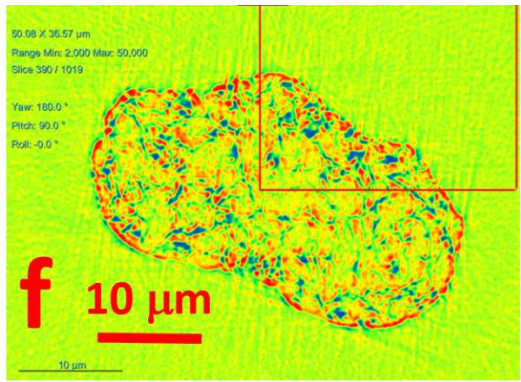
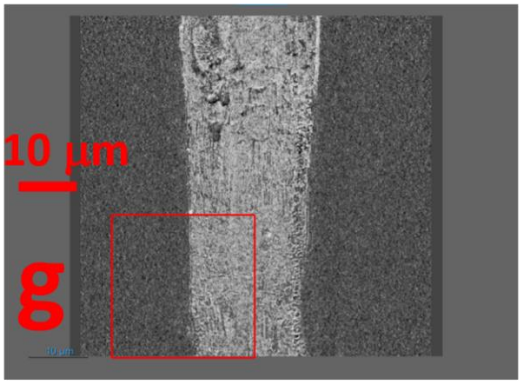
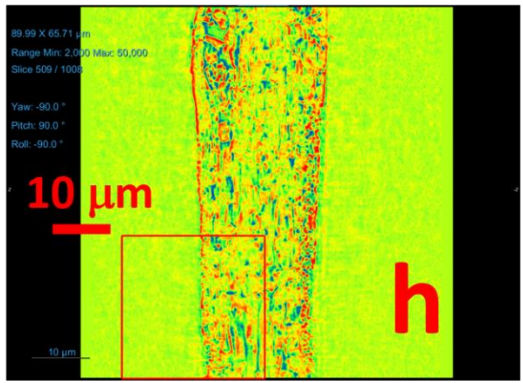

Supplemental figure 38. For the annealed high loading fiber, absorption (a, c, e, g ) and phase (b,d,f,h) 2D cross section and 3D reconstruction. The phase contrasting 3D image (b) has the background subtracted and in both 3D images the red box (25 µmx 25 µm  x25 µm) cuts away data to better view the internal structure. Faces of the red box correspond to the 2D cross sections (c,d,e,f,g,h) where the background has not been subtracted out. At the bottom of the picture are the color scales, the colors normalized to represent the same values across all samples.

Supplemental Section 7: TEM

Now we show representative TEM images of both low and high loading C60 CNT fibers, after thinning lamella sections using the FIB. Regardless of FIB beam current, the lamella bent and bowed substantially with 30kV when below 500 nm thickness. FIB curtaining artifacts were difficult to avoid because of the slow milling rate of carbon, and the relatively large size of the fibers with respect to traditional lifted-out FIB lamella. Varying the ion milling direction did reduce the visibility of the curtaining artifacts in the low-load fiber(figure 3b); however this was only effective in areas where curtaining was already fairly low.

Amorphous surface damage left behind from FIB preparation can be seen in the HRTEM image of the high load fiber, figure 3e. This damage persisted even after milling with a low energy Argon ion mill. This thin damaged layer did not disrupt the fiber structure, as we can resolve the nanotubes in all higher magnification images.

As is, low loading. First, we show TEM images from the as is low loading CNT fiber (supplemental figure 39). This shows small oval inclusions oriented in the direction of fiber alignment, with the small oval inclusions aligned into rows. These inclusions appear between the white triangular inclusions, which are thinned sections of CNTs. In the background there are faint vertical striations that is also in the direction of fiber alignment. The lamella was approximately 200 nm thick, which was more susceptible to curtaining effects (causing bold streaks running left to right) than we will see in the high loading C60 CNT fiber. Supplemental figure 40 and 41 are zoom-ins; some selected inclusions in their largest dimension span 130 to 170 nm in length. Note images here are using an objective aperture to increase contrast . Note the use of the objective aperture did not enhance contrast between the nanotubes and  the inclusions because the inclusions were not Strongly diffracting/crystalline.

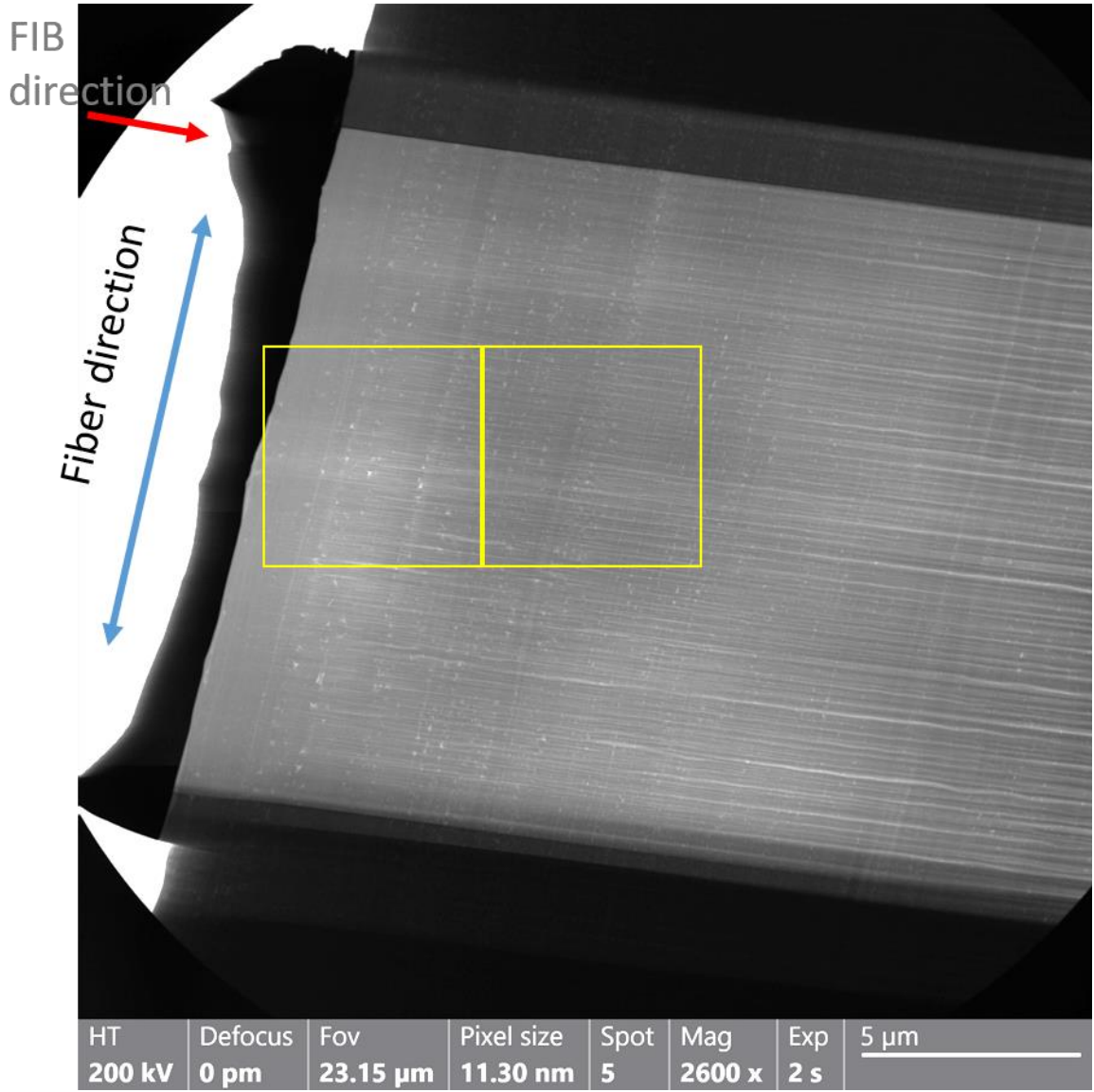

Supplemental figure 39. Bright field TEM photograph of low load C60 CNT fiber. Faint vertical striations run up and down in the direction of fiber alignment; bolder horizontal steaks are from curtaining. Yellow highlighted regions are shown in supplemental figures 40 and 41 below.

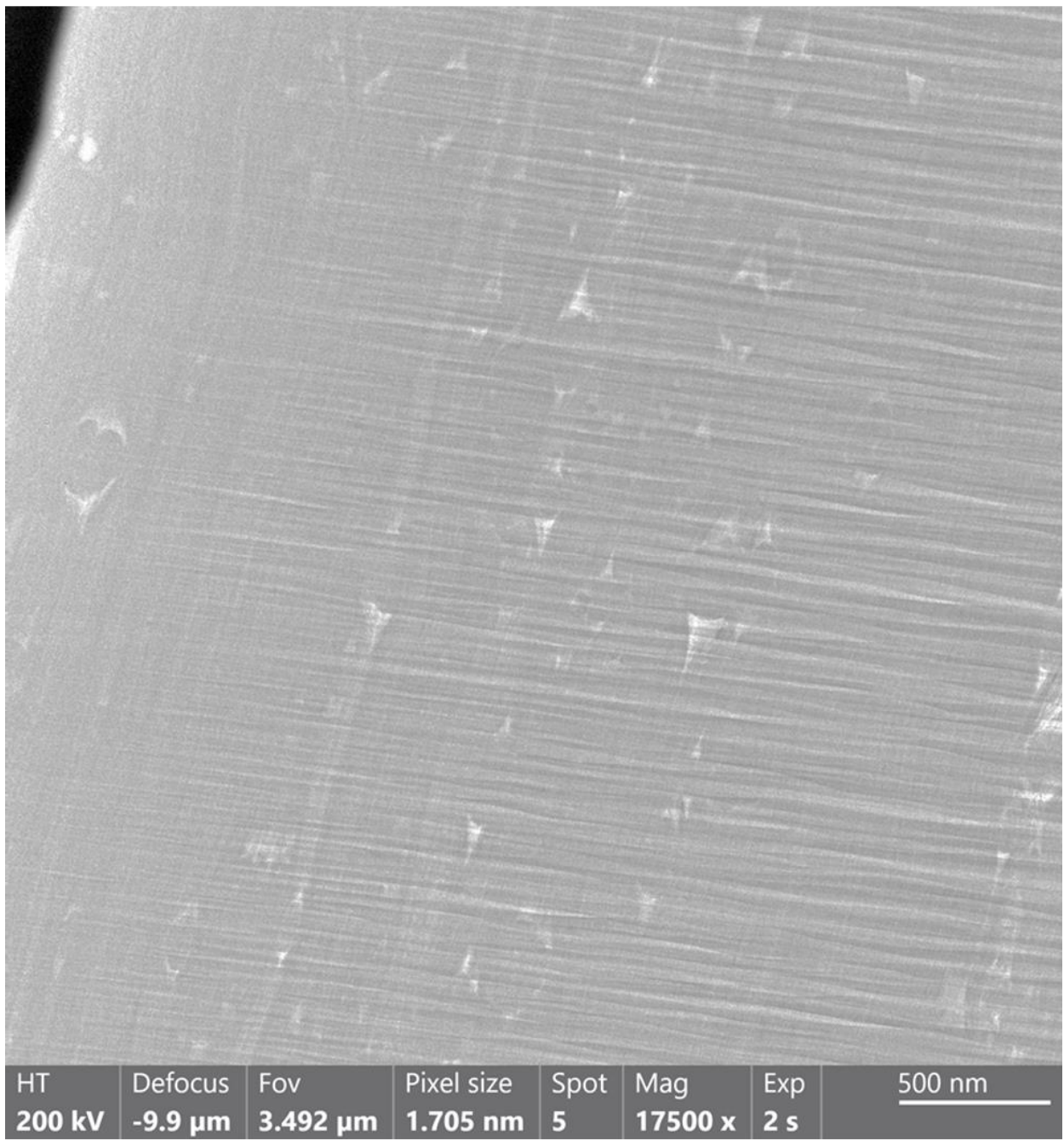

Supplemental figure 40. Bright field TEM photograph of low load C60 CNT fiber—the left highlighted box in Supplemental figure 39.

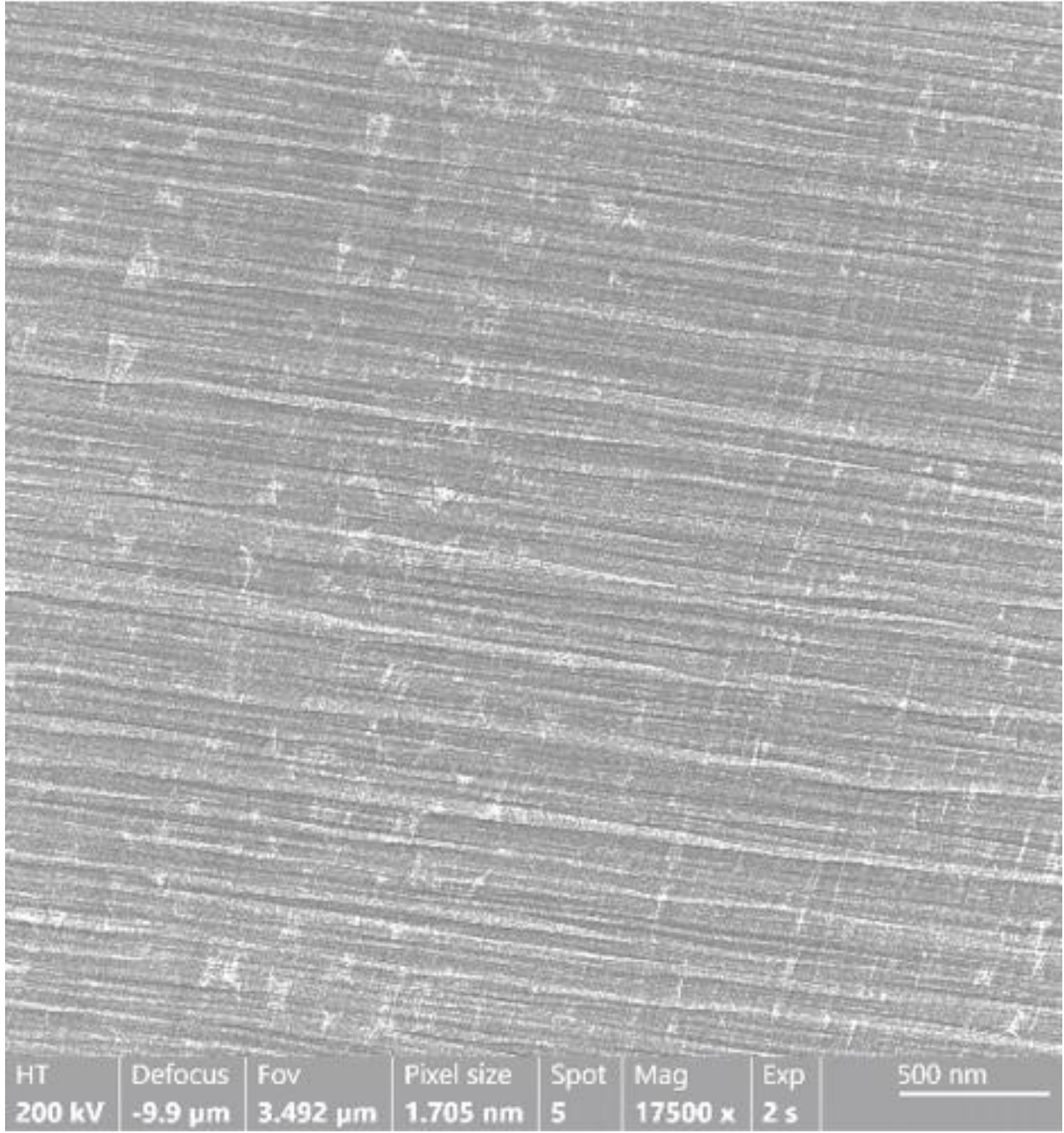

Supplemental figure 41. Bright field TEM photograph of low load C60 CNT fiber—the right highlighted box in Supplemental figure 39.

Now we will discuss the crystalline and amorphous features of the as is low loading C60 CNT fiber. Below we have a representative TEM image illuminating a selected area that contains the inclusions (supplemental figure 42) and the associated diffraction pattern from that selected area (supplemental figure 43). The diffraction pattern shows the familiar 2.2 Å spacing of the carbon carbon bond, as well as the ~4 Å lobes from CNT spacing and alignment in the fiber direction. Considering that there are no other diffraction peaks present, as well as a lack of diffraction contrast as previously noted, it suggests that the inclusions shown in the as is low loading C60 CNT fiber are amorphous. In supplemental figure 44, we show a high-resolution TEM photograph of the inclusion itself. This area was thinned with Ar/O plasma to remove surface damage; however, the plasma treatment still left behind amorphous surface damage all over the area as indicated by the dashed shape in the image. A Fourier transform of just the inclusion region (supplemental figure 45) shows that here the inclusions are amorphous.

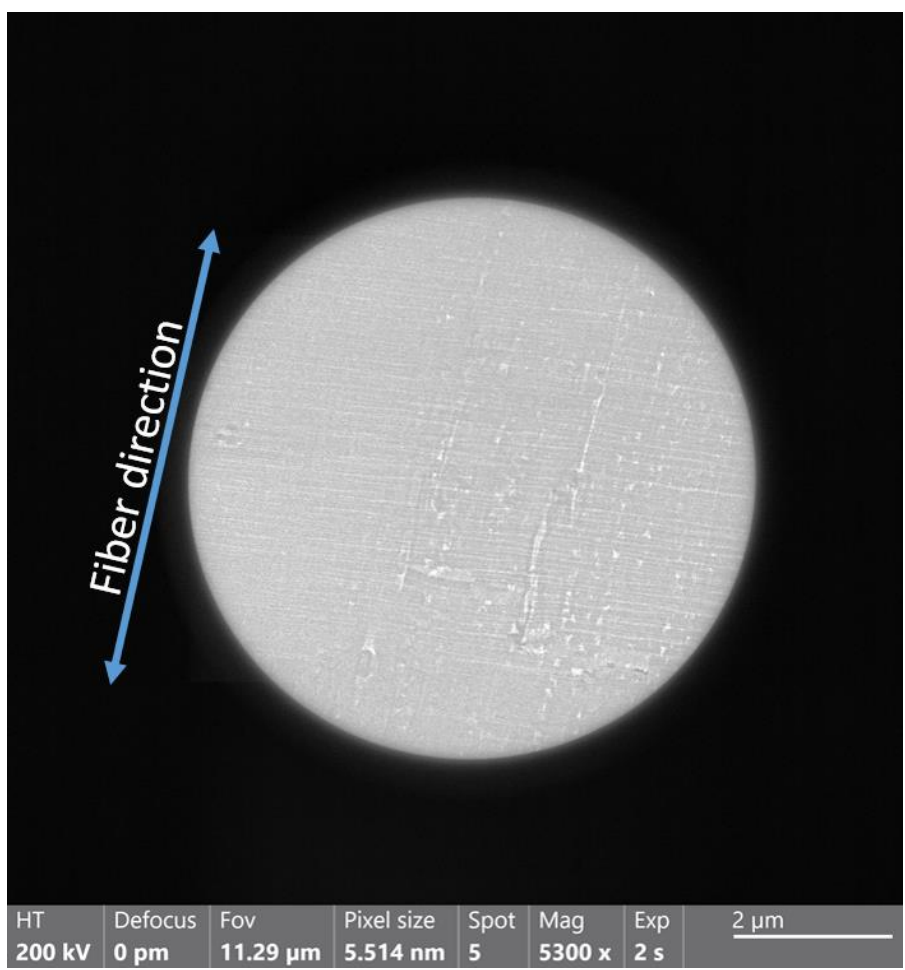

Supplemental figure 42. For the as is low loading C60 CNT fiber, TEM photograph of the region undergoing electron diffraction as shown below.

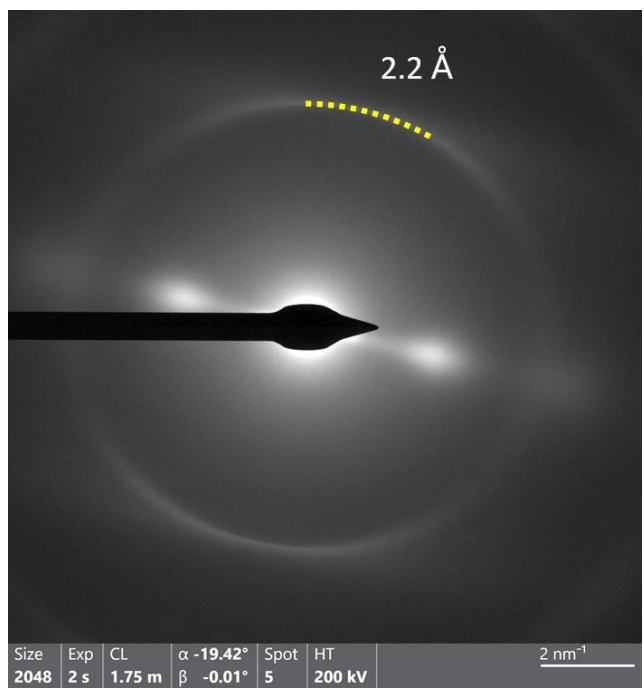

Supplemental figure 43. For the as is low loading C60 CNT fiber, electron diffraction pattern of a region with inclusions and this resembles a typical CNT electron diffraction pattern.

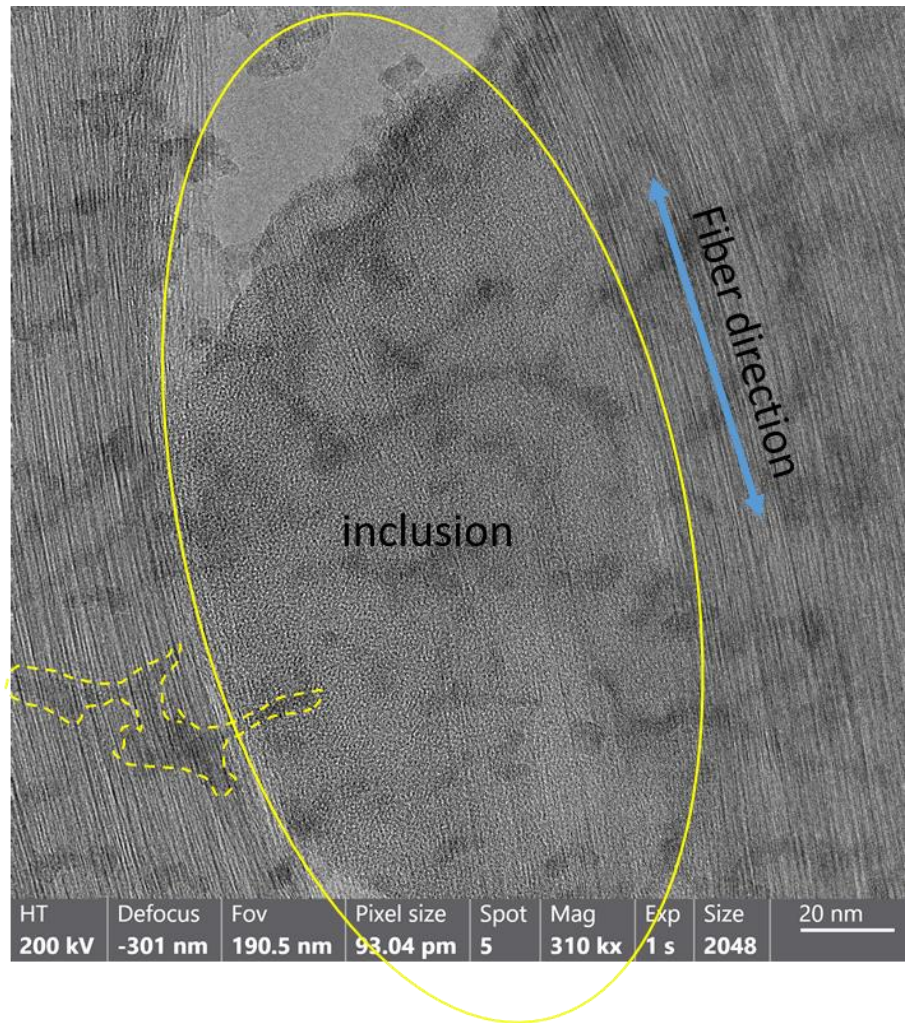

Supplemental figure 44. Zoom on an inclusion as highlighted in the yellow oval. Dashed line indicates an example of amorphous surface damage caused by plasma treatment.

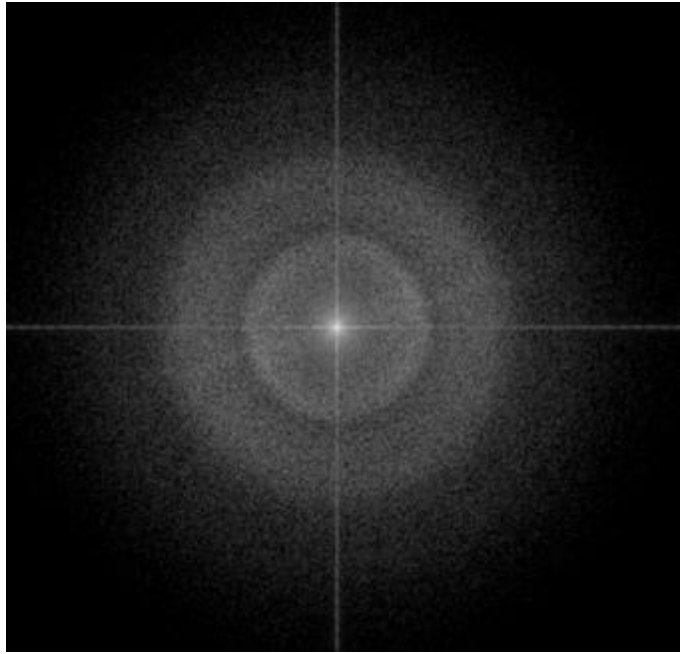

Supplemental figure 45. Fourier transform of the inclusion region, showing it is amorphous with rings around 8 Å and broad ring 4.5 Å. The 8 Å spacing indicates the typical length scale of the detail in the amorphous inclusion.

Annealed, low loading. TEM was accomplished for annealed, low loading C60 CNT fiber. Supplemental figure 46 and 47 are overview images, which look similar to the as is, low loading C60 CNT fiber with the same small inclusions.

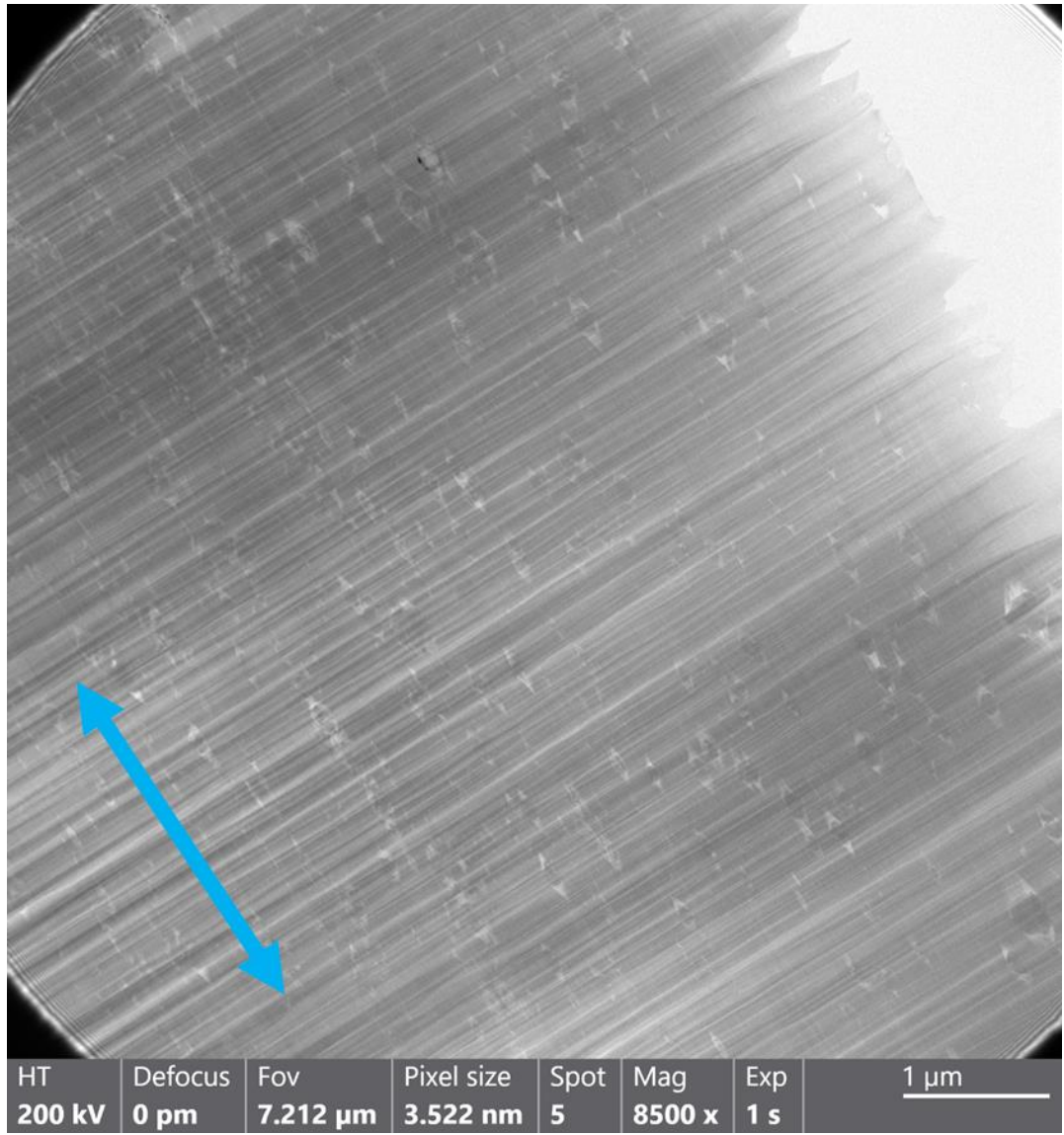

Supplemental figure 46. TEM image of the annealed, low loading C60 CNT fiber. The blue arrow indicates fiber alignment direction. This image shows similar inclusions as the as is case.

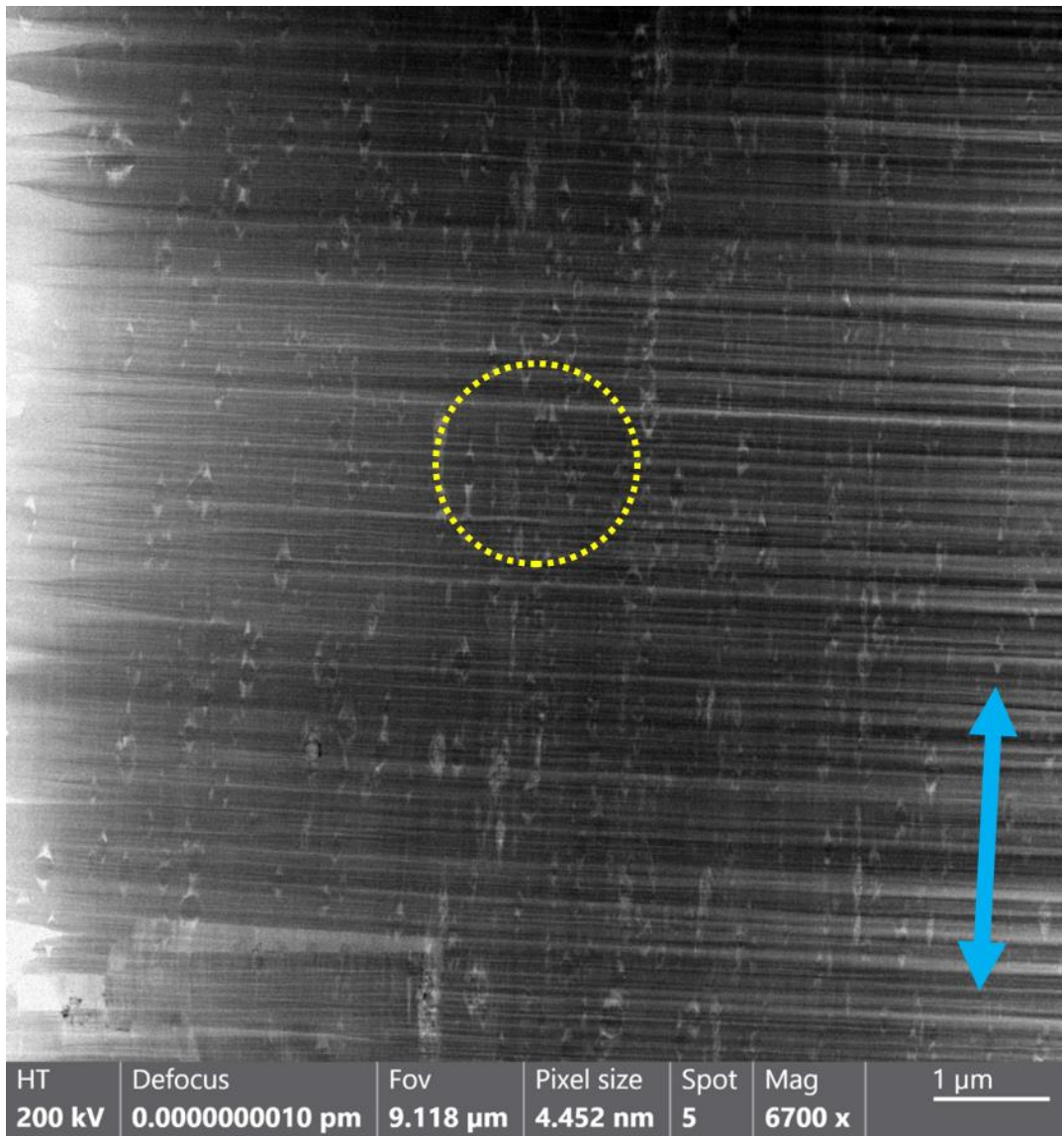

Supplemental figure 47. TEM image of the annealed, low loading C60 CNT fiber. This shows similar inclusions as the as is case. The blue arrow indicates fiber alignment direction and the dashed yellow circle indicate the illuminated region contributing to the next diffraction images. The objective aperture was used to enhance contrast from any crystalline grains. No inclusions are significantly darker than the surrounding matrix meaning there are no large single crystalline grains.

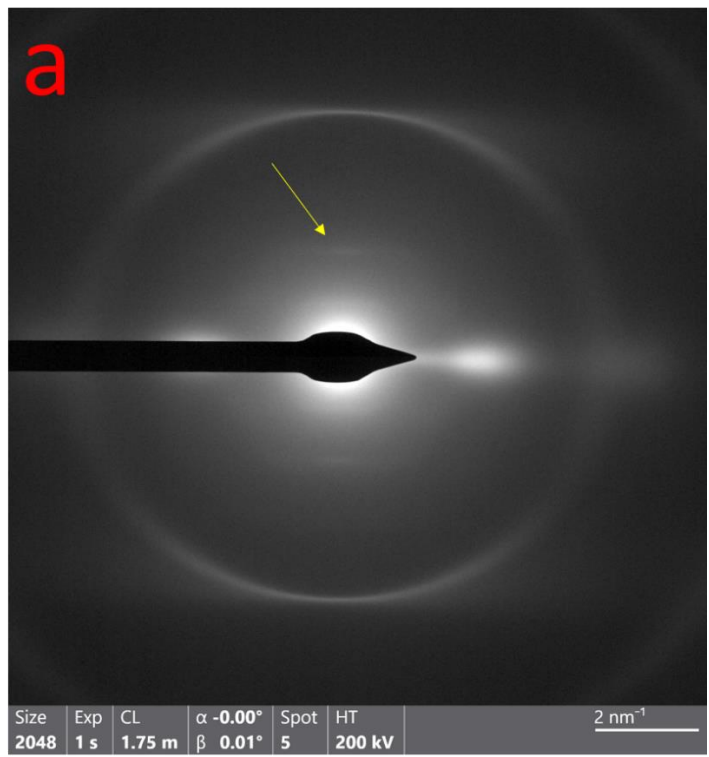

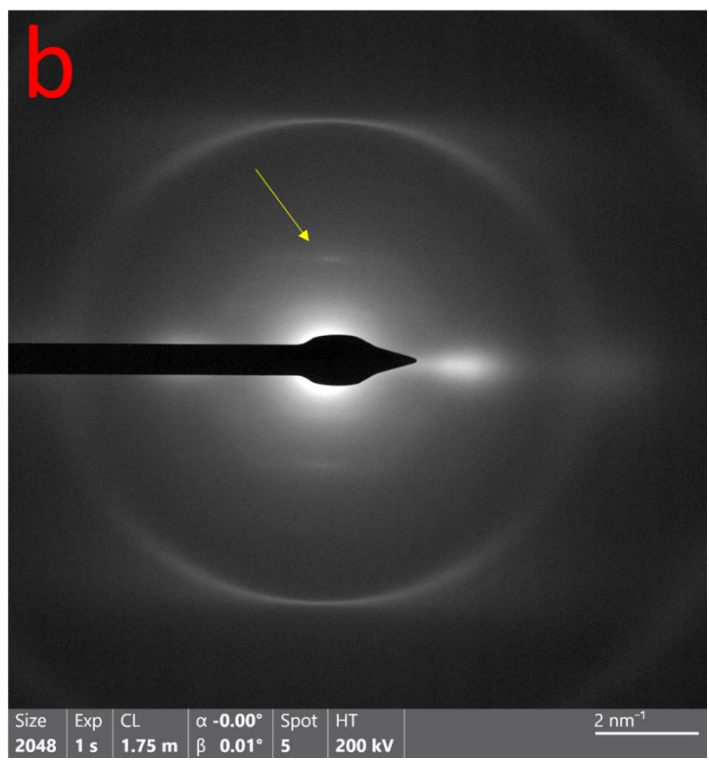

Supplemental figure 48. Electron diffraction for the TEM image above for the annealed, low loading C60 CNT fiber from a, the majority of the lamella and b, for just the selected region from the previous supplemental figure. We see spots and possibly the development of a ring at 0.5 nm spacing (see yellow

arrow). For the selected region of interest, this diffraction feature is more pronounced because the inclusion is a larger relative volume fraction of the selected area vs the entire fiber.

High-angle annular dark-field (HAADF) scanning transmissionelectron microscopy (STEM) is highly sensitive to thickness and density and was used for the annealed low loading C60 CNT fiber. Supplemental figure 49 a shows a similar contrast between the inclusions and the surrounding nanotube bundles. This indicates there is not a substantial difference between thickness or density between the CNTs and C60. Figure 49 b is a high resolution TEM image of one of the inclusions and Figure 49 c is a Fourier transform of this image. The Fourier image shows a diffuse ring at ~0.8 A and ~0.4 A, indicative of short range ordering.

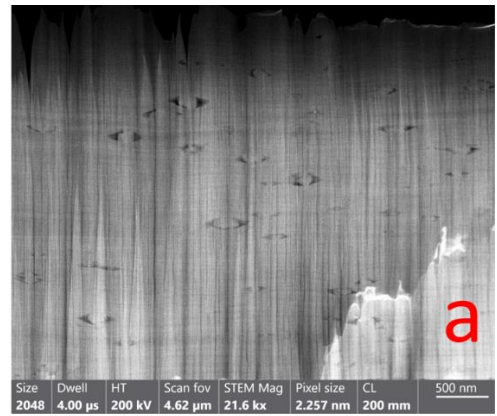

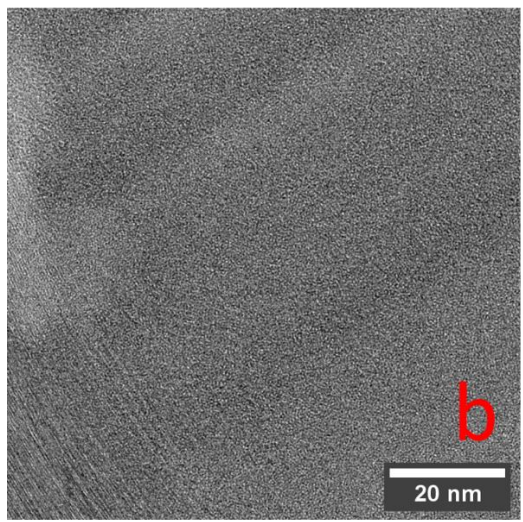

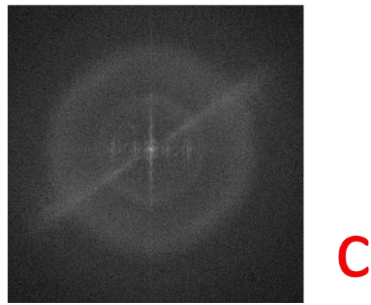

Supplemental figure 49. STEM imaging of the annealed low loading C60 CNT fiber. a, HAADF STEM image. b, High resolution TEM image of an inclusion and c, the Fourier transform.

Dark field TEM filters the electron diffraction information to select specific diffraction features, filter out the unselected data, and construct a TEM image from electrons contributing to the selected diffraction information. Supplemental figure 50 shows the various diffraction features we selected and color coded. Their reconstructed images are in the following supplemental figures. Following are some observations: In red-- the direct beam only (bright field), shows no dark patches where inclusions are located indicating there are no strongly diffracting inclusions. In green—low angle scattering and small part of first diffraction ring (dark field) -- this shows mostly the thickness/ density of the lamella with some enhanced contrast in the inclusions from a small part of the 0.5 nm spacing diffraction ring being caught in the aperture. In blue—centered around the faint inner 0.5 nm diffraction ring (dark field) -- this filtering highlights the inclusions indicating the inclusions contain ordering, albeit mostly weak short range ordering. In yellow—Centered around the outer ring, this just shows CNTs (dark field). In black—centered around the strong lobe, this shows well aligned CNT bundles (dark field).

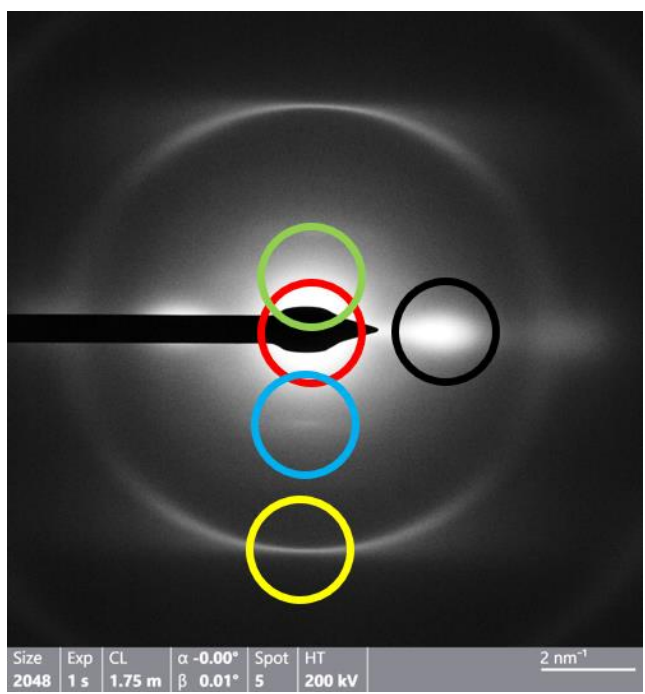

Supplemental figure 50. For the annealed low loading C60 CNT fiber, we select diffraction features as depicted in the color coded circles. An image will be reconstructed from each selected feature with the other diffraction data filtered out. These reconstructed color coded images are below.

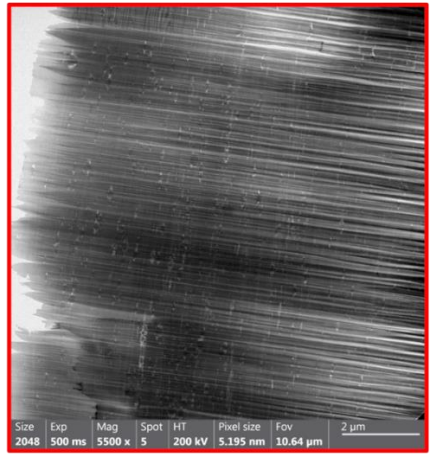
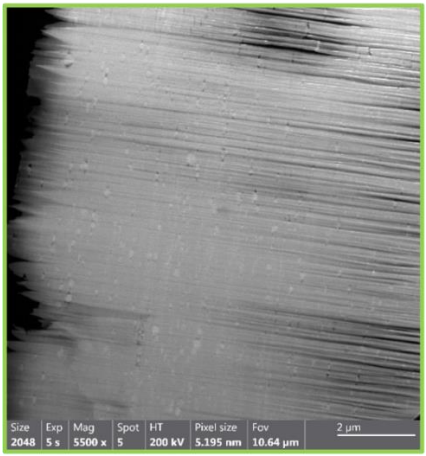
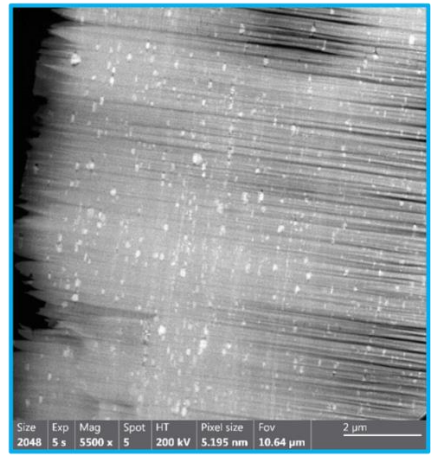
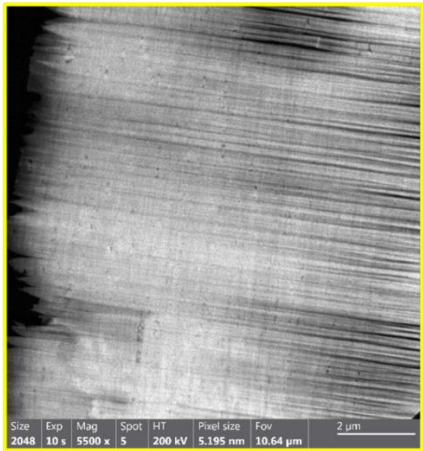
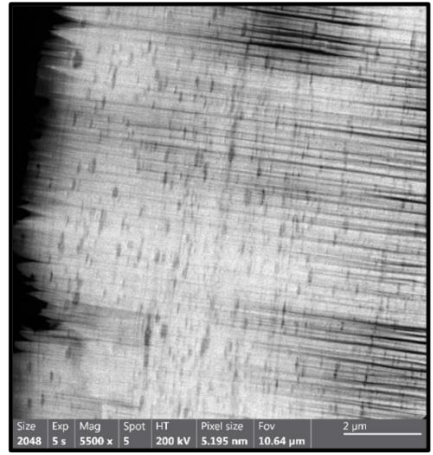

Supplemental figure 51. Reconstruction of the diffraction pattern, where each color coded photograph is reconstructed from a selected diffraction feature with the rest of the diffraction data filtered out.

For comparison, we did similar diffraction feature selection analysis for the as-is low loading C60 CNT fiber. a, HAADF STEM image showing that the inclusions do not have substantially different thickness or density than the surrounding CNTs. b, Image reconstructed from just the direct beam and filtering out the rest of the data (bright field). This shows no strongly diffracting inclusions. c, Image reconstructed from just selecting the weak 0.5 nm spacing diffraction feature and filtering out the rest of the data (dark field). This diffraction feature was too faint to see, although its location was inferred from the annealed fiber. This shows highlighted inclusions and implies they have short range order although to a lesser degree than the annealed case. d, Image reconstructed from just selecting diffraction features around the outer 2.2 nm ring and filtering out the rest of the data (dark field). This shows CNTs and inclusions which both contain 2.2 nm carbon-carbon bonds.

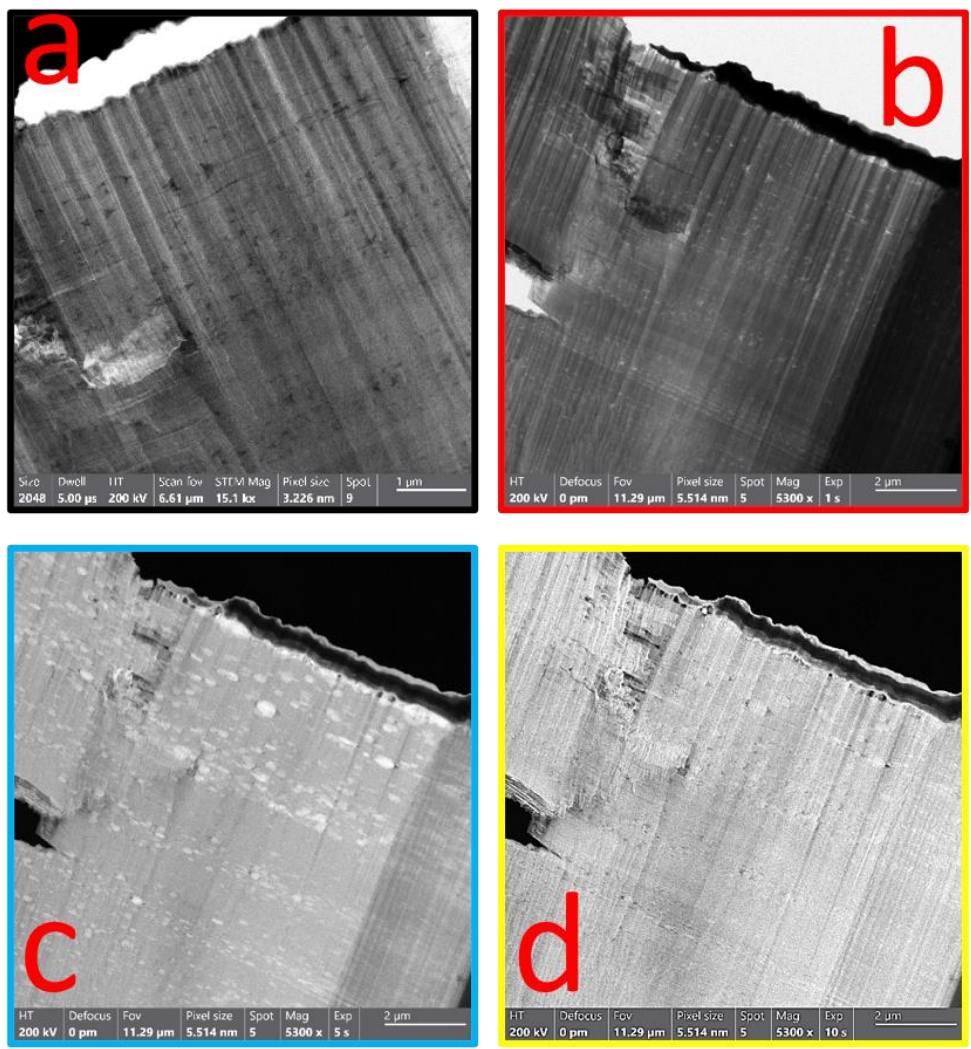

Supplemental figure 52. For the as is low loading CNT fiber, where certain features of the diffraction pattern are exclusively selected to reconstruct the image. a, HAADF STEM. b, image from direct beam only. C, image from inner ring (although was too faint to see). d, image from outer ring.

As is, high loading. Now we show the TEM photographs of the as is, high loading C60 CNT fiber. Supplemental figure 53, 54 and 55 shows bright field zoomed out images of the lamellas cut by the FIB. We see striated CNT bundle regions and other granular regions oriented in the direction of fiber alignment. Some selected granular regions ranged in length (along the fiber axis) 3 to 9 µm; selected widths span 0.75 to 2 µm. Note that in this high loading C60 CNT fiber, we also observed the small inclusions that were present in the previous sample.

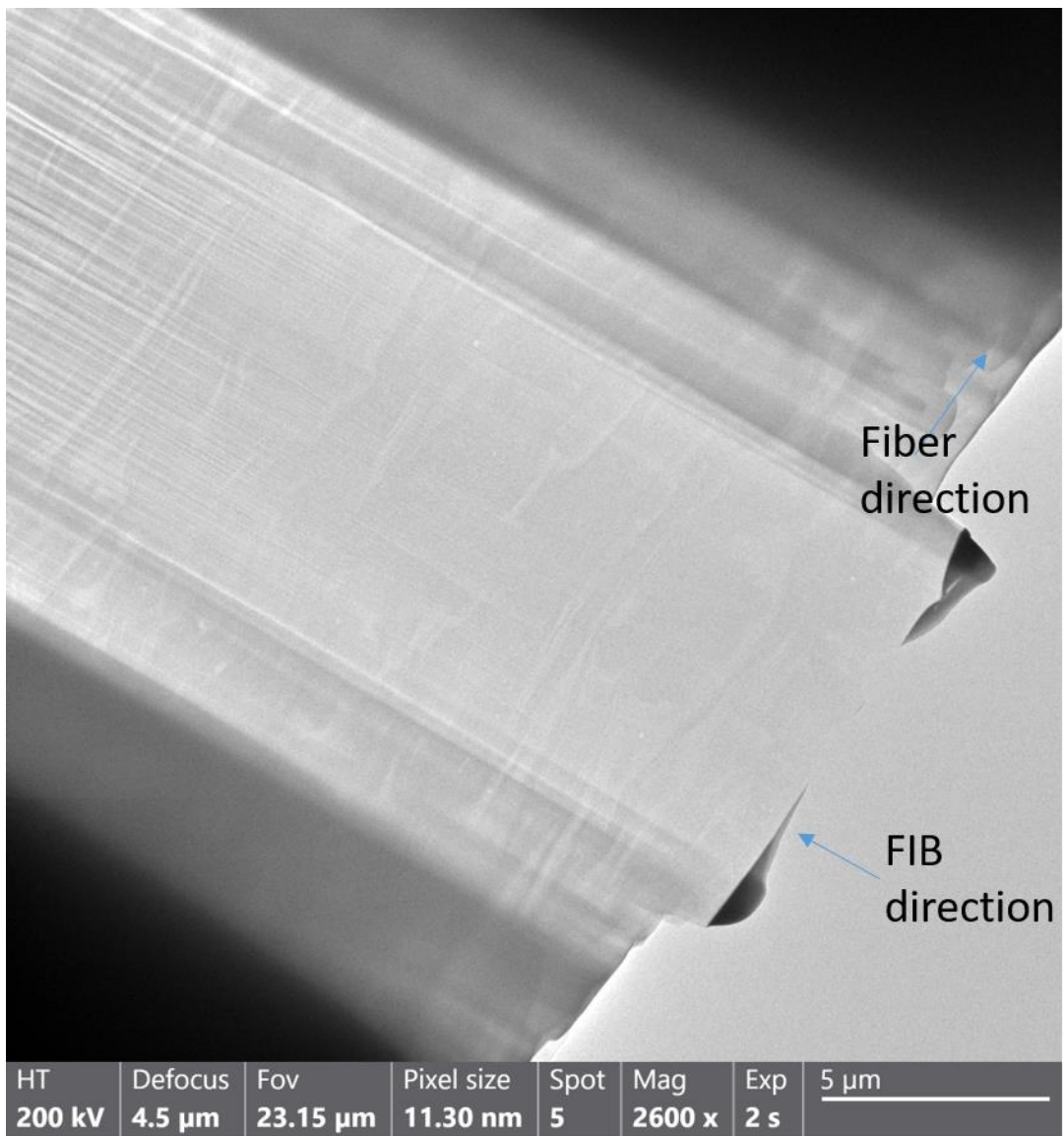

Supplemental figure 53. For the as is high loading C60 CNT fiber, bright field image of the lamella.

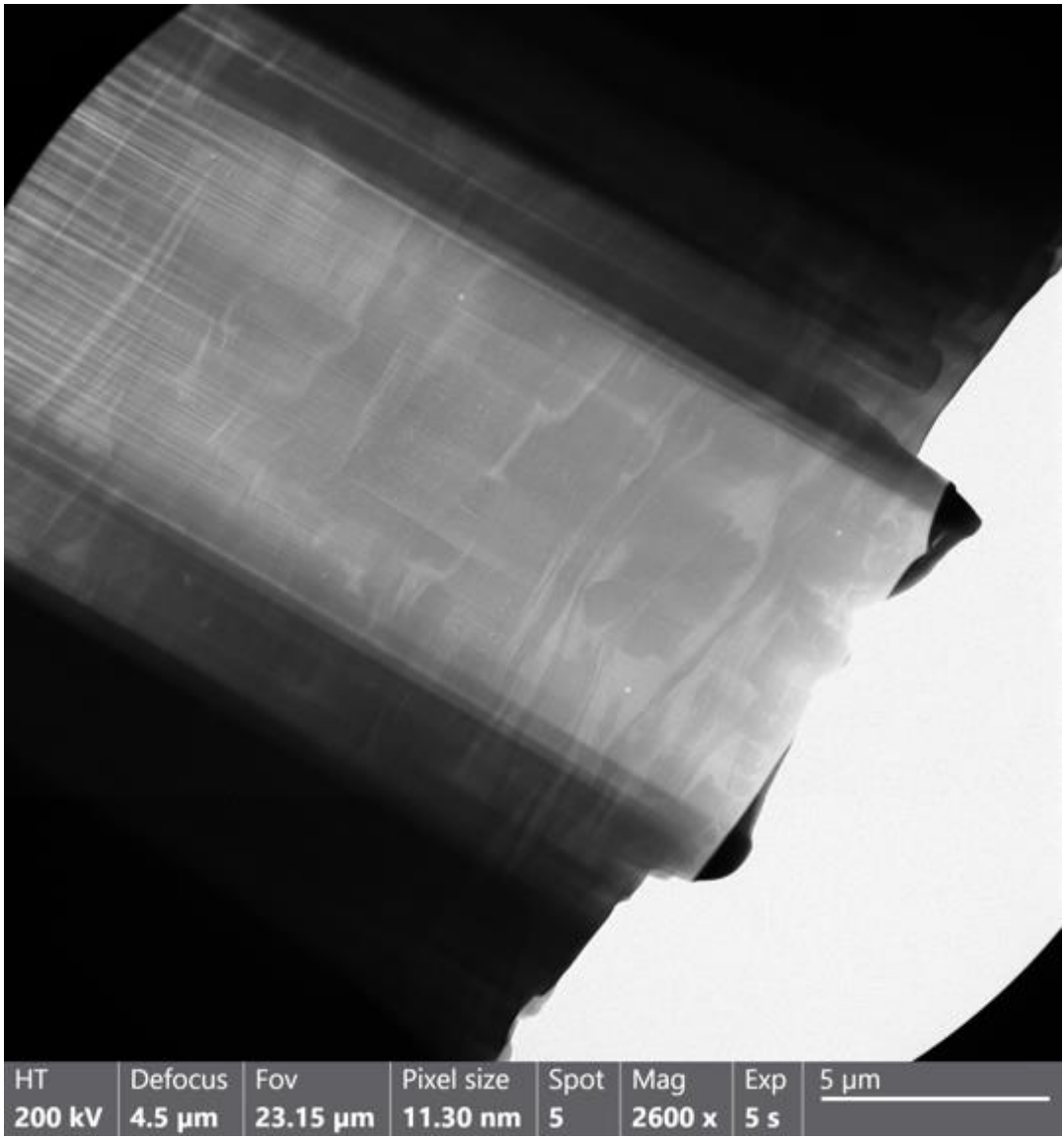

Supplemental figure 54. The same image as Supplemental figure 53, although now with a 30 µm objective aperture to better show the inclusions. Interestingly the granular inclusions do not show different contrast than the CNTs.

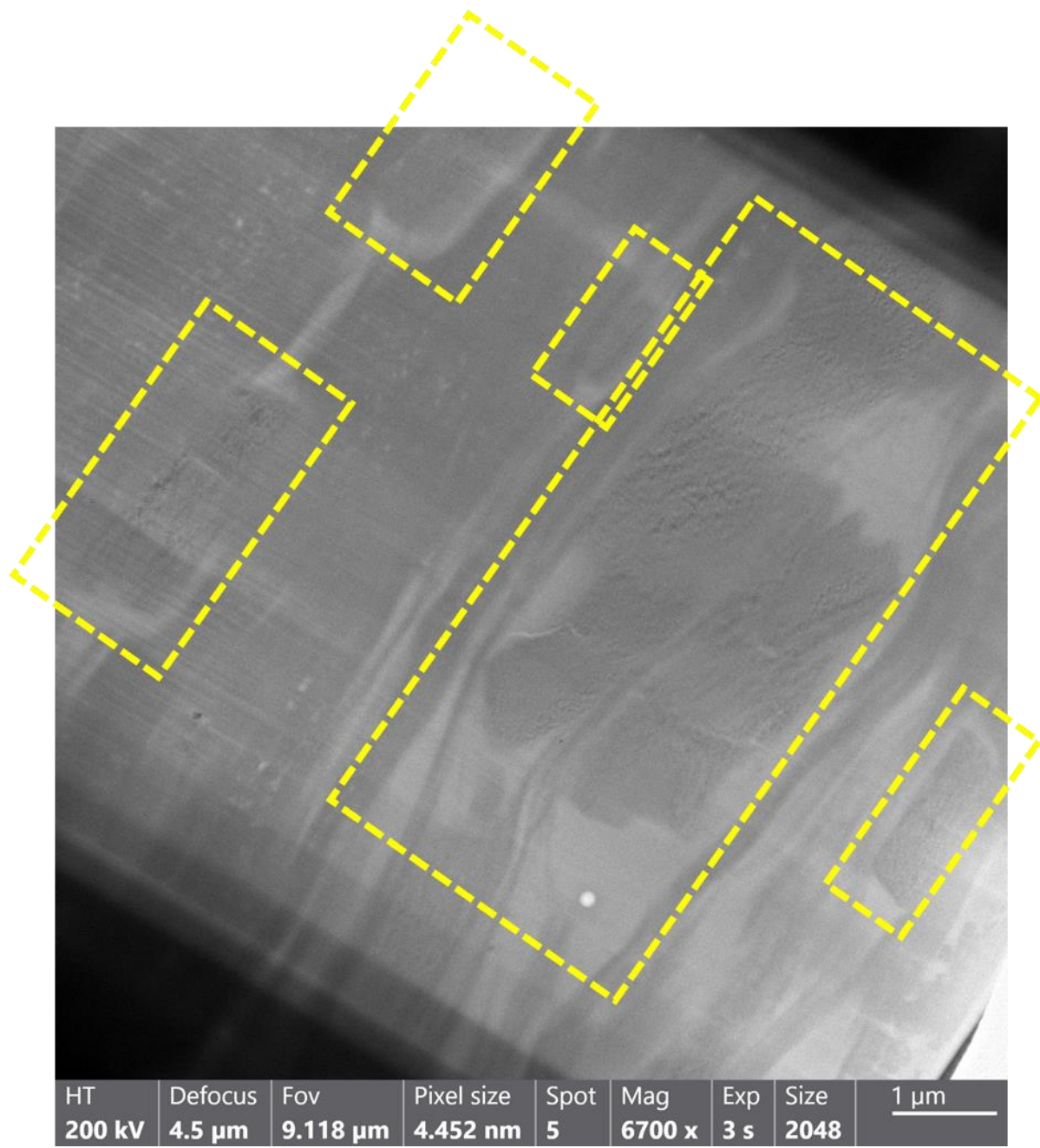

Supplemental figure 55. The same image as Supplemental figure 53, although now with the largest granular inclusions highlighted in yellow.

Supplemental figure 56 shows that the diffraction pattern of these larger granular inclusions are discrete and crystalline (unlike the amorphous response of the small inclusions in the low loading C60 CNT fiber). For the striated regions, the electron diffraction pattern was typical for CNT diffraction. This is better shown in the high-resolution TEM photographs and their Fourier transforms (Supplemental figure 57 - 59). The striated regions of the CNTs become more obvious and a periodic lattice structure is observed for the large granular inclusion. A Fourier transform of this image reveals a six-fold symmetry and a spacing of 5.1 Å.

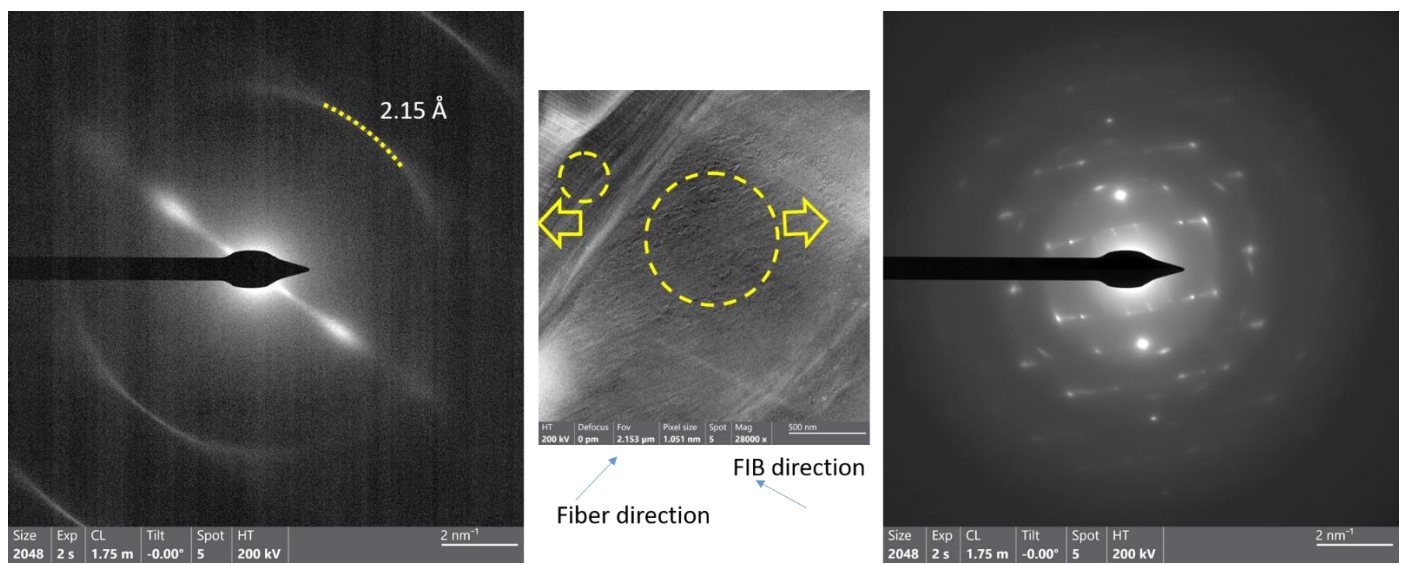

Supplemental figure 56. Electron diffraction patterns of the yellow highlighted regions, showing that the striated regions have the typical CNT diffraction pattern (left) and the large rectangular inclusions have a complex and discrete diffraction pattern (right).

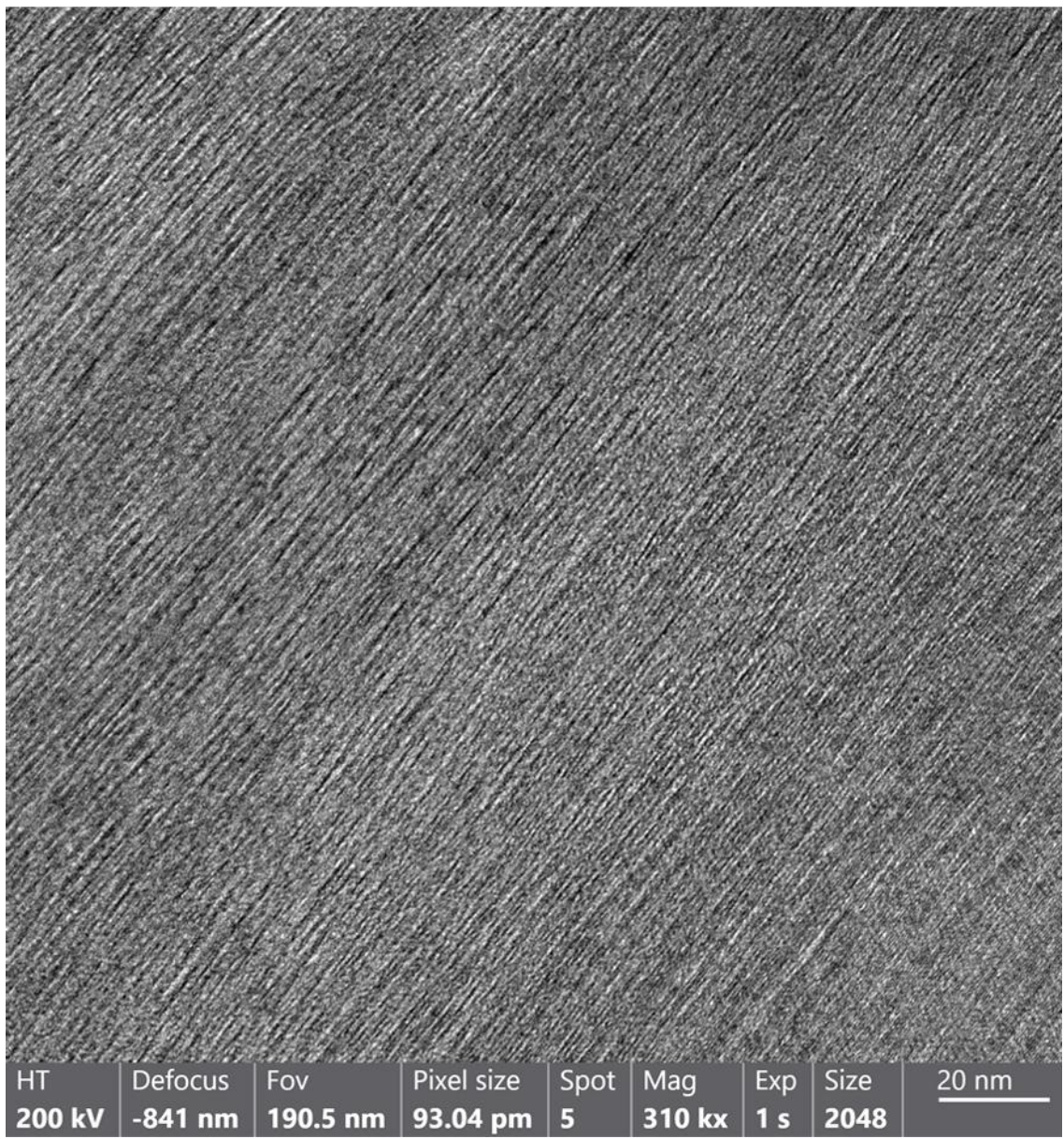

Supplemental figure 57. High resolution TEM image of just a CNT region. The objective aperture was inserted, so all of the images will not contain smaller spacings than about 3 to 4 Å.

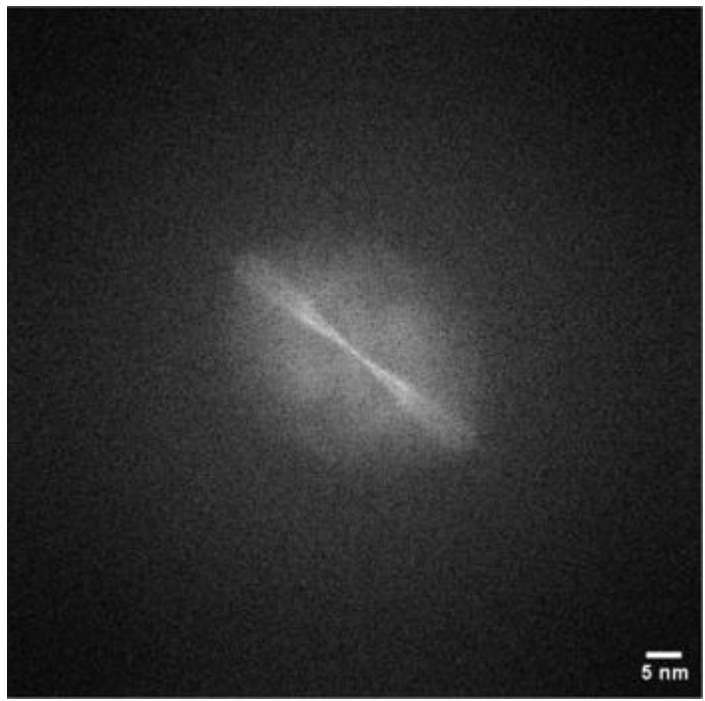

Supplemental figure 58. Fourier transform of the high-resolution TEM photograph of just the striated region, returning the typical aligned CNT response.

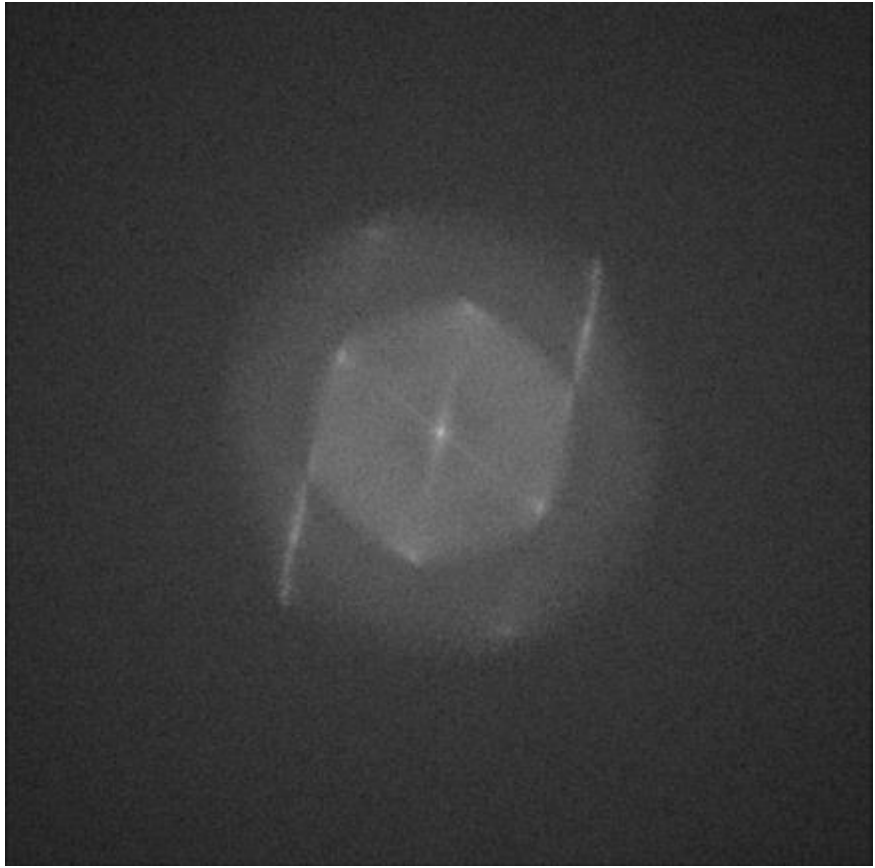

Supplemental figure 59. Fourier transform of the high-resolution TEM photograph of the large granular inclusion.

A plasma cleaning further thinned the lamella of the as is high load C60 CNT fiber. This is shown in supplemental figure 60, where the yellow highlighted area indicates forthcoming diffraction and high resolution TEM analysis. The lamella was sufficiently thinned to show a single crystal diffraction patten showing forming a hexagon (supplemental figure 61). This determined that the crystal had a face centered cubic (FCC) structure with spacings between opposing hexagonal edges 4.8 to 5.0 Å. Spacing between inner diffraction spots ranged from 8.4 to 9.3 Å. Supplemental figure 62 - 64 shows high resolution TEM of this thinned region, showing hexagonal packing and alignment of the C60 into linear rows.

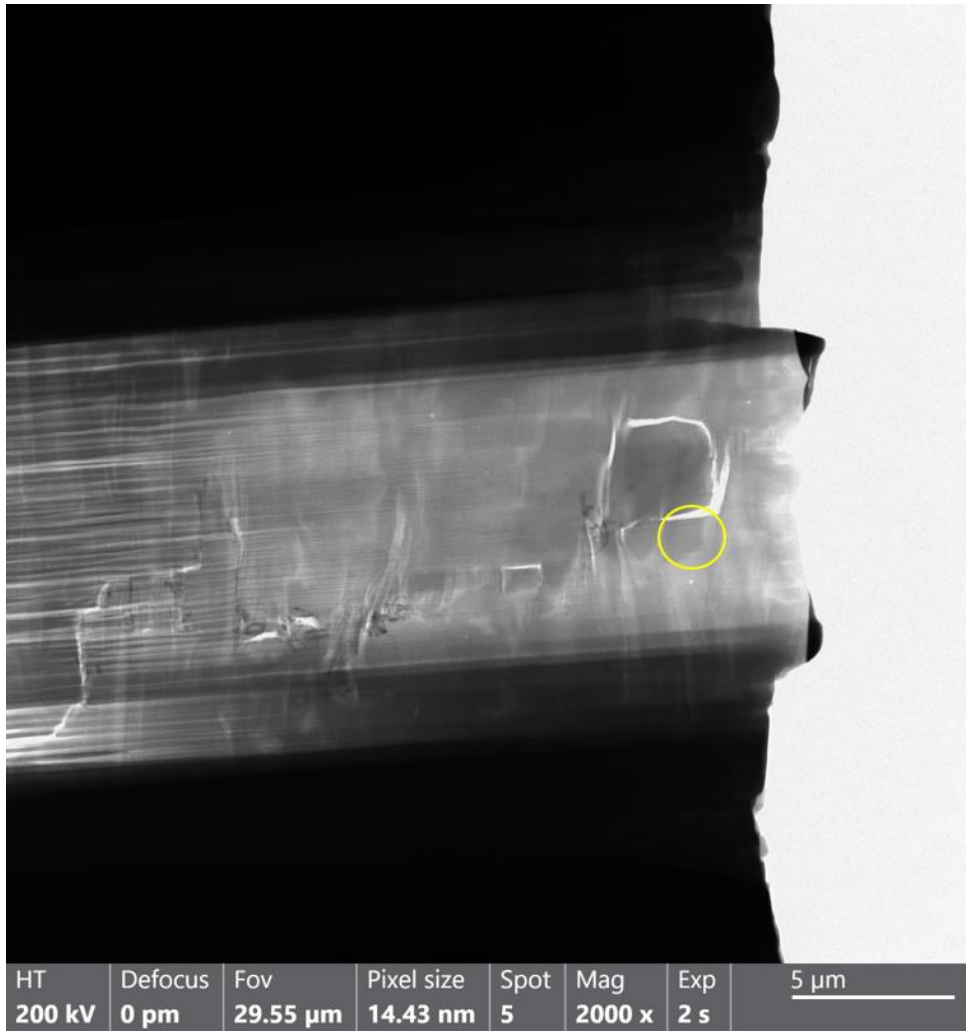

Supplemental figure 60. Highlighted circle indicates thinned region for viewing of the single crystallographic layer of a large granular inclusion.   Diffraction contrast highlights the large inclusions.

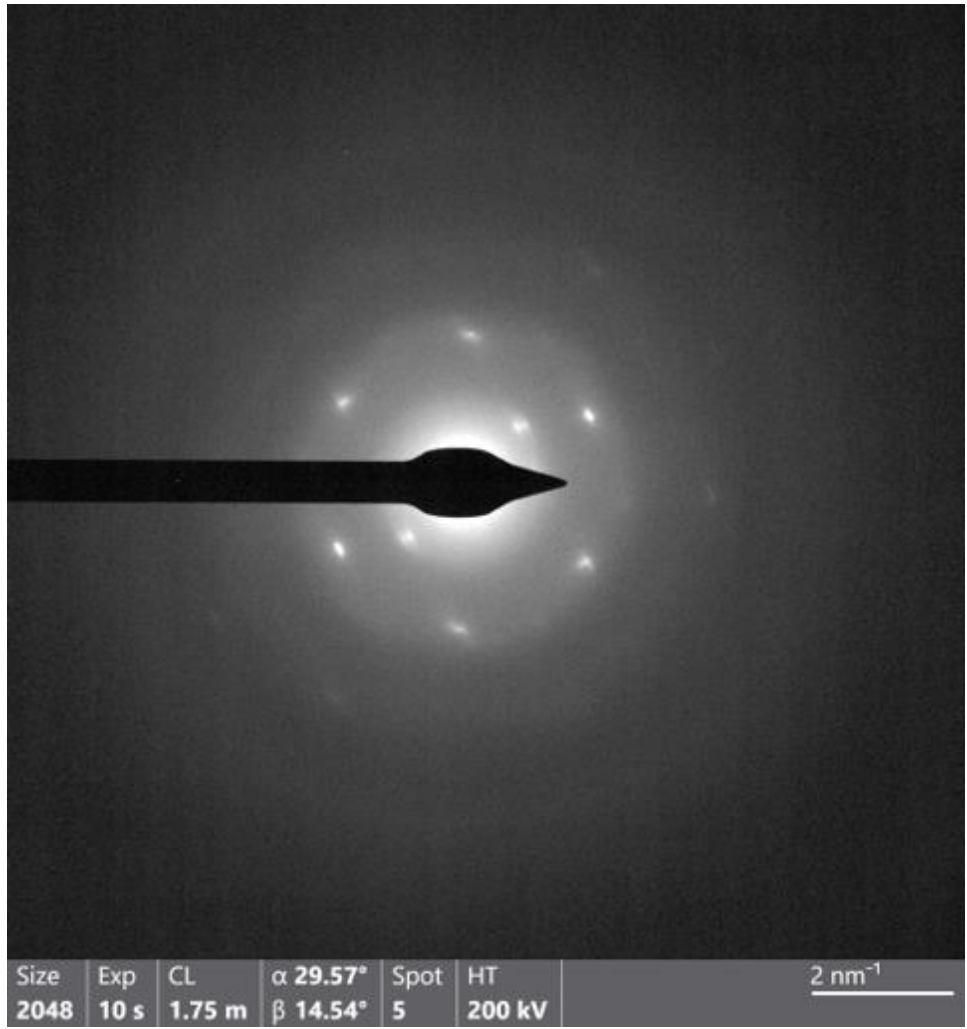

Supplemental figure 61. Single crystal diffraction pattern of the highlighted area in the thinned large granular inclusion. There is specific α and β tilt to align graph onto zone axis.

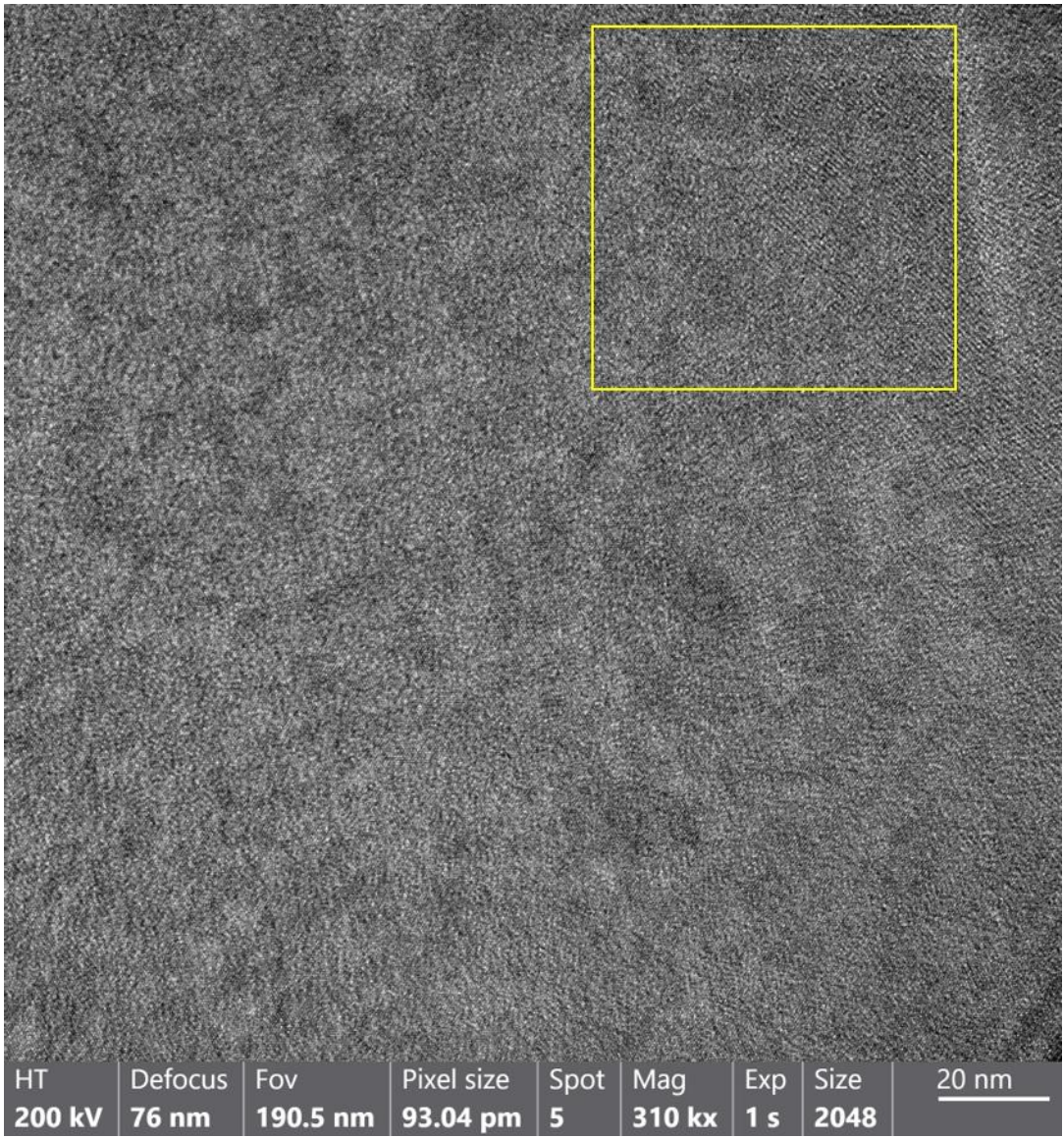

Supplemental figure 62. High resolution TEM of the highlighted thinned large granular inclusion.

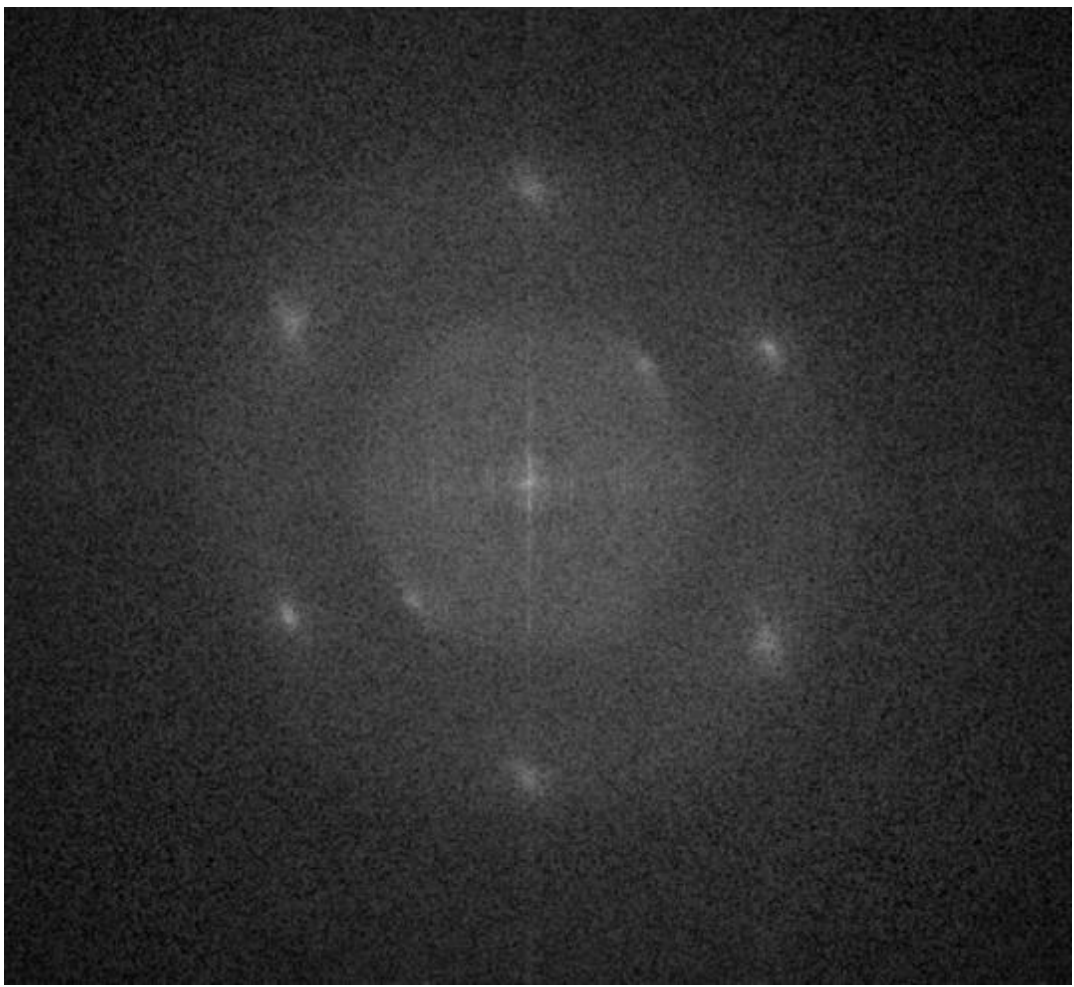

Supplemental figure 63. Fourier transform of the high resolution TEM image above.

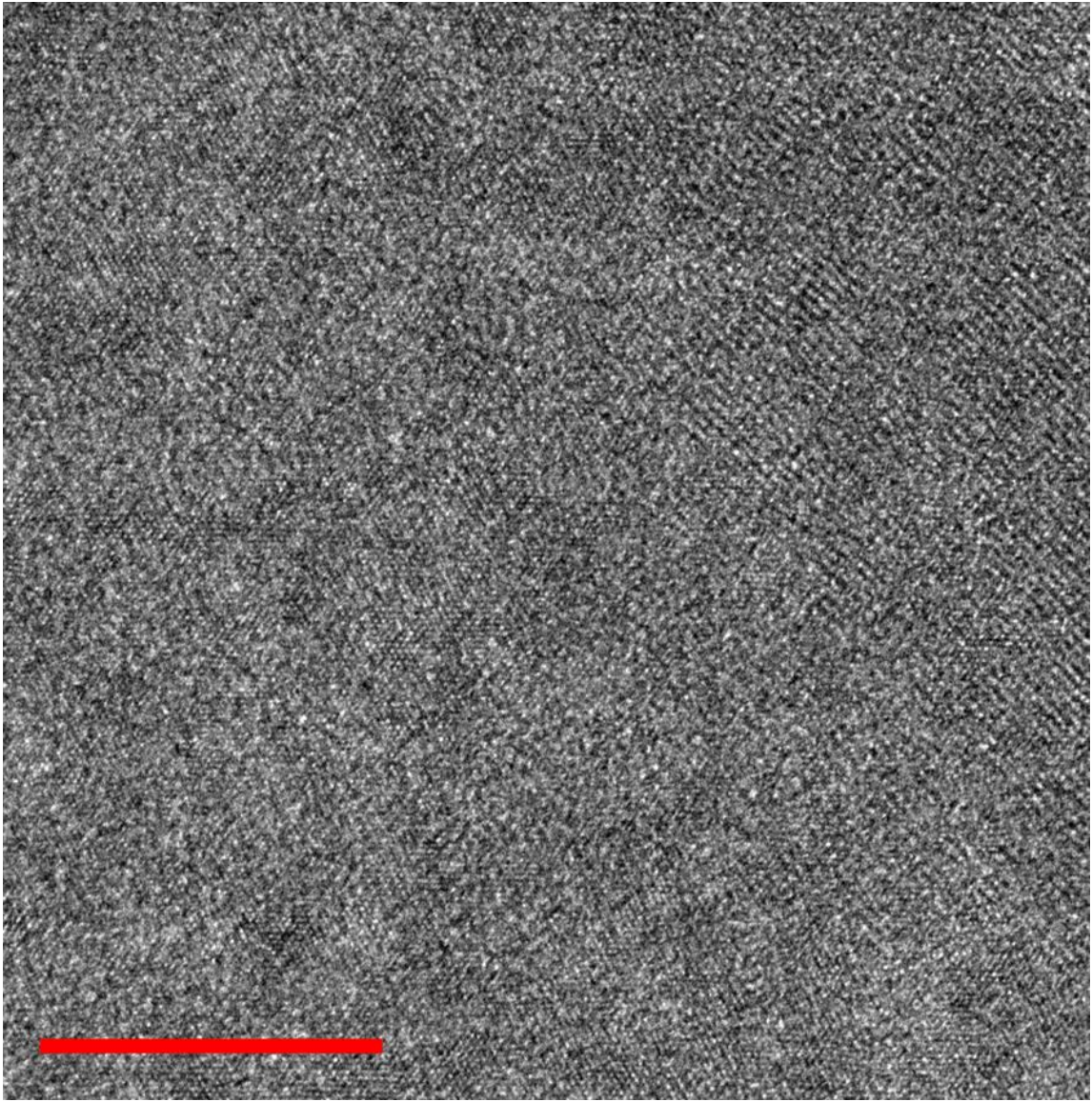

Supplemental figure 64. Further zoom-in on the high resolution TEM image of the thinned large granular inclusion, showing hexagonal packing with ~5 Å spacing and particular linear features in the top right. Surface contamination/damage from preparation limits seeing the hexagonal spacing throughout the entire full region. The red bar is 20 nm across.

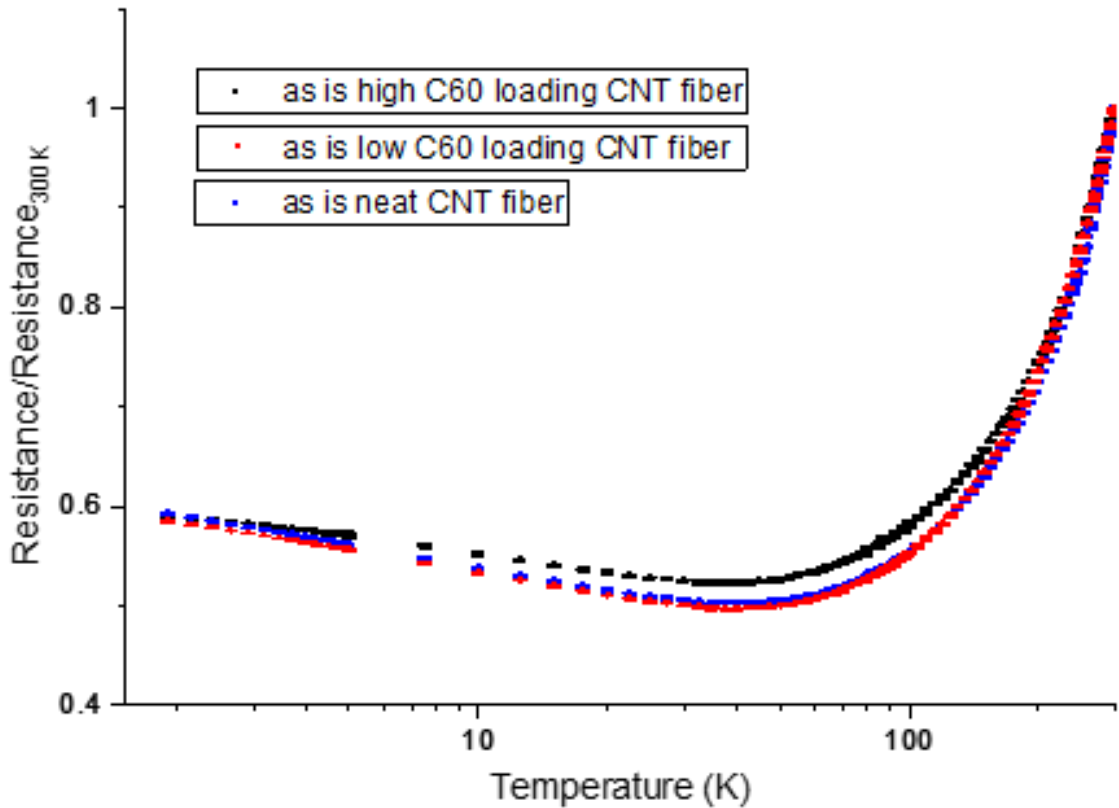

Supplemental figure 65. Four probe resistance measurements of the neat CNT, low loading, and high loading fiber, all in the as is state. Resistance is normalized to the value at 300 K. Measurement was conducted with a Quantum Design Physical Property Measurement System (PPMS) using 20 µA probe current. It was verified sample heating was not an issue. The room temperature, four probe resistance of the neat CNT, low loading, and high loading fiber are 1.2 Ω, 1.6 Ω, 0.8 Ω respectively. The conductivity and specific conductivity are given in the table above. These values can be used to reverse the normalization of these traces as desired by the reader.